\documentstyle[eqsecnum,prd,aps,epsf]{revtex}

\draft
\begin{document}

\newcommand{\beq}{\begin{equation}}
\newcommand{\eeq}{\end{equation}}
\newcommand{\beqn}{\begin{eqnarray}}
\newcommand{\eeqn}{\end{eqnarray}}
\newcommand{\pa}{\partial}
\newcommand{\vp}{\varphi}
\newcommand{\varep}{\varepsilon}
\newcommand{\ep}{\epsilon}

\twocolumn[\hsize\textwidth\columnwidth\hsize\csname
@twocolumnfalse\endcsname

\begin{center}
{\large\bf{Simulation of 
merging binary neutron stars in full general relativity: 
$\Gamma=2$ case}}
~\\
~\\
Masaru Shibata$^{1,2}$ and K\=oji Ury\=u$^3$\\
~\\
$^1${\em Department of Physics, 
University of Illinois at Urbana-Champaign, Urbana, IL 61801, USA} \\
$^2${\em Department of Earth and Space Science,~Graduate School of
 Science,~Osaka University,\\
Toyonaka, Osaka 560-0043, Japan} \\
$^3${\em SISSA, Via Beirut 2/4, 34013 Trieste, Italy} 
\\
\end{center}

\begin{abstract}
~\\
We have performed 3D numerical simulations for merger of equal mass 
binary neutron stars in full general relativity. 
We adopt a $\Gamma$-law 
equation of state in the form $P=(\Gamma-1)\rho\varepsilon$ 
where $P$, $\rho$, $\varep$ and $\Gamma$ are the pressure, rest mass density, 
specific internal energy, and the adiabatic constant with $\Gamma=2$. 
As initial conditions, we adopt models of corotational and irrotational 
binary neutron stars in a quasi-equilibrium state 
which are obtained using the conformal flatness approximation for 
the three geometry as well as an assumption that a helicoidal Killing 
vector exists. In this paper, we pay particular attention to the 
final product of the coalescence. We find that the final product 
depends sensitively on the initial compactness parameter of 
the neutron stars : 
In a merger between sufficiently compact neutron stars, a black hole is 
formed in a dynamical timescale.  As the compactness is decreased, 
the formation timescale becomes longer and longer.  
It is also found that a differentially rotating 
massive neutron star is formed instead of a black hole 
for less compact binary cases, in which 
the rest mass of each star is less than $70-80\%$ of the 
maximum allowed mass of a spherical star. 
In the case of black hole formation, we roughly evaluate the 
mass of the disk around the black hole. For the merger 
of corotational binaries, 
a disk of mass $\sim 0.05-0.1M_*$ may be formed, where $M_*$ 
is the total rest mass of the system. On the other hand, 
for the merger of irrotational binaries, 
the disk mass appears to be very small : $< 0.01M_*$. 
\end{abstract}
\vskip2pc]

\baselineskip 5.7mm

\section{Introduction}

Several neutron star-neutron star binaries are known to exist 
in our galaxy \cite{PULSAR}.  
According to high precision measurements of the general relativistic (GR) 
effects on their orbital motion, 
three of these binaries are going to merge as a result of 
gravitational radiation 
reaction within the Hubble timescale $\sim 10^{10}$ years. 
What are the final fates of these binaries after the mergers? 
Their total gravitational masses are  
is in a narrow range $\sim 2.65-2.85M_{\odot}$ where 
$M_{\odot}$ denotes the solar mass. 
The stars will not be tidally disrupted before the merger, 
since the masses of the two stars in each binary are nearly equal.  
Hence, the mass loss from the binary systems is expected to be small 
during their evolution and the mass 
of the merged object will be approximately equal to the initial mass. 
The maximum allowed gravitational mass for spherical neutron stars is 
in a range $\sim 1.5-2.3M_{\odot}$ depending on 
the nuclear equation of state \cite{ST,EOS}. 
Even if we take into account the effect 
of {\it rigid} rotation, it is increased by at most 
$\sim 20\%$ \cite{FIP}. Judging 
from these facts, the compact objects formed just 
after the merger of 
these binary systems seem bound to collapse to a black hole.  

However, this is not the case if the merged object rotates 
{\it differentially}. 
The maximum allowed mass can be increased by a larger factor 
($> 50\%$) due to the differential rotation \cite{BSS0}, 
which suggests that 
the merged objects of $\sim 2.5-3M_{\odot}$ may be dynamically stable 
against gravitational collapse to a black hole. 
Such differentially rotating stars could be 
secularly unstable, since viscosity or magnetic field 
could change the differential rotation 
into rigid rotation. A star with a high ratio 
of rotational energy to the gravitational binding energy 
could also be secularly unstable to gravitational wave emission 
\cite{SF}. 
These processes might dissipate or redistribute 
the angular momentum, and 
induce eventual gravitational collapse to a black hole. 
However, the timescales 
for such secular instabilities are in general 
much longer than the dynamical timescale of the system. 
Hence, the merged objects may not 
collapse to a black hole promptly, but remain as a massive 
neutron star supported by differential rotation 
at least for these secular timescales. 
These facts imply that the final product of the merger of binary 
neutron stars is an open question depending not only 
on the nuclear equation of state for high density neutron matter 
but also on the rotational profile of the merged object. 

Interest in the final product of binary coalescence 
has been stimulated by the prospect 
of future observation of extragalactic binary 
neutron stars by gravitational wave detectors. A statistical study 
shows that mergers of binary neutron stars 
may occur at a few events 
per year within a distance of a few hundred Mpc \cite{phinney}. 
This suggests that binary merger is a 
promising source of gravitational waves. 
Although the frequency of gravitational waves in the 
merging regime will be larger than $1$kHz and 
lies beyond the upper end of the frequency 
range accessible to laser interferometers such as LIGO \cite{LIGO} 
for a typical event at a distance $\sim 200$Mpc,  
it may be observed using specially designed 
narrow band interferometers or resonant-mass detectors \cite{KIP}. 
Such future observations will provide valuable information about 
the merger mechanism of binary neutron stars and the final products. 

Interest has also been stimulated by a hypothesis about 
the central engine of $\gamma$-ray bursts (GRBs) \cite{piran}. Recently, 
some of GRBs have been shown to be of cosmological 
origin \cite{grb}. In cosmological GRBs, 
the central sources must supply 
a large amount of the energy $\agt 10^{51}$ ergs in a 
very short timescale of order from a millisecond to a second. 
It has been suggested that 
the merger of binary neutron stars is a likely candidate 
for the powerful central source \cite{piran}. 
In the typical hypothetical scenario, 
the final product should be a system composed of a rotating 
black hole surrounded by a massive disk of mass $>0.1M_{\odot}$, 
which could supply the large amount of energy by neutrino processes 
or by extracting the rotational energy of the black hole. 

To investigate the final product of the merger theoretically, 
numerical simulation appears to be the unique promising approach. 
Considerable effort has been made for this in the 
framework of Newtonian and post-Newtonian gravity 
\cite{NO,SNO,RS,Centrella,ruffert,davis}. 
Although these simulations have clarified a wide variety 
of physical features which are important during the 
coalescence of binary neutron stars, a fully 
GR treatment is obviously necessary for determining the final product 
because GR effects are crucial. 

Intense efforts have been made for constructing a reliable 
code for 3D hydrodynamic simulation 
in full general relativity in the past decade \cite{supple,ON,waimo,gr3d}. 
Recently, Shibata presented a wide variety of
numerical results of test problems for his fully GR code and
showed that simulations for many interesting problems are now
feasible \cite{gr3d}.
 
To perform a realistic simulation, we also need 
realistic initial conditions for the merger, i.e., a realistic 
density distribution and velocity field for the component stars. 
Since the timescale of gravitational wave emission 
is longer than the orbital period, we may consider the 
binary neutron stars to be in a quasi-equilibrium state even 
just before the merger. 
The velocity field in the neutron stars has been turned out to 
be nearly irrotational because (1) the shear viscosity is 
too small to redistribute angular momentum to produce a 
corotational velocity field 
during the timescale of gravitational wave emission and (2) 
the initial vorticity of each star is negligible 
as long as the rotational period of the neutron stars is not 
$\sim$ milliseconds \cite{KBC}. 
Therefore, as realistic initial conditions, we 
should prepare a quasi-equilibrium state of binary 
neutron stars with an irrotational velocity field. 
Recently, several groups have developed 
various numerical methods to compute GR irrotational 
binary neutron stars in quasi-equilibrium in a framework with 
the appropriate approximations that a helicoidal Killing 
vector (see Eq. (\ref{helic}) for definition) 
exists and that the three geometry is conformally flat \cite{BGM,UE,MMW}.  
Their numerical results have been compared and found to agree to within 
a few percent error in the gravitational mass and the central density as 
a function of orbital separation. 

In this paper, 
we perform simulations for the merger of binary neutron stars of 
equal mass 
by using these new numerical implementations developed recently. 
As a first step, 
we adopt a simple $\Gamma$-law equation of state with  
$\Gamma=2$ as a reasonable qualitative approximation 
to the high density nuclear equation of state. 
Although microscopic effects such as 
cooling due to neutrino emission or heating due to bulk viscosity may 
become important for discussing the merging process in detail 
\cite{ruffert}, we neglect them here for simplicity. 
The purpose of this paper is to investigate the dynamical nature 
of the mergers, the final products after the mergers and 
the dependence of these outcomes on the initial velocity field and 
the compactness parameter of the binary neutron stars. 
Simulations are performed not only for irrotational binaries but also 
for corotational ones to clarify the difference due to the initial 
velocity field. 

Throughout 
this paper, we adopt the units $G=c=M_{\odot}=1$ where $G$ and $c$
denote the gravitational constant and speed of light, respectively.  
Latin and Greek indices denote spatial components ($1-3$) 
and space-time components ($0-3$), respectively. 
$\delta_{ij}(=\delta^{ij})$ denotes the Kronecker delta. 
We use Cartesian coordinates $x^k=(x, y,
z)$ as the spatial coordinates with $r=\sqrt{x^2+y^2+z^2}$; $t$ denotes
coordinate time.

\section{Methods}

\subsection{Summary of formulation}

We have performed numerical simulations using the same formulation as
in~\cite{gr3d}, to which the reader may refer for details about 
the basic equations, the gauge conditions and the computational method. 
The fundamental variables used in this paper are:
\beqn
\rho &&:{\rm rest~ mass~ density},\nonumber \\
\varep &&: {\rm specific~ internal~ energy}, \nonumber \\
P &&:{\rm pressure}, \nonumber \\
u^{\mu} &&: {\rm four~ velocity}, \nonumber \\
v^i &&= {u^i \over u^0},\nonumber \\
\alpha &&: {\rm lapse~ function}, \nonumber \\
\beta^k &&: {\rm shift~ vector}, \nonumber \\
\gamma_{ij} &&:{\rm metric~ in~ 3D~ spatial~ hypersurface},\nonumber \\ 
\gamma &&
={\rm det}(\gamma_{ij})=e^{12\phi}=\psi^{12}, \nonumber \\
\tilde \gamma_{ij}&&=e^{-4\phi}\gamma_{ij}, \nonumber \\
K_{ij} &&:{\rm extrinsic~curvature}.\nonumber 
\eeqn
These variables (together with auxiliary functions $F_i$ and the
trace of the extrinsic curvature $K_k^{~k}$) are numerically evolved 
as an initial value problem (see \cite{gw3p,gw3p2} for details of 
the numerical method for handling the evolution equations and 
initial value equations). Several test
calculations, including spherical collapse of dust, stability 
of spherical neutron stars, and the evolution of rotating neutron
stars as well as corotational binary systems have been 
presented in~\cite{gr3d}.  
Violations of the Hamiltonian constraint \cite{footH} and conservation of
rest mass and angular momentum \cite{footJ} are monitored to check 
the accuracy, 
and we stop simulations before the accuracy becomes too poor. 
Black holes which are formed during the last phase of the
merger are located with an apparent horizon finder described
in~\cite{S}.

We also define a density $\rho_*(=\rho \alpha u^0 e^{6\phi})$ from
which the total rest mass of the system can be integrated as
\beq
M_*=\int d^3 x \rho_*. 
\eeq
We have performed the simulations using a fixed uniform grid with 
a typical size of $233 \times 233 \times 117$ for the $x-y-z$ 
directions, respectively, and assuming 
reflection symmetry with respect to the $z=0$ plane. 
All of the results shown in Sec. IV are obtained from simulations 
with this grid size. 
We have also performed a number of test simulations 
with a smaller grid size of $193 \times 193 \times 97$ 
changing the grid spacing and location of the outer boundaries 
to confirm that the results do not change significantly. 
For one model (I2) (see Sec. III and Table I), we performed 
a simulation with a larger grid size of 
$293 \times 293 \times 147$, widening the 
computational domain to investigate the effect of the outer 
boundary on the gravitational waveforms. 

The slicing and spatial gauge conditions which we use in this paper 
are basically the same as those adopted in~\cite{gw3p,gw3p2,gr3d}; 
i.e., we impose an ``approximate'' maximal slicing condition 
($K_k^{~k} \simeq 0$)
and an ``approximate'' minimum distortion gauge condition ($\tilde D_i
(\pa_t \tilde \gamma^{ij}) \simeq 0$ where $\tilde D_i$ is the
covariant derivative with respect to $\tilde \gamma_{ij}$). 
However,
for the cases when a merged object collapses to form a black hole, we
slightly modify the spatial gauge condition in order to improve the
spatial resolution around the black hole forming region.
The method of the modification is described in Sec. II.B.

Throughout this paper, we assume a $\Gamma$-law equation of state in
the form
\beq
P=(\Gamma-1)\rho \varep,
\eeq 
where $\Gamma$ is the adiabatic constant. For the hydrostatic problem, 
which appears in solving for initial value configurations, 
the equation of state can be rewritten in the polytropic form
\beq
P = K \rho^{\Gamma}, \mbox{~~~~~} \Gamma = 1 + \frac{1}{n} \label{eos},
\eeq
where $K$ is a constant (different from $K_k^{~k}$)
and $n$ is the polytropic index.  We
adopt $\Gamma=2$ ($n = 1$) as a qualitative approximation
to realistic, moderately stiff equations of state for neutron stars.

Physical units enter only through the constant
$K$, which can be chosen arbitrarily or completely scaled out of
the problem.  In the following, 
we quote values for $K = 200/\pi$, for which in
our units ($G = c = M_{\odot} = 1$) the radius of a spherical star 
is $R = (\pi K/2)^{1/2} =10 (\sim 15{\rm km})$ 
in the Newtonian limit. 
Since
$K^{n/2}$ has units of length, dimensionless variables can be
constructed as
\beqn \label{rescale}
\bar M_* = M_* K^{-n/2}, & \bar M_g =  M_g K^{-n/2}, & 
\bar R = R K^{-n/2}, \nonumber  \\
\bar J = J K^{-n},  & \bar P_{\rm orb} = P_{\rm orb} K^{-n/2}, &
\bar{\rho} = \rho K^n,
\eeqn
where $M_g$, $J$ and $P_{\rm orb}$  
denote the gravitational mass, angular momentum 
and orbital period.  All results can be rescaled for
arbitrary $K$ using Eqs.~(\ref{rescale}). 
(For example, the maximum value of $M_*$ for spherical stars is 
$1.435M_{\odot}$ for $K=200/\pi(G^3M_{\odot}^2/c^4)$ 
with the rest mass density 
$\simeq 3.1\times 10^{15}{\rm g/cm}^3$. 
If we want to choose $M_*$ as $2M_{\odot}$, we should change $K$ to 
$123.7(G^3M_{\odot}^2/c^4)$ and the corresponding density is then
$\simeq 1.6\times 10^{15}{\rm g/cm}^3$; {\it cf}. Fig. 1. )
In other words, the invariant quantities are only the 
dimensionless quantities such as $M_*/M_g$, $M_g/R$, $M_g/P_{\rm orb}$ 
and $J/M_g^2$. 

\subsection{Spatial gauge condition}

When a black hole is not formed as a final product of the merger, 
we adopt the approximate minimum 
distortion gauge condition as our spatial gauge condition (henceforth
referred to as the AMD gauge condition). However, as pointed out in
previous papers \cite{gr3d,gw3p2}, during black hole formation
(i.e., for an infalling radial velocity field), the expansion of the
shift vector $\pa_i \beta^i$ and $\pa_t \phi \simeq \pa_i \beta^i/6$ 
become positive using this gauge condition together with our 
slicing $K_k^{~k}\simeq 0$. 
Accordingly, the coordinates diverge outwards and the
resolution around the black hole forming region becomes worse and
worse during the collapse.  This undesirable property motivates us to
modify the AMD gauge condition when we treat black hole formation.
Following \cite{gw3p2}, 
we modify the AMD shift vector as 
\beq
\beta^i=\beta^i_{\rm AMD}-f(t,r){x^i \over r+\epsilon}
\beta^{r'}_{\rm AMD}.
\eeq
Here $\beta^i_{\rm AMD}$ denotes the shift vector 
for the AMD gauge condition, $\beta^{r'}_{\rm AMD}
\equiv x^k\beta^k_{\rm AMD}/(r+\epsilon)$, $\epsilon$ is a 
small constant much smaller than the grid spacing, 
and $f(t,r)$ is a function chosen specifically as
\beq
f(t,r)=f_0(t){1 \over 1+(r/3M_{g0})^6}.
\eeq 
where $M_{g0}$ denotes the gravitational mass of a system 
at $t=0$.  
We determine $f_0(t)$ from $\phi_0=\phi(r=0)$. 
Taking into account the fact that the resolution around $r=0$ 
deteriorates when $\phi_0$ becomes large, we choose $f_0$ as 
\beq
f_0(t)=\left\{
\begin{array}{ll}
\displaystyle 
1 & {\rm for}~\phi_0 \geq 0.8,\\
2.5\phi_0 -1& {\rm for}~0.4 \leq \phi_0 \leq 0.8,  \\
0 & {\rm for}~\phi_0 < 0.4, \\
\end{array}
\right. \label{type1}
\eeq
Note that for spherical collapse with $f_0 = 1$, $\pa_i \beta^i \simeq 
0$ and $\pa_t \phi \simeq 0$ at $r = 0$.  
We employ this modified gauge condition whenever
a merged object collapses to form a black hole.

It is worth mentioning that with this modification, 
the coordinate radius of the apparent horizon 
(when it is formed) becomes larger than 
without the modification. This implies that 
more grid points are located along the radius of the apparent horizon 
and accuracy for determination of the apparent horizon is improved. 

\section{Initial conditions}

Even just before the merger, the binary neutron stars are 
considered to be in a quasi-equilibrium state because 
the timescale of gravitational radiation reaction 
$\sim 5/\{64\Omega(M_g\Omega)^{5/3}\}$ \cite{ST}, 
where $\Omega$ denotes the orbital angular 
velocity of the binary neutron stars, 
is several times longer than the orbital period. Thus, 
for performing a realistic simulation of the merger, 
we should prepare a quasi-equilibrium state as the initial condition. 
In this paper, we construct such initial conditions as follows. 

First, we assume the existence of a helicoidal Killing vector 
\beq
\ell^{\mu}=(1, -y\Omega, x\Omega, 0).\label{helic}
\eeq
Since emission of gravitational waves violates the 
helicoidal symmetry, this assumption does not strictly hold 
in reality. However, as mentioned 
above, the emission timescale of gravitational waves 
is several times 
longer than the orbital period even just before the merger 
({\it cf}. Table I) so that this 
assumption can be acceptable for obtaining an approximate 
quasi-equilibrium state. 
In addition to this assumption, 
we adopt the so-called conformal flatness approximation 
in which the three geometry is assumed to be conformally flat, 
for simplicity. 

In this paper, we consider irrotational and corotational 
binary neutron stars. Then, 
the geometric \cite{BCSST} and hydrostatic equations \cite{irre}
for solutions of the quasi-equilibrium states are described as
\beqn
&& \Delta \psi = -2\pi (\rho h w^2-K\rho^{\Gamma}) \psi^5 -{\psi^5 \over 8}
\delta^{ik}\delta^{jl}L_{ij} L_{kl}, \label{eq302} \\
&& \Delta (\alpha\psi) = 2\pi \alpha \psi^5 
[\rho h (3w^2-2)+5K \rho^{\Gamma}] \nonumber \\
&& \hskip 1.5cm +{7\alpha\psi^5 \over 8}
\delta^{ik}\delta^{jl} L_{ij} L_{kl} ,\\
&& \delta_{ij} \Delta \beta^i + {1 \over 3} \beta^k_{~,kj}
-2L_{jk}\delta^{ki} \Bigl(
\pa_i \alpha - {6\alpha \over \psi} \pa_i \psi \Bigr) 
\label{quasibet}\nonumber \\
&& \hskip 3cm =16\pi\alpha \rho h w u_j,\label{eq304}\\
&& {\alpha h \over w} +h u_k V^k 
={\rm const.},\label{eq305}
\eeqn
where 
\beqn
&&w=\alpha u^0=\sqrt{1+\psi^{-4}\delta^{ij}u_iu_j}, \\
&&h=1+K\Gamma\rho^{\Gamma-1}/(\Gamma-1), \\
&&L_{ij}={1 \over 2\alpha}
\biggl(\delta_{jk} \pa_i \beta^k + \delta_{ik} \pa_j \beta^k 
- {2 \over 3}\delta_{ij}
\pa_k \beta^k\biggr),\\
&&V^k=-\beta^k + \delta^{kl} {\alpha u_l \over w\psi^4} - \ell^k. 
\eeqn
and $\Delta $ denotes the flat Laplacian in the three space. 
Eqs. (\ref{eq302}--\ref{eq304}) are the geometric equations and 
Eq. (\ref{eq305}) is the so-called Bernoulli equation. 
$V^k$ can be regarded as the coordinate three velocity in the 
corotating frame rotating with angular velocity $\Omega$. 

In the case of corotational binaries in which $V^k=0$, 
$u_i$ is written as
\beq
u_i=w\psi^4(\epsilon_{izk}\Omega x^k+\delta_{ij}\beta^j)/\alpha. 
\eeq
In the case of irrotational binaries, on the other hand, 
$u_i$ is written as 
\beq
u_i=h^{-1}\pa_i \Phi
\eeq
where $\Phi$ denotes the velocity potential which 
satisfies an elliptic PDE \cite{irre}
\beq
\delta^{ij} 
\pa_i(\rho\alpha \psi^2 h^{-1} \pa_j \Phi)
-\pa_i[\rho\alpha h^{-1}\psi^6(\ell^i+\beta^i)]=0, 
\eeq
with the following boundary condition 
at the stellar surface; 
\beq
\left. V^i\pa_i\rho\right|_{\rm surf} = 0. 
\eeq

The above Poisson type equations such as 
Eqs. (\ref{eq302}--\ref{eq305}) and (3.12) as well as the 
Bernoulli equation (\ref{eq305}) are solved iteratively 
with appropriate boundary conditions. 
Corotational binaries are calculated
using the same numerical method as adopted in \cite{gr3d}. 
Irrotational binaries are calculated
using the method developed recently by Ury\=u and Eriguchi 
(\cite{UE}, to which the reader may refer for details). 

For the corotational case, we prepare binaries with several compactness
parameters, with the surfaces of the two stars coming into contact. 
As shown in \cite{BCSST}, such binaries with $\Gamma=2$ 
are located near to the energy minimum along the 
sequence of corotational binaries of constant rest mass. 
Therefore, they are expected to be located near to a marginally stable 
point for hydrodynamic \cite{RS} or GR orbital instability. 

For the irrotational case, 
the sequence of binaries of constant rest mass ends when 
cusps (i.e., Lagrange (L1) points) 
appear at the inner edge of the stars \cite{BGM,UE}. 
This is the case for any compactness parameter. 
If the stars in the binary system approach further, 
mass transfer will begin and the resulting state is not clear. 
As shown in \cite{BGM}, 
the closest binaries with cusps 
are far outside the energy minimum for $\Gamma=2$, 
which indicates that 
they are stable against hydrodynamic and GR orbital instability. 
Furthermore, the gravitational 
radiation reaction timescale is several times longer than the 
orbital period ({\it cf}. Table I). 
Thus, if we choose such a binary as the initial condition 
for a simulation, 
a few orbits are maintained stably before the merger 
starts, decreasing the orbital separation and changing 
the shapes in a quasi-adiabatic manner. 

It is still difficult to perform an accurate simulation for such 
a quasi-adiabatic phase. 
It is desirable to choose a binary state 
which is located near to the unstable point 
against hydrodynamic or GR orbital instability and 
starts merging soon; 
i.e., a state after the nearly adiabatic phase. 
However, a method for obtaining 
such a state has not yet been developed. 
Hence, in this paper, we prepare the following initial conditions 
modifying the quasi-equilibrium state slightly. 
First, we prepare a binary in which 
cusps appear at the surfaces. Then, we reduce 
the angular momentum by $\simeq 2.5\%$ from the 
quasi-equilibrium state to destabilize 
the orbit and to induce the merger promptly \cite{footnote}. 
We deduce that such an initial condition can be acceptable 
for the investigation of the final products after the merger 
since the decrease factor is still small. 
We performed test simulations changing the decrease factor 
slightly in the range $2-3\%$, 
and we indeed found that the results shown in Sec. IV 
are only weakly dependent on this parameter. 

In the numerical computation of the quasi-equilibrium states 
as initial conditions, 
we typically adopt a grid spacing in which the major diameter of each star 
is covered by $\sim 35$ grid points. 
With this resolution, the error for the estimation of $M_g$ and $J$ 
is less than $1\%$ (e.g., \cite{UE}), and 
the gravitational radius of the system defined as $GM_{g0}/c^2$ 
is covered with $\sim 4-7$ grid points ({\it cf}. Table I). 
Keeping this grid spacing, 
the outer boundaries of the computational domain 
in the simulation are located at $\alt 0.3\lambda_{\rm gw}$ 
with $233 \times 233 \times 117$ grid points, 
where $\lambda_{\rm gw}(\equiv \pi/ \Omega)$ 
denotes the characteristic wavelength of gravitational 
waves emitted from the binaries in a quasi-equilibrium state. 
With this setting, gravitational waveforms are not accurately evaluated 
because the outer boundaries are not located in the wave zone. 
In this paper, we pay particular 
attention to the merger process, final products, and 
dependence of these outcomes on initial parameters of the binaries, 
but we do not treat the accurate extraction 
of gravitational waveforms and hence the accurate computation of 
gravitational radiation back reaction. 
As mentioned above, we start with binaries in almost 
dynamically unstable orbits. This implies that 
the effect of radiation reaction is not very important in the 
early phase of the merger. In the later phase of the merger, 
the dynamical timescale seems to 
be shorter than the emission timescale of gravitational waves and 
the evolution of the merged object due to gravitational radiation 
is secular. Hence, we deduce that the effect due to 
the error in evaluating the 
radiation reaction is small throughout the evolution. 

In Fig.~1, we show the relation between the rest mass $M_*$ 
and the maximum density $\rho_{\rm max}$ of each star for binaries 
in quasi-equilibrium states. The solid 
line denotes the relations for spherical neutron stars. 
The crosses and filled circles denote those for the
corotational and irrotational binary neutron stars, respectively. 
The binaries which are used 
in the following simulation as initial conditions are marked 
with (C1), (C2), (C3), (I1), (I2) and (I3) (C and I denote 
``corotational'' and ``irrotational'', respectively). 
The relevant quantities for these 
initial conditions are shown in Table I. 
We note that the orbital period is calculated as 
\beq
P_{\rm orb}=1.5{\rm msec}\biggl({C_i \over 0.15}\biggr)^{-3/2}
\biggl({M_{g0} \over 2.8M_{\odot}}\biggr).
\eeq

In a realistic situation, each star of the binary has a small 
approaching velocity because of gravitational radiation 
reaction. We approximately add this to the above quasi-equilibrium state 
in setting the initial conditions. According to the 
quadrupole formula with Newtonian equations of motion, 
the absolute value of the 
approaching velocity of each star is written as \cite{ST}
\beq
v_{\rm a}=1.6(M_{g0} \Omega)^2. 
\eeq
Thus, in giving initial conditions, we change $u_i$ to be 
\beq
u_x=(u_x)_{\rm eq}-v_{\rm a}{|x| \over x},
\eeq
where $(u_x)_{\rm eq}$ denotes $u_x$ of the quasi-equilibrium state. 
Here, we implicitly assume that the 
center of mass of each star is initially located 
on the $x$-axis (see Figs. 2--4 and 9--11). 

For models (C1), (C2) and (C3), we performed simulations 
without the approaching velocity and found that the 
outcomes such as the final products and the disk mass depend only 
weakly on this approaching velocity. 
Thus, it does not seem to affect the following results significantly. 

\section{Numerical results}

\subsection{corotational cases}

In Figs. 2--4, 
we show snapshots of the density contour lines for 
$\rho_*$ and velocity field $(v^x, v^y)$ 
in the equatorial plane at selected times for models 
(C1), (C2), and (C3), respectively. 
For (C1), a new massive neutron star is 
formed, while for other cases, a black hole is formed. 
We note that for model (C2), we could 
not determine the location of the apparent horizon before the 
simulation crashed, because the grid spacing was too wide to 
satisfactorily resolve the black hole forming region. 
However, the central value of the lapse function 
is small enough $< 0.01$ at the crash so that we may judge that 
a black hole is formed in this simulation. On the other hand, 
for model (C3), we can determine the location of the 
apparent horizon 
(see the thick solid line in the last snapshot of Fig. 4). 

Irrespective of the compactness parameters, 
the orbital distance gradually decreases 
due to the initial approach velocity. 
When it becomes small enough to destabilize the 
orbit due to the hydrodynamic or GR orbital instability, 
the orbital distance begins to quickly decrease and in the outer part, 
spiral arms are formed. For more compact binaries, the 
decrease rate of the orbital separation is larger because the 
initial approach velocity is larger, 
and the orbit soon becomes unstable. 
We deduce that the neutron stars for model (C3) 
are initially located near to the innermost stable circular 
orbit against GR orbital instability because their initial compactness 
$C_i\equiv (M_{g0}\Omega)^{2/3} \sim M_{g0}/a$, 
where $a$ denotes the orbital 
separation, is nearly equal to 1/6 (see Table I). 
Indeed, they begin merging soon after the simulation is started. 
Once merger begins, the spiral arms continue to develop 
transporting angular momentum outward in the outer part of the 
merged object. 

For model (C1), the inner part first 
contracts after the orbit becomes unstable, but subsequently it bounces 
due to the pressure and centrifugal force. 
The shape of the merged object changes 
from ellipsoidal to spheroidal, 
redistributing the angular momentum as well as dissipating it 
by gravitational radiation.  
Eventually, it forms a new rapidly rotating neutron star.
In Figs. 5 and 6, we show the density contour lines for $\rho_*$ in 
the $x-z$ plane and the angular velocity 
$\Omega\equiv (x v^y - y v^x)/(x^2+y^2)$ 
along the $x$ and $y$-axes in the equatorial plane 
at $t=2.07P_{\rm orb}$. 
It is found that the new neutron star is highly 
flattened and differentially rotating \cite{foot2}. 
Note that the mass inside $r\simeq 7.5M_{g0}$ 
which appears to constitute the merged object is 
$\sim 0.97M_* (\simeq 2.16)$. 
Since the maximum allowed mass of a spherical star with $K=200/\pi$ is 
$M_{*~\rm max}^{\rm sph} \simeq 1.435$, 
the mass of the new neutron star is 
$\sim 50\%$ larger than this \cite{BSS}. 
We point out that we have monitored the evolution of 
$K'(x^{\mu})\equiv P/\rho^{\Gamma}$ 
which is initially equal to $K$ anywhere in the star and can be 
regarded as a measure of the entropy distribution. 
Since shock heating is not very effective in the merging, 
we have found that the value of $K'$ increases by at most $\sim 10\%$ 
in the regions of high density. 
(Note that in the low density regions 
such as near to the surface of the merged object, $K'$ is slightly 
larger.) 
Thus, the role of thermal energy increase is not significant for 
supporting the large mass in contrast with the case of head-on collision 
\cite{Stu,waimo}. The effect of differential rotation is important 
in the present case.

We note that the new rotating neutron star has a non-axisymmetric 
structure at the time when we stopped the simulation. 
Therefore it will evolve further as a result of 
gravitational wave emission, 
and may become unstable against gravitational collapse to become a 
black hole after a substantial amount of angular 
momentum is carried away \cite{NNS}. 

For models (C2) and (C3), after the orbit becomes 
unstable, 
the inner part contracts due to self-gravity without bouncing 
because the pressure and centrifugal force are not 
strong enough to balance the self-gravity. Subsequently it 
collapses to form a black hole. Since the compactness is sufficiently 
large for model (C3), the inner part 
quickly collapses to form a black hole without a significant bounce.
On the other hand, in the case of model (C2), 
the formation timescale of the black hole is longer 
because the compactness is smaller. 

To show the features of the collapse around the central region, 
we show $\alpha$ at $r=0$ as a function of 
$t/P_{\rm orb}$ in Fig. 7. For model (C3), $\alpha(r=0)$ quickly 
approaches to zero, but for model (C2), the decrease rate 
becomes small at $t \sim 1.2P_{\rm orb}$. This difference indicates 
that the collapse is decelerated 
by the pressure and/or centrifugal force. 

We note that for model (C2), the initial value of $J/M_g^2$ 
is larger than unity. Nevertheless, a black hole appears to be 
formed after the merger. This indicates that some mechanisms
for angular momentum transfer or dissipation act 
to decrease $J/M_g^2$ to less than unity during 
the merger. We can expect that the following two mechanisms 
are effective. 
(a) In the case of corotational binaries, the outer 
part has a large amount of the angular momentum and 
spreads outwards forming the spiral arms. As a result, 
the specific angular momentum in the inner part which finally 
forms a black hole is smaller than that of the 
outer part and $J/M_g^2$ can be smaller 
than unity in the inner region. 
(b) The effect of gravitational radiation can 
reduce the magnitude of $J/M_g^2$ which is estimated as follows: 
If the system has a characteristic angular velocity $\Omega_c$, 
the relation between the energy loss $\delta E(>0)$ 
and the angular momentum loss $\delta J(>0)$ 
due to gravitational radiation can be written as 
$\Omega_c \delta J \simeq \delta E$. 
If we assume $\delta E \ll M_{g0}$ and $\delta J \ll J_0$ 
where $J_0$ denotes the initial value of $J$, 
the resulting $J/M_g^2$ becomes
\beqn
&&{J_0 - \delta J \over (M_{g0}-\delta E)^2} 
\simeq {J_0 \over M_{g0}^2}
\biggl(1-{\delta J \over J_0}+{2\delta E \over M_{g0}}\biggr)
\nonumber \\
\simeq &&{J_0 \over M_{g0}^2}\biggl[1+{\delta J \over J_0 }
\biggl\{-1+\biggl({2J_0 \over M_{g0}^2 }\biggr)
(\Omega_c M_{g0})\biggr\}\biggr]. \label{qqq}
\eeqn
Here, $J_0 /M_{g0}^2 \sim 1$, and 
because of the fact that gravitational waves are efficiently 
emitted in the early phase of merger, we may set 
$\Omega_c \sim \Omega$ and consequently 
$\Omega_c M_{g0} \ll 1$ (see Table I). 
Thus, $2J_0\Omega_c/M_{g0}$ 
(the second term in $\{~~\}$ of Eq.~(\ref{qqq}))
is much less than unity, and 
Eq. (\ref{qqq}) is approximately 
$(J_0/M_{g0}^2)(1-\delta J/J_0)$. 
Using the quadrupole formula and the Newtonian expression for 
the angular momentum, $\delta J/J_0$ in one orbital period 
for a binary system of point masses is 
\cite{ST}
\beq
{\delta J \over J_0}={16 \pi \over 5}(M_{g0} \Omega)^{5/3}
=0.0876\biggl( {C_i \over 0.15} \biggr)^{5/2}. 
\eeq
Therefore, $J/M_g^2$ can decrease by $\sim 10\%$.

Since the gradient of the metric becomes very steep in 
the high density region of the merged object, the simulation 
could not be accurately 
continued for models (C2) and (C3) after $\alpha$ at $r=0$ becomes
less than $\sim 10^{-2}$. Although we cannot strictly calculate the 
final states of the disks around the black holes for these models, 
we may extrapolate the final state from the evolution 
of the central region as follows. 
In Fig. 8, we show time evolution of 
the fraction of the rest mass inside a 
coordinate radius $r$, defined as
\beq
{M_*(r) \over M_*}=
{1 \over M_*} \int_{|x^i| < r} d^3x \rho_*,
\eeq
for models (C2) and (C3). 
We choose $r=1.5, 3$ and $4.5M_{g0}$ as coordinate radii. 
It is found that 
more than $95\%$ of the total rest mass is inside $r=4.5M_{g0}$, and 
a small fraction 
$< 3-5\%$ of the total rest mass can be in a disk around the 
black hole at $r \geq 4.5M_{g0}$. 
For model (C3), the location of the 
apparent horizon is at $r \sim 1.2M_{g0}$ 
at $t=1.08P_{\rm orb}$, 
so that most of the matter inside $r =1.5M_{g0}$ seems to 
be swallowed by the black hole eventually. 
On the other hand, since the newly formed black hole seems to be 
rotating rapidly ($J/M_g^2 \sim 0.8-0.9$, see Table I), 
the innermost stable circular orbit is located near the event horizon 
and so even some of the matter located 
between $r = 1.5M_{g0}$ and $3M_{g0}$ 
may go to form the disk around the black hole. 
Hence, it may still be possible that a very compact disk of mass 
$\sim 0.05M_*$ and radius $\sim 3M_{g0}$ is formed eventually for 
model (C3). For model (C2), we cannot discuss details 
because we could not determine the apparent horizon. 
As shown in Fig. 8, the fraction of matter inside $r=1.5M_{g0}$ is 
still increasing at the time when we terminated the simulation, 
so that the mass fraction of the compact disk at $r \sim 3M_{g0}$ 
seems to be at most $0.05M_*$. 
Thus, a disk of mass at most $\sim 0.05-0.1M_*$ may be 
formed around black holes in an optimistic estimation.

\subsection{irrotational cases}

In Figs. 9--11, 
we show snapshots of the density contour lines for 
$\rho_*$ and velocity field $(v^x, v^y)$ 
in the equatorial plane at selected times for models 
(I1), (I2), and (I3), respectively. 
For model (I1), a new massive neutron star is 
formed, while for the other cases, a black hole is formed. 
We note that we could not determine the location of the 
apparent horizon for model (I2) before the 
simulation crashed. However, the central value of the lapse function 
is small enough $< 0.01 $ at the crash, so that we judge that 
a black hole is formed in this simulation 
as in the case (C2). On the other hand, 
we could determine the location of the apparent horizon for model (I3). 

As in the corotational case, the orbital distance decreases 
gradually in the initial stages, and then when the orbital 
instability is triggered, it quickly decreases leading to merger. 
However, the behavior of the merger is different from 
that in the corotational cases. For the irrotational binary, 
the initial distribution of 
angular velocity around the center of mass 
is a decreasing function of the 
distance from the center (and the absolute value of the 
velocity $|v^i|$ is almost independent of position; 
{\it cf}. Figs. 9--11 at $t=0$). 
Hence, the centrifugal force in the outer region of the 
merged objects is not as strong as 
that in the corotational cases. 
Consequently, spiral arms are not 
formed in a significant way. On the other hand, 
the magnitude of the centrifugal force in the inner region is 
stronger than that in the corotational cases. 
As a result, two oscillating 
cores are formed in the inner region, 
and this structure is maintained for a short while.  These features 
have been found also in Newtonian simulations \cite{SNO,ruffert}.

In the case of model (I1), the two cores bounce after 
their first collision, and then they merge to form an oscillating 
new neutron star. 
In Figs. 12 and 13, we show the density contour lines for $\rho_*$ in 
the $x-z$ plane and the angular velocity defined by 
$\Omega\equiv (x v^y - y v^x)/(x^2+y^2)$ 
along the $x$ and $y$-axes in the equatorial plane at 
$t=1.81P_{\rm orb}$. 
We find that the new neutron star has a 
toroidal structure which is sustained by differential rotation 
\cite{foot2}. 
Note that $99.5\%$ of the total rest mass 
is inside $r\simeq 6M_{g0}$ and appears to 
constitute the merged object at $t=1.81P_{\rm orb}$ in this case. 
Thus, the rest mass of the 
new neutron star is $\simeq 45\%$ larger than 
$M_{*~\rm max}^{\rm sph}$ \cite{BSS}. As in the corotating case, 
$K'$ increases only by a small factor (at most $10\%$) 
in the high density region, so that the 
role of the thermal pressure increase for supporting the large mass 
is not significant.

In this model (I1), the initial value of $J/M_g^2$ is less than unity and 
the final value should be even smaller as argued in Sec. IV. A. 
Nevertheless, the merged object does not collapse to form a black hole. 
Axisymmetric simulations of stellar core collapse 
\cite{Nakamura} have indicated that a black hole is formed 
for $J/M_g^2 < 1$ in most cases if the mass is large enough, 
and that the angular momentum parameter 
is a good indicator for predicting the final product. 
The present simulation suggests that this is not always the case 
for the merger of binary neutron stars. 

We note again that the new rotating neutron star was non-axisymmetric 
when we stopped the simulation. Therefore it will evolve secularly 
and may become unstable against gravitational collapse 
to a black hole after a substantial amount of the angular 
momentum has been carried away by gravitational radiation \cite{NNS}.

For model (I3), 
the inner part contracts due to self-gravity without bouncing 
because the pressure and centrifugal force are not 
strong enough to balance the self-gravity. Consequently, it quickly 
collapses to form a black hole. 
For model (I2), on the other hand, the self-gravity 
is weaker than that for model (I3), 
so that the two cores bounce 
at the first collision for $t \sim 1.2P_{\rm orb}$. 
Then, they approach again redistributing 
the angular momentum as well as dissipating it 
by gravitational radiation, and finally the merged object 
forms a black hole. 
To demonstrate this feature, we show $\alpha$ at $r=0$ as a function of 
$t/P_{\rm orb}$ in Fig. 14. For model (I3), $\alpha(r=0)$ monotonically 
approaches zero, but for model (I2), 
it increases again after it reaches a first minimum at 
$t \sim 1.2P_{\rm orb}$. 
These numerical results indicate that 
the merging process towards the final state depends considerably 
on the initial compactness of the neutron stars. 

In Fig. 15, we show the time evolution of 
the fraction of the rest mass inside a 
coordinate radius $r$, 
$M_*(r)/ M_*$, for models (I2) and (I3). 
We again choose $r=1.5, 3$ and $4.5M_{g0}$ as coordinate radii. 
It is found that 
more than $99\%$ of the total rest mass was inside $r=4.5M_{g0}$ 
for both models when we stopped the simulations.
Thus, in contrast with the corotational cases, 
only a tiny fraction 
of the total rest mass ($< 1\%$) can form a disk around the 
black hole at $r \geq 4.5M_{g0}$. 
For model (I3), $\agt 99\%$ of the total rest mass is 
inside $r=1.5M_{g0}$ which almost coincides with the location of 
the apparent horizon at the final snapshot of Fig. 11. 
Hence, we can conclude that 
the disk mass is very small ($<0.01M_*$) for model (I3). 
For model (I2), we could not determine the location of 
the apparent horizon before the simulation crashed, and so we cannot 
make any strong conclusion. However, Fig.~15 shows that 
the mass fraction outside $r=3M_{g0}$ is $<0.01M_*$ and 
that inside $r=1.5M_{g0}$ is quickly increasing 
at $t\sim 1.8P_{\rm orb}$. Hence, the final disk mass again appears 
to be very small as in the case (I3).

\subsection{gravitational waves}

To extract gravitational waveforms, we define 
non-dimensional variables 
\beqn
&&h_+ \equiv r(\tilde \gamma_{xx} - \tilde \gamma_{yy})/(2M_{g0}),\\
&&h_{\times} \equiv  r \tilde \gamma_{xy}/M_{g0},
\eeqn
along the $z$-axis. 
Since we adopt the AMD gauge condition and have 
prepared initial conditions for which 
$\delta^{ij}\pa_i \tilde \gamma_{jk}=0$, $\tilde \gamma_{ij}$ 
is approximately transverse and traceless 
in the wave zone \cite{gw3p2}. 
As a result, $h_+$ and $h_{\times}$ are expected to be  
appropriate measures of gravitational waves. 

In Figs.~16, we show waveforms 
for corotational models (C1) (the solid lines) and 
(C2) (the dashed lines), and in Figs.~17,
for irrotational models (I1) (the solid lines) and 
(I2) (the dashed lines) 
as a function of retarded time $(t-z_{\rm obs})/P_{\rm orb}$ where 
$z_{\rm obs}$ denotes the point along the $z$-axis 
at which the waveforms are extracted. 

For both corotational and irrotational cases, the 
amplitude gradually rises with decreasing orbital separation, 
but after the amplitude reaches the maximum, 
the waveforms for the two cases have different characters. 
In the corotational cases, 
the amplitude soon becomes small after the maximum, while in 
the irrotational cases, it does not become small very quickly, but 
has a couple of fairly large peaks. 
The reason is that the double core structure which enhances 
the amplitude is preserved for a short while 
after the merger in the irrotational cases. Such a feature has been 
found also in Newtonian simulations \cite{SNO,ruffert}, 
indicating that 
the Newtonian simulations are helpful for investigation of the 
qualitative outcome of gravitational waveforms. 
In particular, the waveforms for models (C1) and (I1) 
in which new neutron stars are formed are qualitatively 
similar to the corresponding Newtonian models \cite{SNO,ruffert}, 
although quantitative features such as amplitude and wavelength 
are different. Therefore, 
the Newtonian simulation is useful as a guideline for 
fully GR simulations particularly 
when the final product is a neutron star. 

The maximum amplitude for $h_{+}$ and $h_{\times}$ is typically 
$0.1$ as shown in Figs. 16 and 17. 
This implies that the typical maximum amplitude of gravitational 
waves from a source at the distance $r$ is 
\beq
\sim 1.4 \times 10^{-22} \biggl({M_{g0} \over 2.8M_{\odot}} \biggr)
\biggl({100 {\rm Mpc} \over r}\biggr)
\biggl({h_{+,\times} \over 0.1}\biggr). 
\eeq

As we mentioned above, the outer boundaries of the computational 
domain with $233 \times 233 \times 117$ grid points 
are located at $\alt 0.3\lambda_{\rm gw}$ on each axis. 
This implies that the 
waveforms extracted are not accurate asymptotic 
waveforms. For example, a slight unrealistic modulation (the wave 
amplitude deviates gradually with time in the positive direction) 
is found in the waveform 
for $h_+$ in every case which seems due to numerical error.  

To estimate the magnitude of the error, 
we performed one large simulation for model (I2) 
with grid size $293 \times 293 \times 147$, fixing 
the grid spacing but widening the computational region. 
Even in this case, the outer boundaries are located at 
$\sim 0.35\lambda_{\rm gw}$ on each axis. In Fig. 18, we show the 
waveforms for $293 \times 293 \times 147$ (the solid lines), 
$233 \times 233 \times 117$ (the dashed lines), 
and $193 \times 193 \times 97$ (the dotted lines). 
For the early phase of merging, 
the magnitude of the modulation is smaller and smaller with increasing 
number of grid points, which implies that this effect is spurious due to 
the restricted computational region. 
For the very late phase of merging, on the other hand, 
the magnitude of the modulation does not change even 
with widening the computational domain. 
This suggests that the resolution of the 
central regions of the merged object is not sufficient in that phase 
to compute accurate waveforms. 
We also find that the wave amplitude increases slightly with widening the 
computational region. 
This indicates that the amplitudes shown in Figs. 16 and 17 
might underestimate the asymptotic one 
by several tens of percent especially in the early phase of the merger. 
All of these facts indicate that 
we need a larger scale computation 
to improve the accuracy of the gravitational waveforms.

Even in the case of black hole formation, 
the shapes of the waveforms are similar to those in the 
neutron star formation case before the gravitational collapse to a 
black hole has occurred (compare the waveforms for models (I1) and (I2)). 
The difference in the waveforms 
will appear after the gravitational collapse. However, since 
we could not continue simulations 
for a long time at this stage, 
we cannot describe the features of the waveform in detail. 
In the following, we speculate on the expected outcome and 
discuss the significance of the waveforms 
from the observational point of view. 

According to the standard scenario, the quasi normal 
modes of the black hole are excited 
in the final phase of black hole formation, 
and the amplitudes of these modes subsequently damps. 
As we showed in the previous two subsections, 
the formation timescale of the black hole is different depending on 
the compactness of the neutron stars before the merger. 
This implies that the time duration from 
the moment when the amplitude of the gravitational waves becomes 
a maximum to the moment when 
the amplitude of the waves from the merged object 
damps depends on the initial 
compactness parameter of the neutron stars (see Fig. 19, 
which shows a schematic 
picture for the expected gravitational waveforms): 
In the case of neutron star formation ({\it cf}. Fig. 19 (a)), 
the damping time for quasi-periodic gravitational waves 
of small amplitude emitted from non-axisymmetric deformation of 
the new neutron star is the timescale of gravitational 
radiation reaction which is much longer than the dynamical timescale. 
In the case of black hole formation, 
we have a number of possibilities: If the 
compactness parameter of the neutron stars before the merger is not 
very large ({\it cf}. Fig. 19 (b)), 
the timescale for the formation process is fairly 
long and the quasi-periodic oscillations due to non-axisymmetric 
deformation of the merged object will be seen for a short while 
after the merger. If the 
neutron stars are sufficiently compact ({\it cf}. Fig. 19 (c)), 
the black hole is formed quickly 
and the amplitude of gravitational waves will also damp quickly. 
Therefore, the time duration from the gravitational wave burst to 
its damping 
(note that we do not need here the detail of the waveforms) 
will constrain the initial compactness of the neutron stars, and, 
consequently, the equation of state for high density neutron 
matter \cite{foot5}.

\section{Discussion}

As we found in Sec. IV, the final products of the merger depend 
sensitively on the initial compactness of the neutron stars. 
In the corotational case, 
(1) the final product is a massive neutron star 
when the ratio of the rest mass of each star to 
$M_{*~\rm max}^{\rm sph}$ ($C_{\rm mass}$) 
is $\alt 0.8$ ; 
(2) the final product is a black hole 
when $C_{\rm mass}$ is $\agt 0.9$. 
If it is at most $\sim 0.9$, 
the formation timescale is longer than the dynamical 
timescale (or the oscillation period of the merged object, 
$P_{\rm osc}$). 
On the other hand, if $C_{\rm mass}$ is $\sim 1$, 
the formation timescale of the black hole is as 
short as the dynamical timescale ($\alt P_{\rm osc}$). 
In the irrotational case, 
(3) the final product is a massive neutron star 
when $C_{\rm mass}$ is $\alt 0.7$ ; 
(4) the final product is a black hole when $C_{\rm mass}$ is $\agt 0.8$. 
If it is at most $\sim 0.8$, the formation timescale is longer than 
$P_{\rm osc}$. 
On the other hand, if $C_{\rm mass}$ is larger than $\sim 0.9$, 
the formation timescale is $\alt P_{\rm osc}$.

Let us consider the case where two irrotational 
neutron stars of rest mass $1.6M_{\odot}$ 
(i.e., the gravitational mass is $\sim 1.4M_{\odot}$) merge. 
The numerical results in this paper indicate that (a) if 
$M_{*~\rm max}^{\rm sph}$ is less than $\sim 1.8M_{\odot}$, 
the merged object forms a black hole on the dynamical timescale 
$\sim P_{\rm osc}$ ; (b) 
if $M_{*~\rm max}^{\rm sph}$ is $\sim 2M_{\odot}$, the 
final product is also a black hole, but the formation timescale is 
longer than $P_{\rm osc}$; (c) if $M_{*~\rm max}^{\rm sph}$ is larger than 
$\sim 2.2M_{\odot}$, the final product will be a 
massive neutron star. 
This fact provides us the following interesting possibility. 
Suppose that we will be able to find the mass of each neutron star 
during the inspiraling 
phase by means of the matched filtering method \cite{CF} with 
the aid of the post-Newtonian template \cite{BIWW}. Then, if we  
observe the merger process to the final products, 
in particular the timescale for formation of a black hole, 
we can constrain the maximum allowed neutron star mass, and 
consequently the nuclear equation of state. 

Unfortunately, the frequency of gravitational waves after the merger 
will be so high (typically $P_{\rm osc}^{-1} \sim 5 P_{\rm orb}^{-1} 
\sim 2-3$kHz) that laser interferometers such as 
LIGO \cite{LIGO} will not be able to detect them. 
To observe such high frequency gravitational waves, 
specially designed narrow band interferometers or resonant-mass 
detectors are needed \cite{KIP}. We should keep in mind that 
such future gravitational wave detectors would have the possibility 
to provide us with important information about 
the neutron star equation of state. 

Another important outcome of the present simulations concerns
the mass of the disk around a black hole 
formed after the merger.
A disk of mass $\sim 0.05-0.1M_*$ may be formed 
around a black hole after the merger of corotational binaries. 
However, for the merger of irrotational binaries, 
the mass of the disk appears to be very small $< 0.01M_*$. 
An irrotational velocity field is considered to be a 
good approximation for realistic binary neutron stars 
before merger \cite{KBC}. 
Therefore, a massive disk of mass $> 0.1M_{\odot}$ 
may not be formed around 
a black hole after the merger of binary neutron stars of 
nearly equal mass. 
This is not very promising for some scenarios for GRBs, 
in which a black hole--toroid system formed after the merger of 
nearly equal mass binary neutron stars 
is considered to be its central engine.

We have performed simulations using a modified form of the 
ADM formalism for the Einstein field equation with the AMD gauge and 
approximate maximal slicing conditions \cite{gr3d}. 
Needless to say, simulations 
by other groups using different formulations, gauge conditions and 
numerical implementations \cite{ON,waimo} are necessary 
to reconfirm the present results. 

In this paper, we have performed simulations only for the case $\Gamma=2$. 
As Newtonian simulations have indicated \cite{RS,Centrella}, 
the merging process and final products may also depend sensitively 
on the stiffness of neutron star matter. 
In a forthcoming paper, we will perform simulations 
changing $\Gamma$ to investigate the dependence 
on the stiffness of the equation of state and to clarify whether the 
present conclusions are modified or not.

\acknowledgments

We thank T. Baumgarte, E. Gourgoulhon, T. Nakamura, K. Oohara, and 
S. Shapiro for helpful conversations and discussions. 
We also thank J. C. Miller for careful reading on the manuscript and 
useful suggestions. For warm hospitality, 
M.S. thanks the Department of Physics of the University of Illinois 
and K.U. thanks D.W. Sciama at SISSA. 
Numerical computations were performed on the 
FACOM VPP 300R and VX/4R machines 
in the data processing center of the 
National Astronomical Observatory of Japan. 
This work was supported by a Grant-in-Aid (Nos. 08NP0801) of 
the Japanese Ministry of Education, Science, Sports and Culture, 
and JSPS (Fellowships for Research Abroad).

\clearpage
\onecolumn

{\bf Table I.~} A list of several quantities for 
initial conditions of binary neutron stars. 
The maximum density, 
total rest mass $M_*$, gravitational mass $M_{g0}$, 
$J/M_{g0}^2$, compactness $C_i\equiv (M_{g0}\Omega)^{2/3}(\sim 
M_{g0}/a$ where 
$a$ is orbital separation), 
ratio of the emission timescale of gravitational waves to 
the orbital period $R_{\tau}=5(M_{g0}\Omega)^{-5/3}/128\pi$, 
the ratio of the rest mass of each star to 
the maximum allowed mass for a spherical star 
$C_{\rm mass}\equiv M_*/2M_{*~\rm max}^{\rm sph}$, 
the ratio of $M_{g0}$ to the grid spacing in the simulation 
($M_{g0}/\Delta x$), 
the type of velocity field and final products are shown. 
Here, $M_{*~\rm max}^{\rm sph}$ denotes the maximum allowed mass 
for a spherical star ($\simeq 1.435$). Here, we quote values for 
$K=200/\pi$. The mass and density can be scaled 
by the rules $M_*(K\pi/200)^{1/2}$, $M_{g0}(K\pi/200)^{1/2}$,  
and $\rho_{\rm max} (K\pi/200)^{-1}$ for arbitrary $K$, while 
other non-dimensional quantities are invariant. 

\vskip 5mm
\noindent
\begin{center}
\begin{tabular}{|c|c|c|c|c|c|c|c|c|c|c|} \hline
\hspace{1mm} $\rho_{\rm max}(10^{-3}) $ \hspace{1mm} &
\hspace{1mm} $M_*$ \hspace{1mm} & \hspace{1mm} $M_{g0}$\hspace{1mm} 
& \hspace{1mm} $J/M_{g0}^2$ \hspace{1mm} & \hspace{1mm} 
$C_i$ \hspace{1mm}
& \hspace{2mm} $R_{\tau}$ \hspace{2mm}  & $C_{\rm mass}$ 
& $M_{g0}/\Delta x$ &velocity field & Final product & model 
\\ \hline
 1.50  & 2.22 & $2.06$ & 1.10 & 0.10 & 3.7  &0.77 & 4.26
& corotational &neutron star  & C1 \\ \hline
 2.00  & 2.52 & $2.31$ & 1.04 & 0.12 & 2.3  & 0.88 & 5.27
& corotational &black hole &C2 \\ \hline
 3.00  & 2.84 & $2.56$ & 0.98 & 0.15 & 1.4  & 0.99 & 6.84
& corotational &black hole &C3 \\ \hline
 1.14 & 2.08 & $1.93$ & $0.98^{*}$ & 0.09 & 4.9  &0.72 & 3.95
& irrotational &neutron star &I1\\ \hline
 1.88 & 2.34 & $2.15$ & $0.93^{*}$ & 0.11 & 3.2  & 0.82 & 4.78
& irrotational &black hole &I2\\ \hline
 2.79 & 2.65 & $2.40$ & $0.88^{*}$ & 0.14 & 1.9  & 0.92 & 6.13
& irrotational &black hole &I3 \\ \hline
\end{tabular}
\end{center}
\noindent
$^{*}$ We initially 
reduced the angular momentum from the 
corresponding quasi-equilibrium states by $\simeq 2.5\%$. 

\vskip 1cm

\begin{figure}[t]
\begin{center}
\epsfxsize=3.5in
\leavevmode
\epsffile{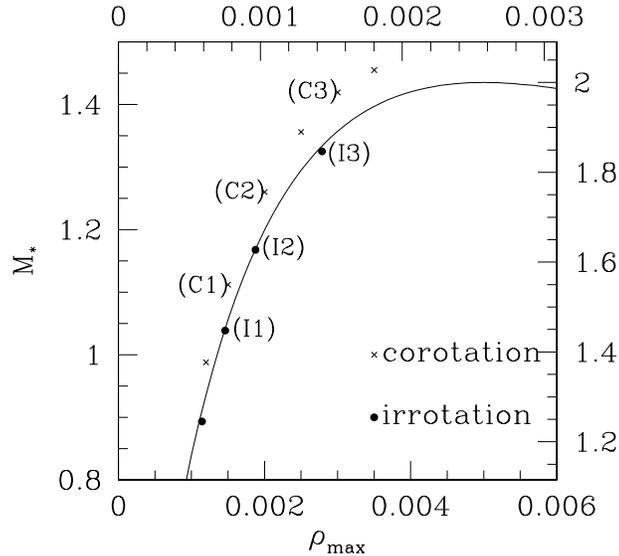}
\end{center}
\caption{Rest mass $M_*$ 
as a function of maximum density $\rho_{\rm max}$ for each star 
in the binary for $K=200/\pi$. 
The binaries which are used in the simulations are marked with 
(C1), (C2), (C3), (I1), (I2), and (I3). 
The solid line denotes the relation for the spherical stars. 
We note that the mass and the density can be scaled by the 
rules $M_*(K\pi/200)^{1/2}$ and $\rho_{\rm max} (K\pi/200)^{-1}$ 
for arbitrary $K$. Scales for the top and right axes are shown 
for $K=123.7$ in which the maximum rest mass for spherical stars 
is $2M_{\odot}$. 
}
\end{figure}

\clearpage
\begin{figure}[t]
\begin{center}
\epsfxsize=2.6in
\leavevmode
\epsffile{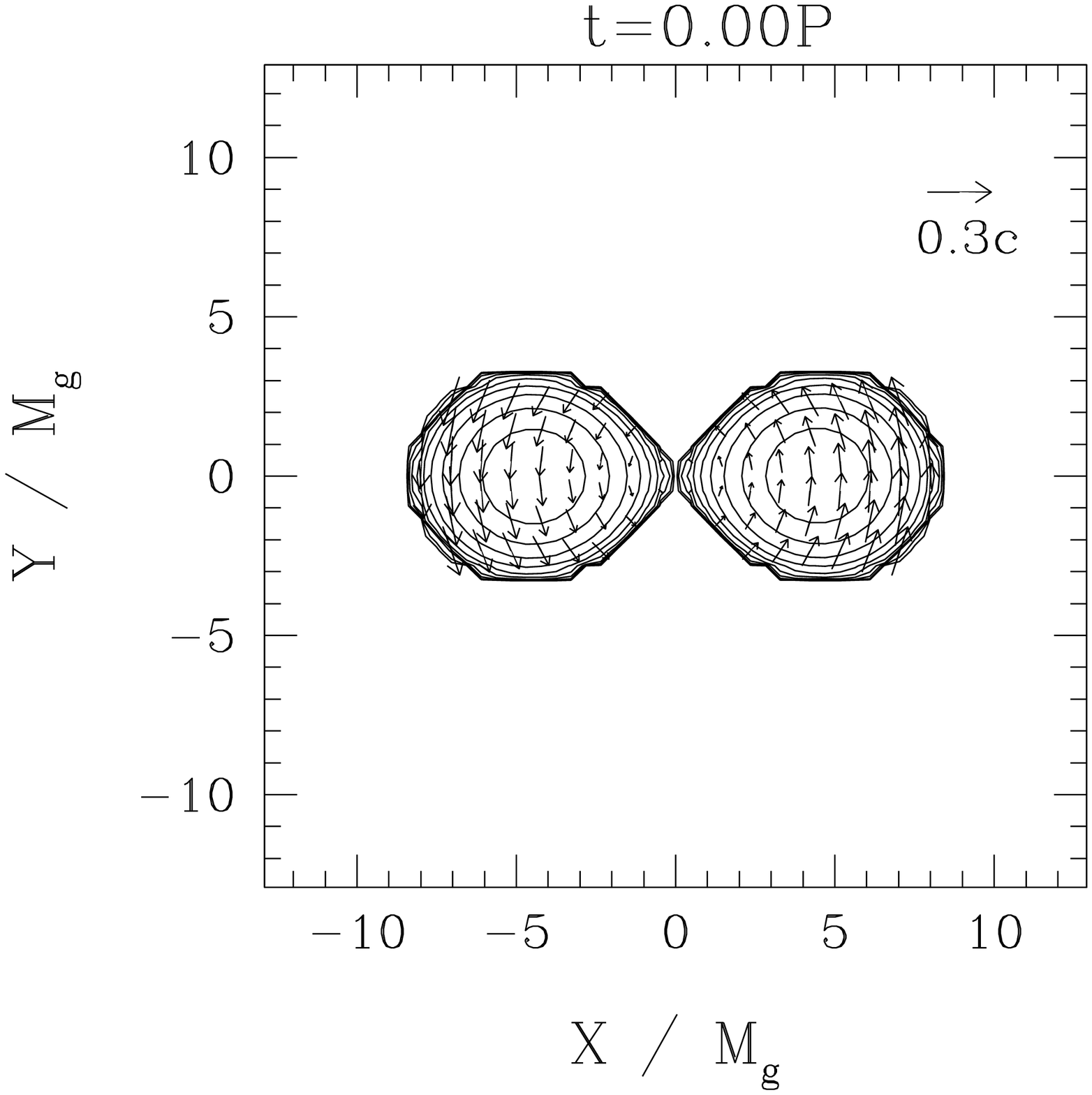}
\epsfxsize=2.6in
\leavevmode
\epsffile{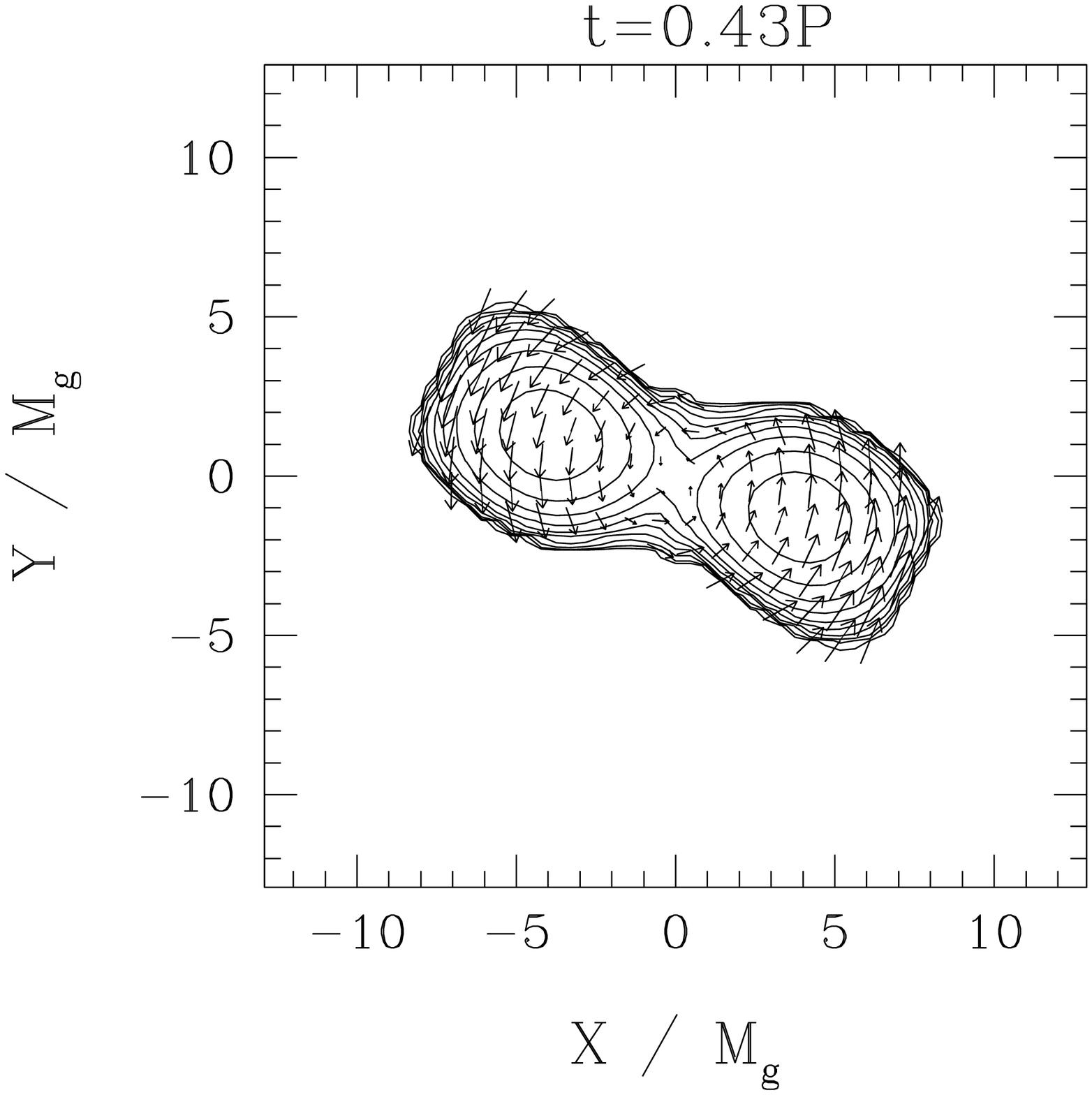}\\
\epsfxsize=2.6in
\leavevmode
\epsffile{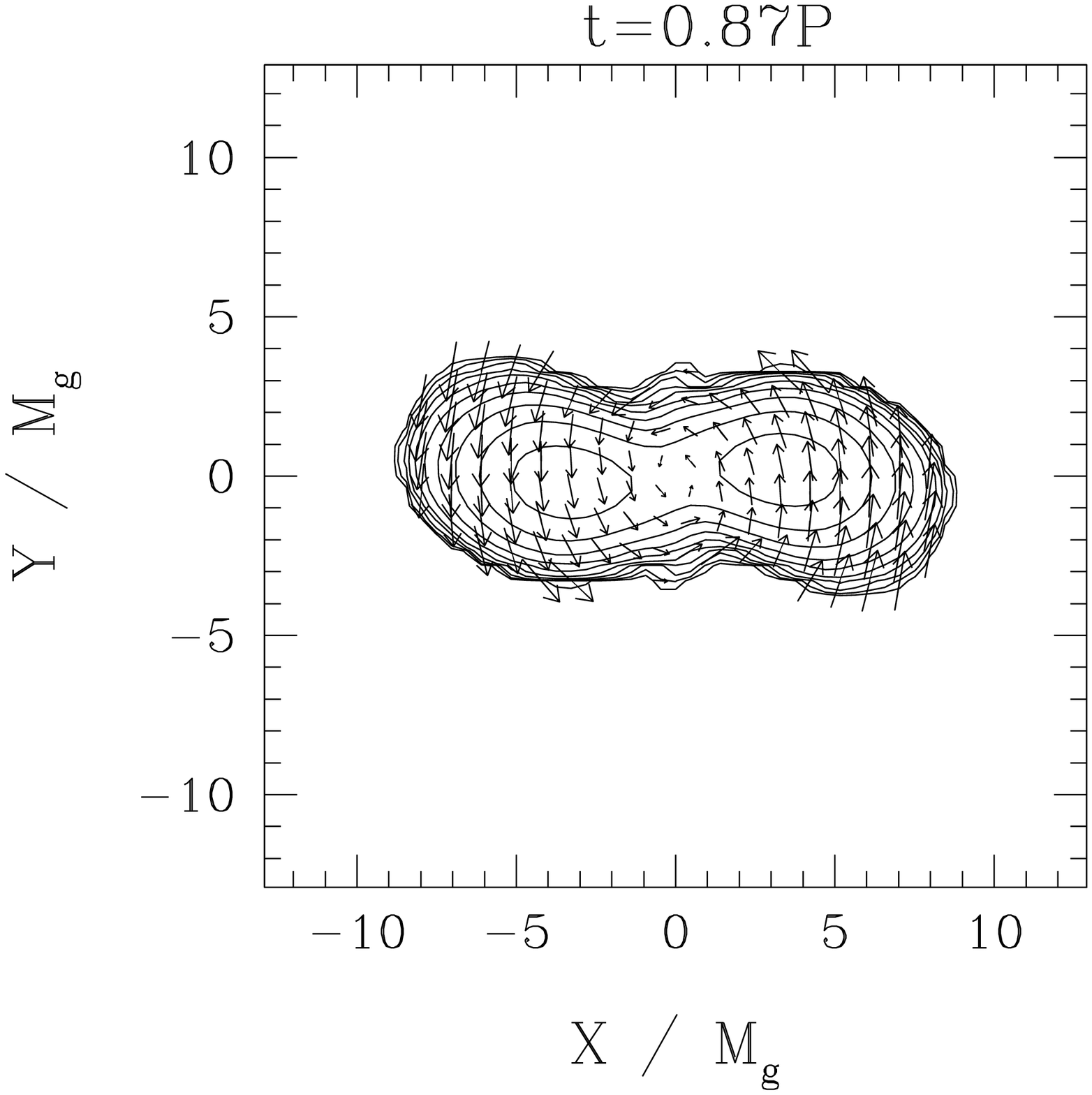}
\epsfxsize=2.6in
\leavevmode
\epsffile{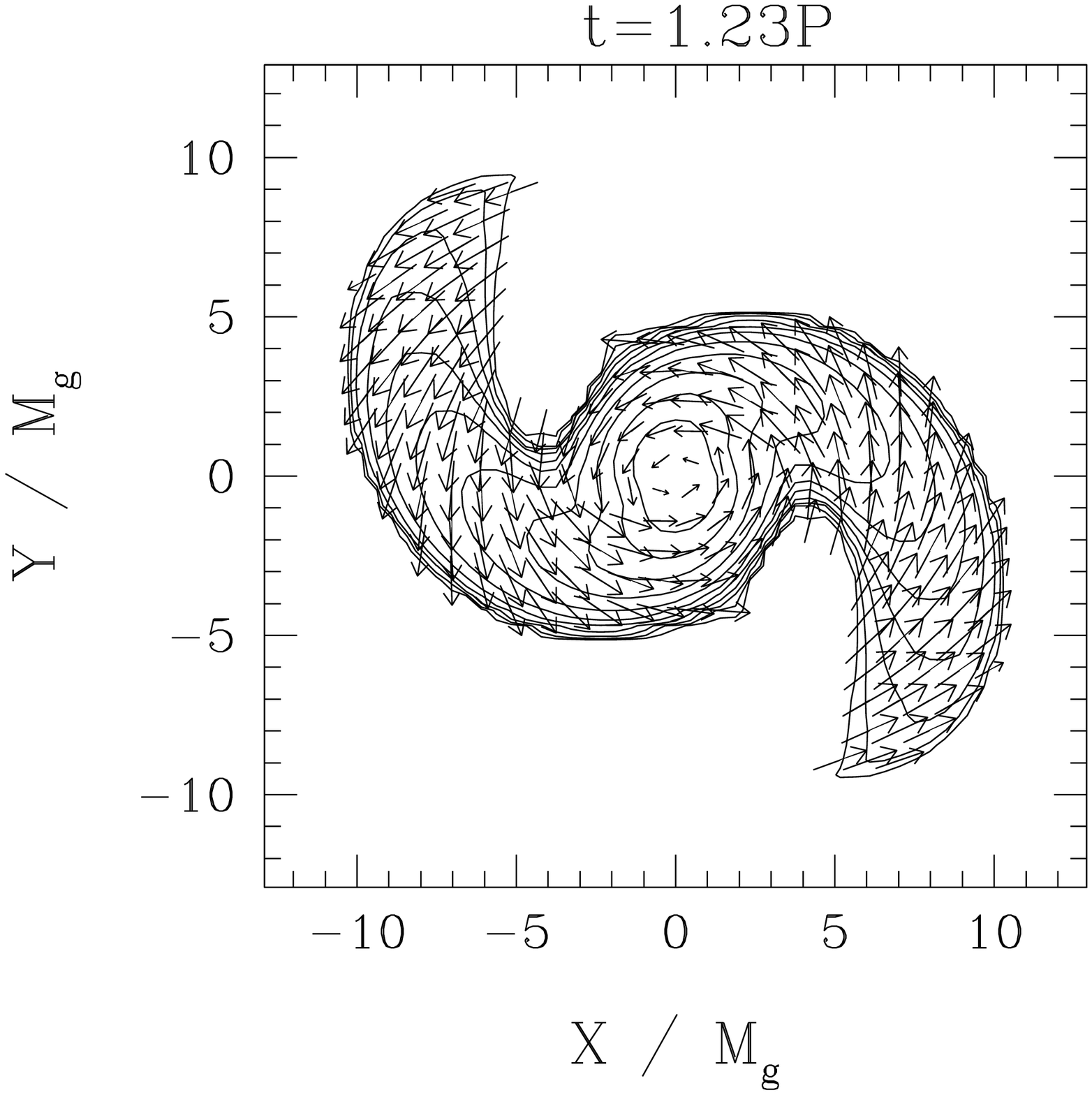} \\
\epsfxsize=2.6in
\leavevmode
\epsffile{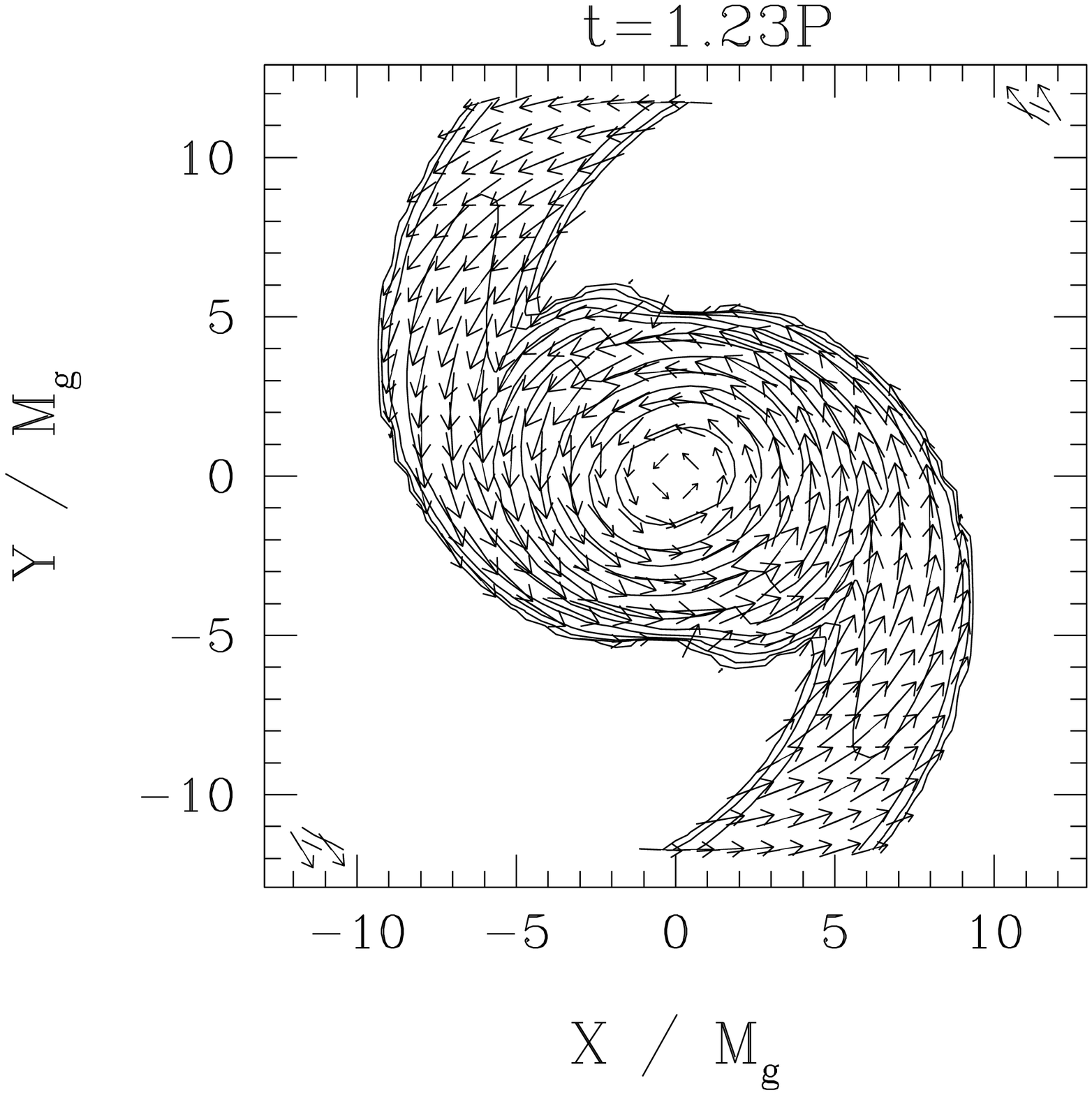}
\epsfxsize=2.6in
\leavevmode
\epsffile{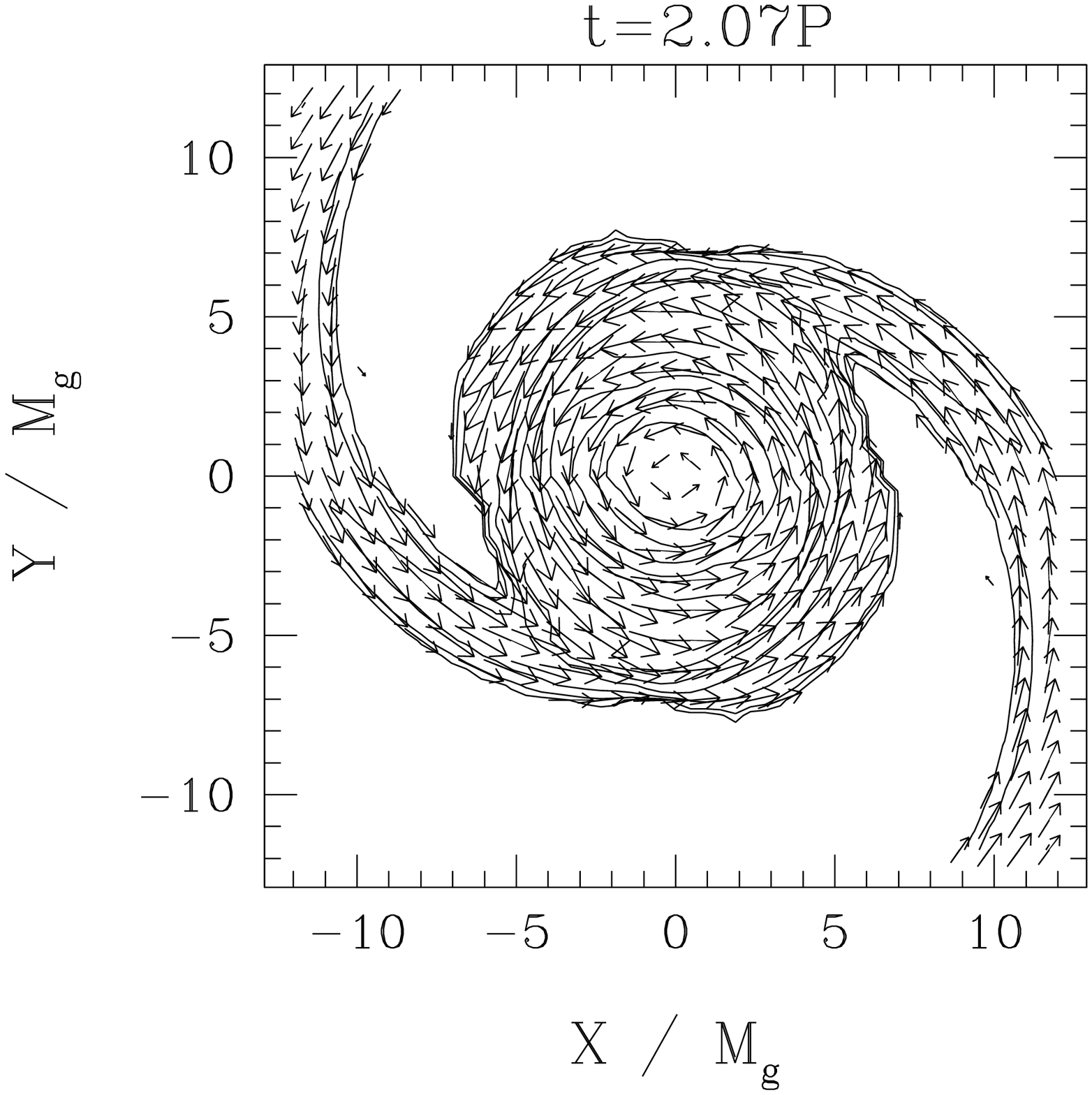} 
\caption{
Snapshots of the density contour lines for $\rho_*$ and 
the velocity field $(v^x,v^y)$ in the equatorial plane for 
model (C1). The contour lines are drawn for 
$\rho_*/\rho_{*~{\rm max}}=10^{-0.3j}$, 
where $\rho_{*~{\rm max}}$ denotes the maximum value of $\rho_*$ at 
$t=0$ (here it is $0.00441$), for $j=0,1,2,\cdots,10$. 
Vectors indicate the local velocity field and the scale 
is as shown in the top left-hand frame. $P$ denotes the 
initial orbital period $P_{\rm orb}$. The length scale is 
shown in units of $GM_{g0}/c^2$. 
}
\end{center}
\end{figure}

\clearpage
\begin{figure}[t]
\begin{center}
\epsfxsize=2.6in
\leavevmode
\epsffile{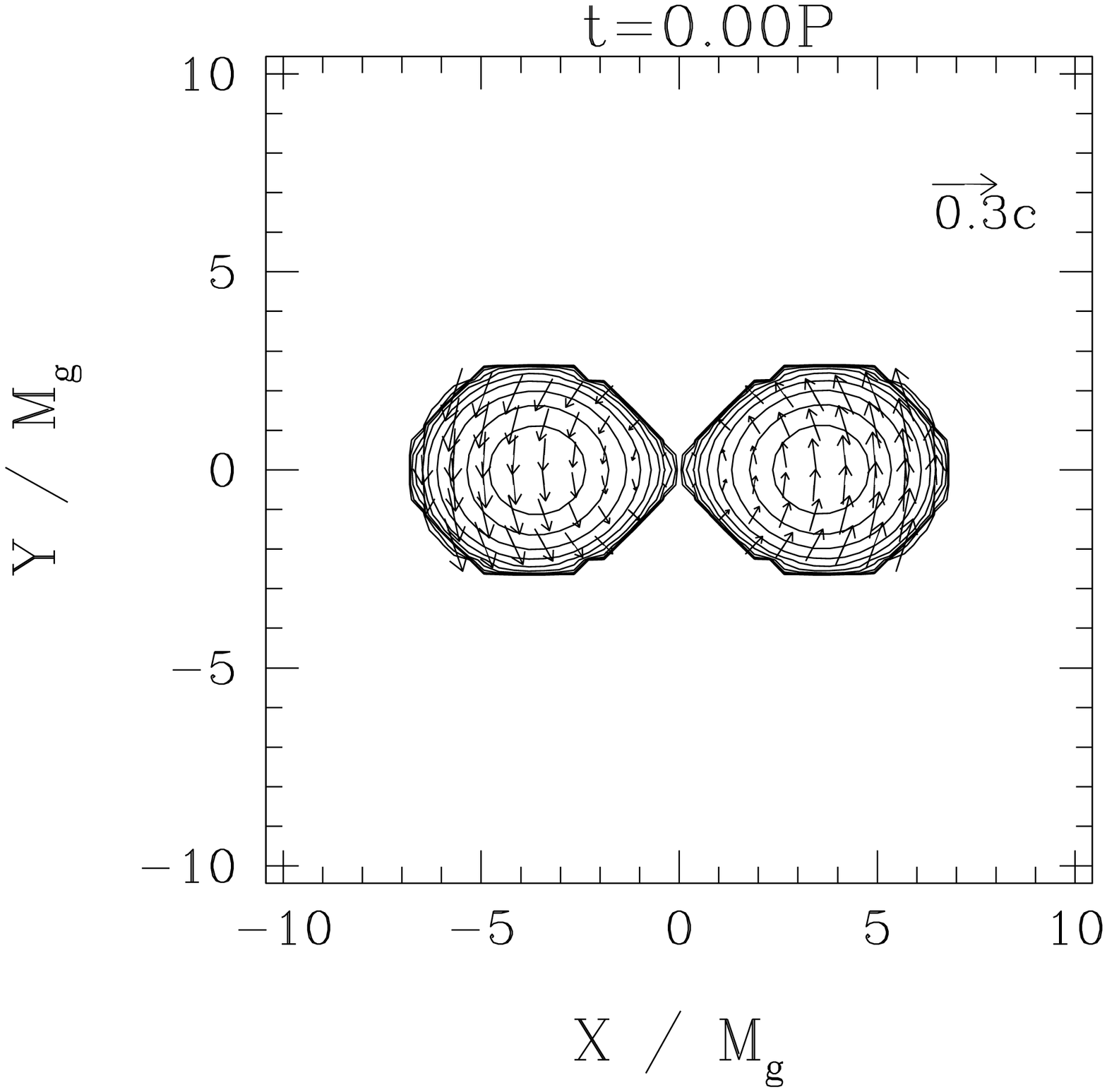}
\epsfxsize=2.6in
\leavevmode
\epsffile{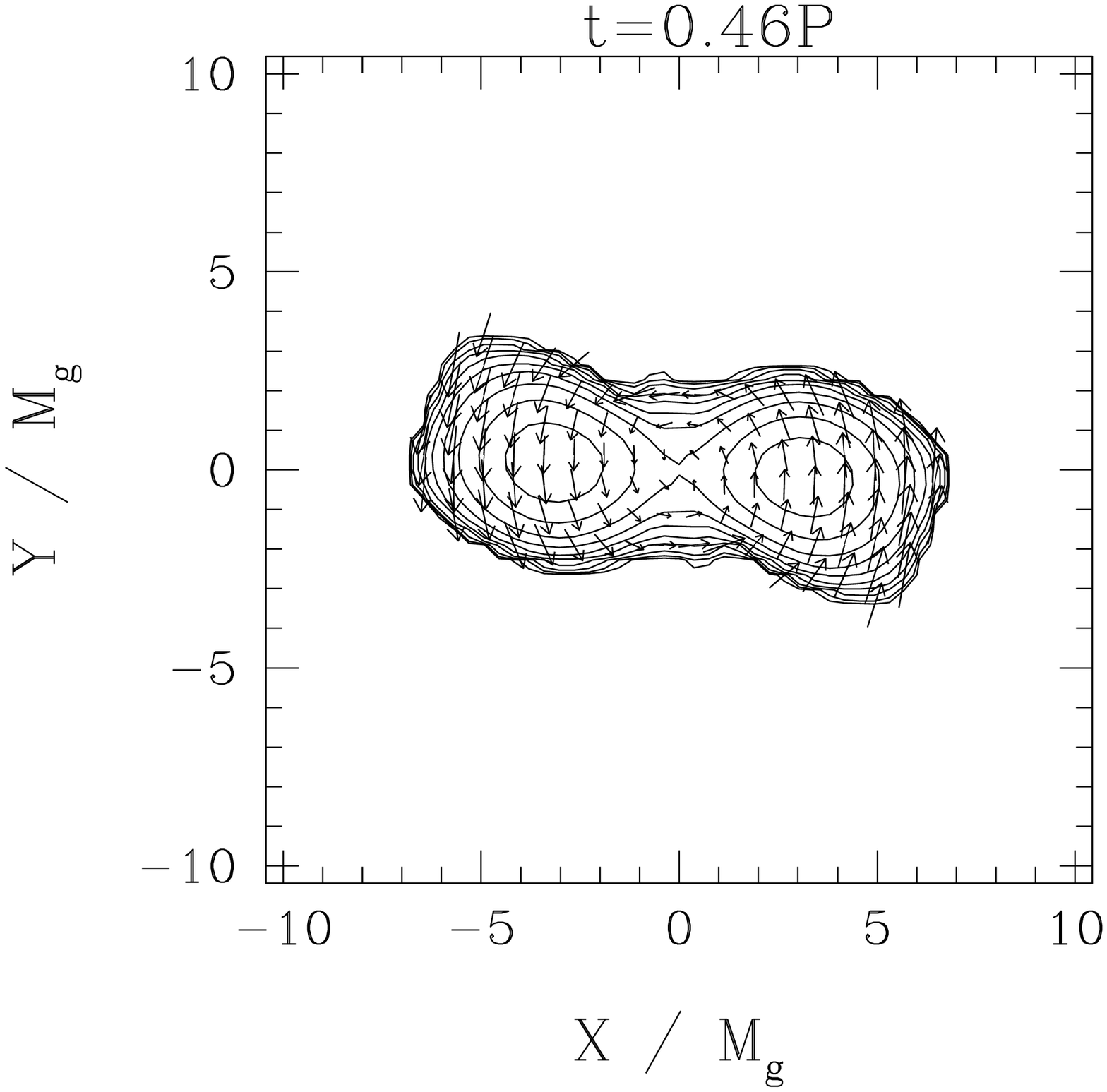}\\
\epsfxsize=2.6in
\leavevmode
\epsffile{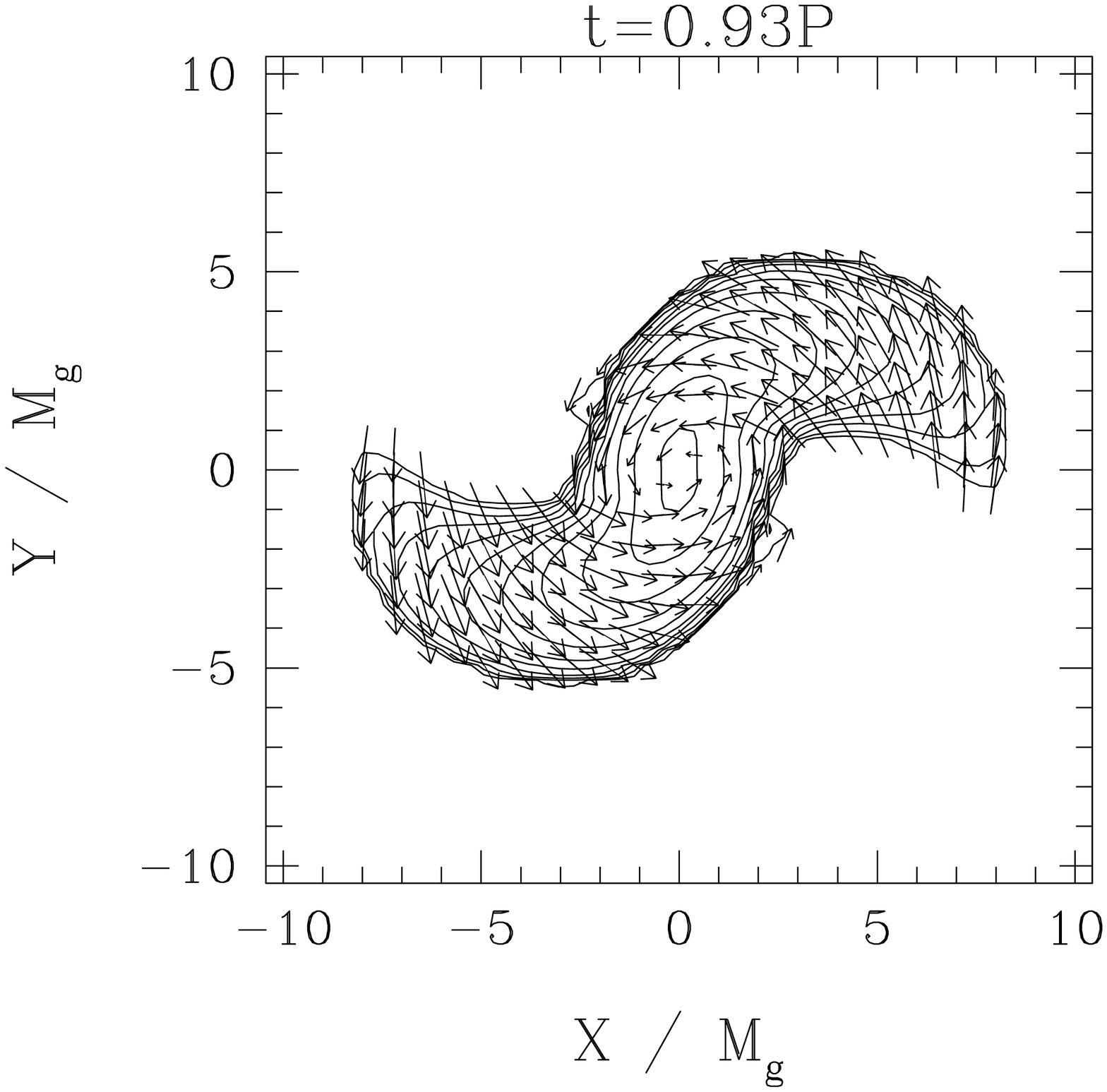}
\epsfxsize=2.6in
\leavevmode
\epsffile{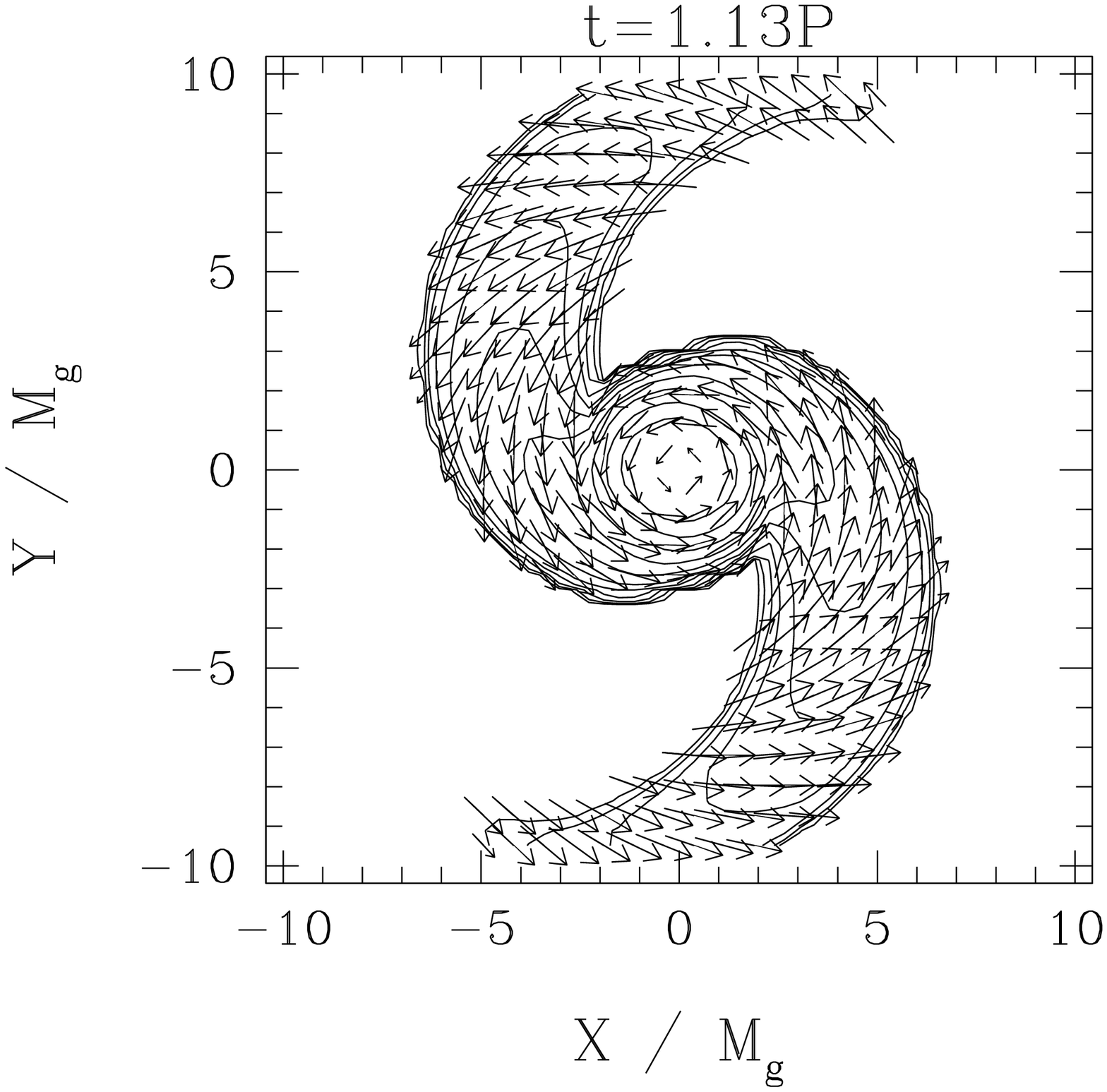} \\
\epsfxsize=2.6in
\leavevmode
\epsffile{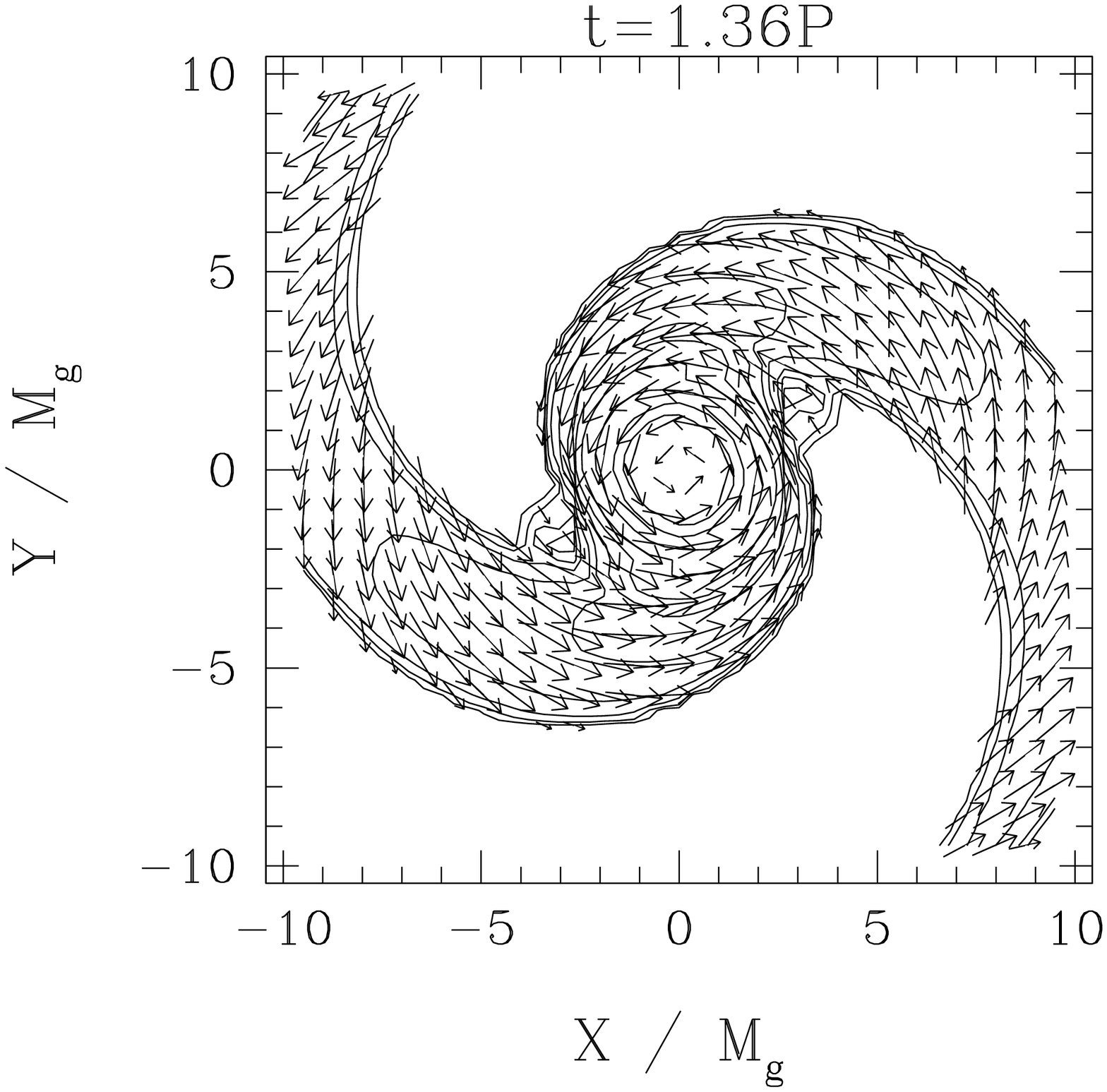}
\epsfxsize=2.6in
\leavevmode
\epsffile{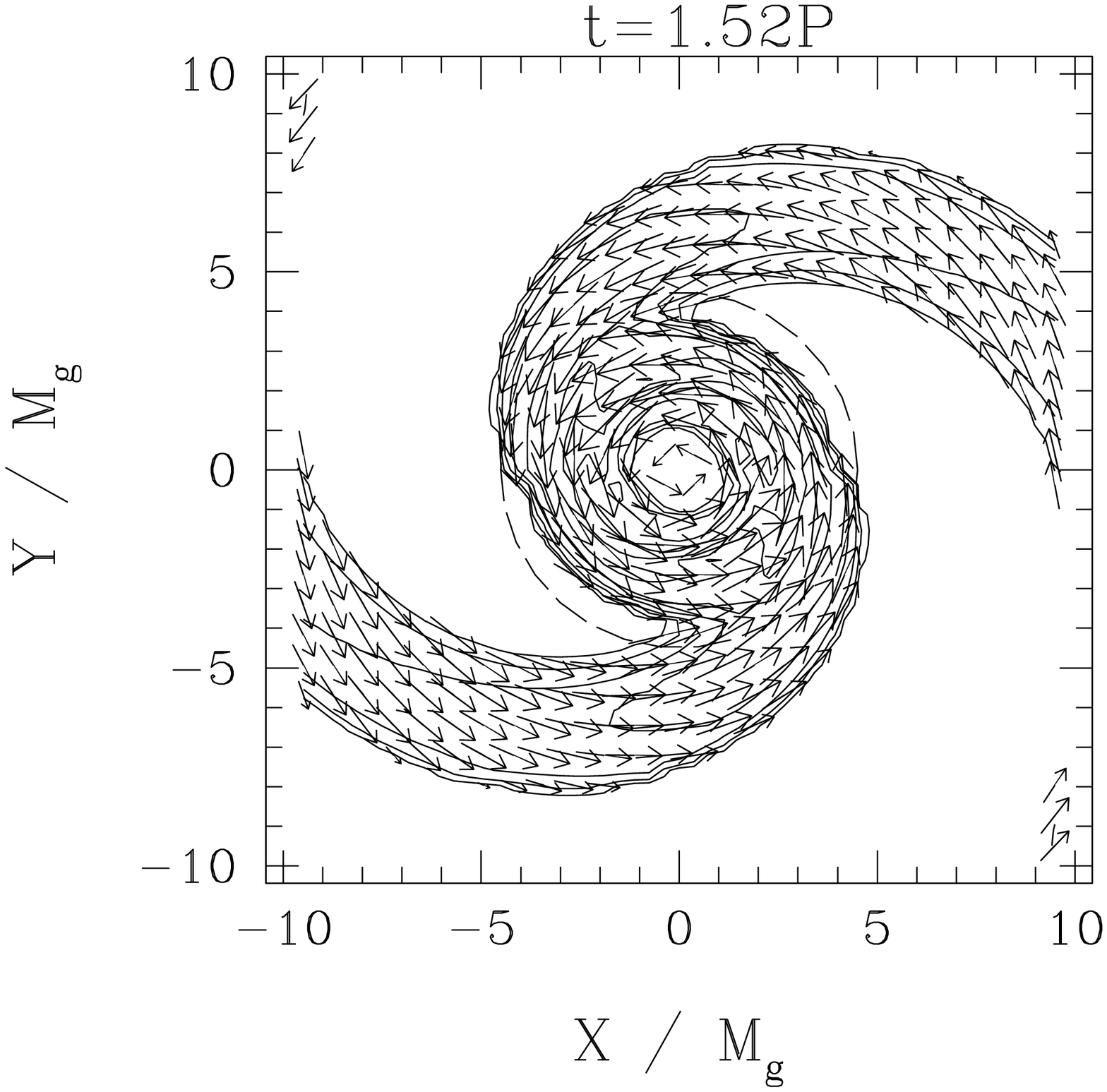} 
\caption{
The same as Fig. 2, but for model (C2). 
The contour lines are drawn for 
$\rho_*/\rho_{*~{\rm max}}=10^{-0.3j}$, 
where $\rho_{*~{\rm max}}=0.00757$, for $j=0,1,2,\cdots,10$. 
The dashed line in the last figure denotes the circle with 
$r=4.5M_{g0}$ 
within which $\sim 95\%$ of the total rest mass is included. 
}
\end{center}
\end{figure}

\clearpage
\begin{figure}[t]
\begin{center}
\epsfxsize=2.6in
\leavevmode
\epsffile{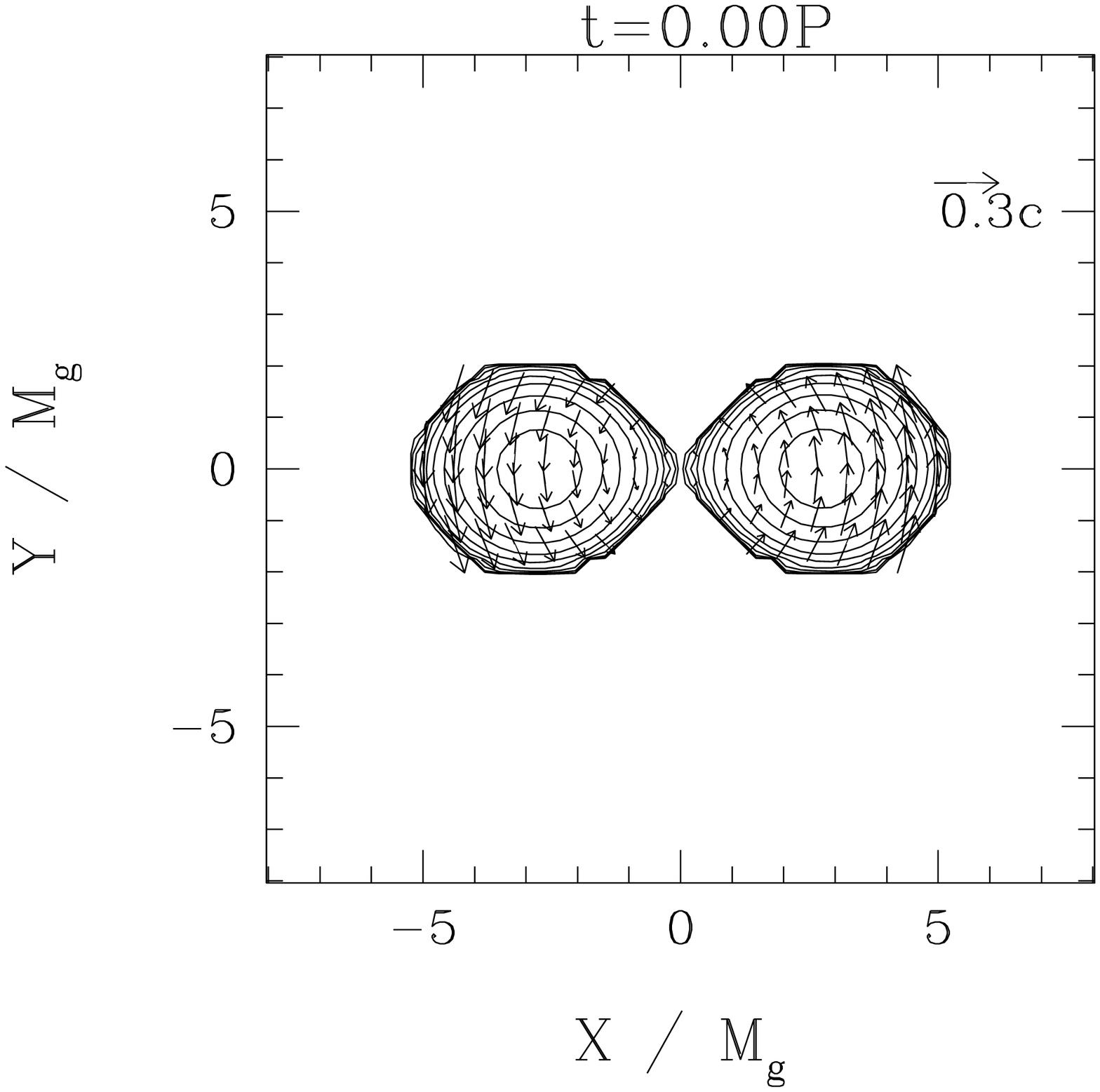}
\epsfxsize=2.6in
\leavevmode
\epsffile{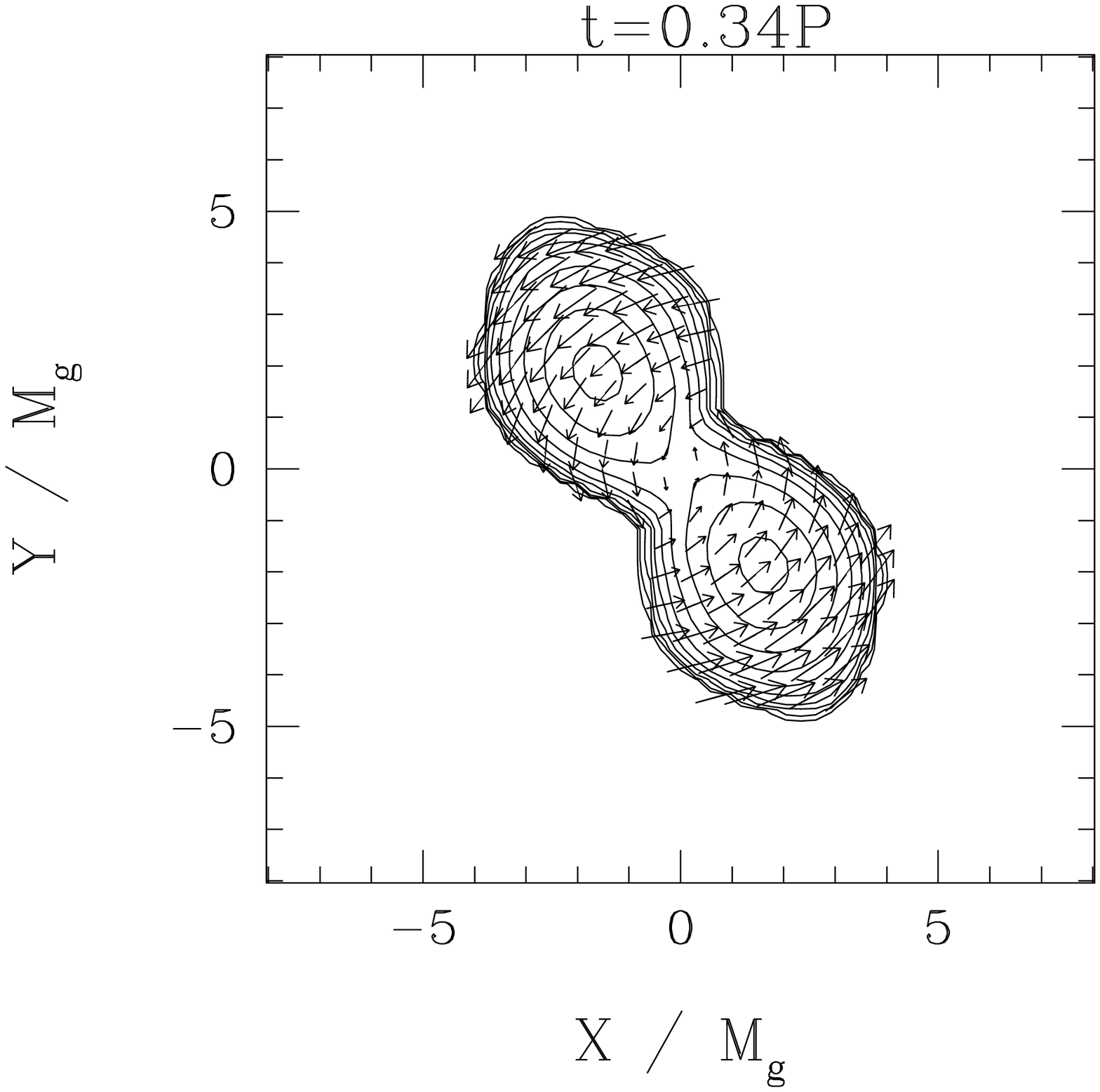}\\
\epsfxsize=2.6in
\leavevmode
\epsffile{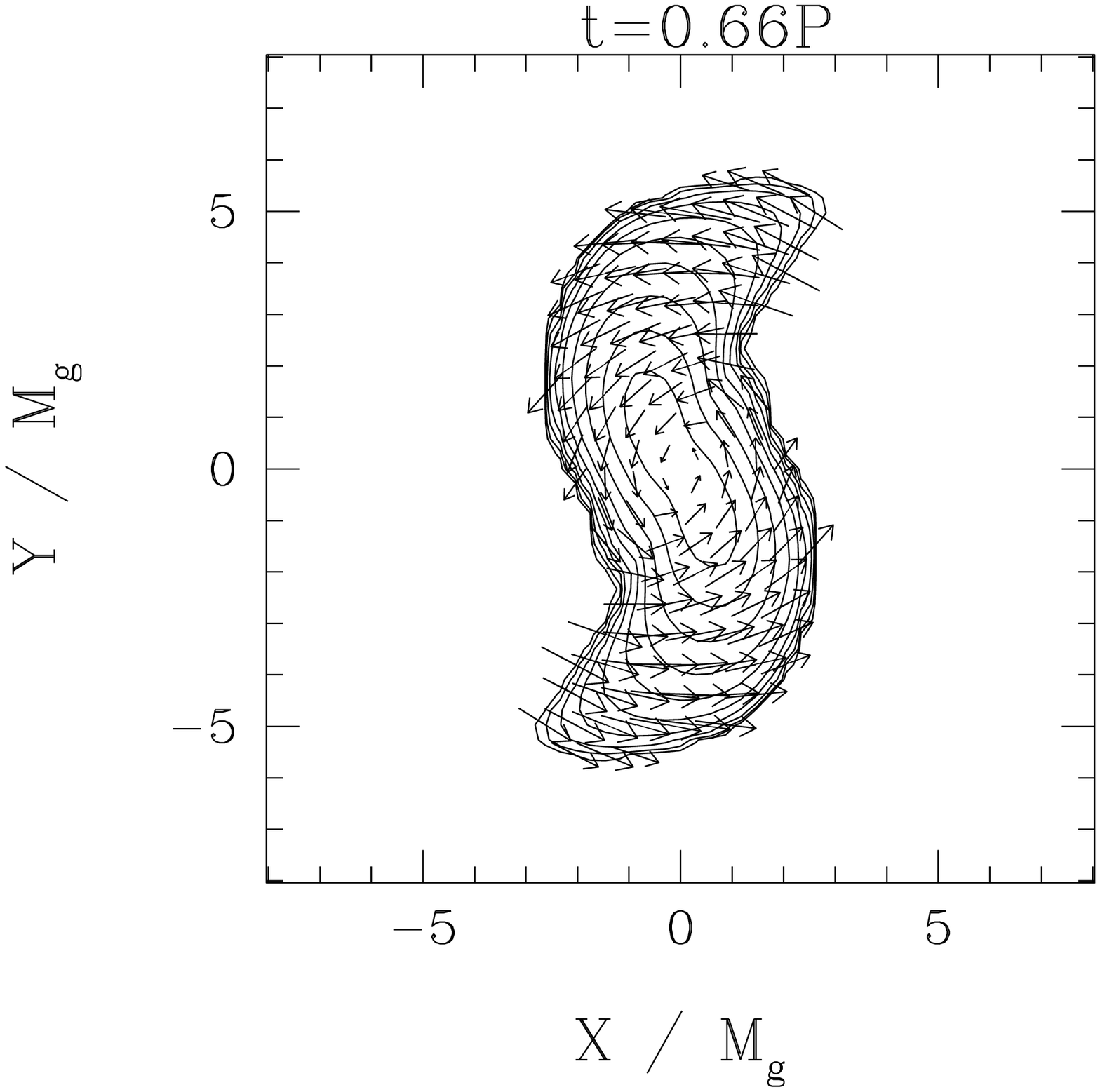}
\epsfxsize=2.6in
\leavevmode
\epsffile{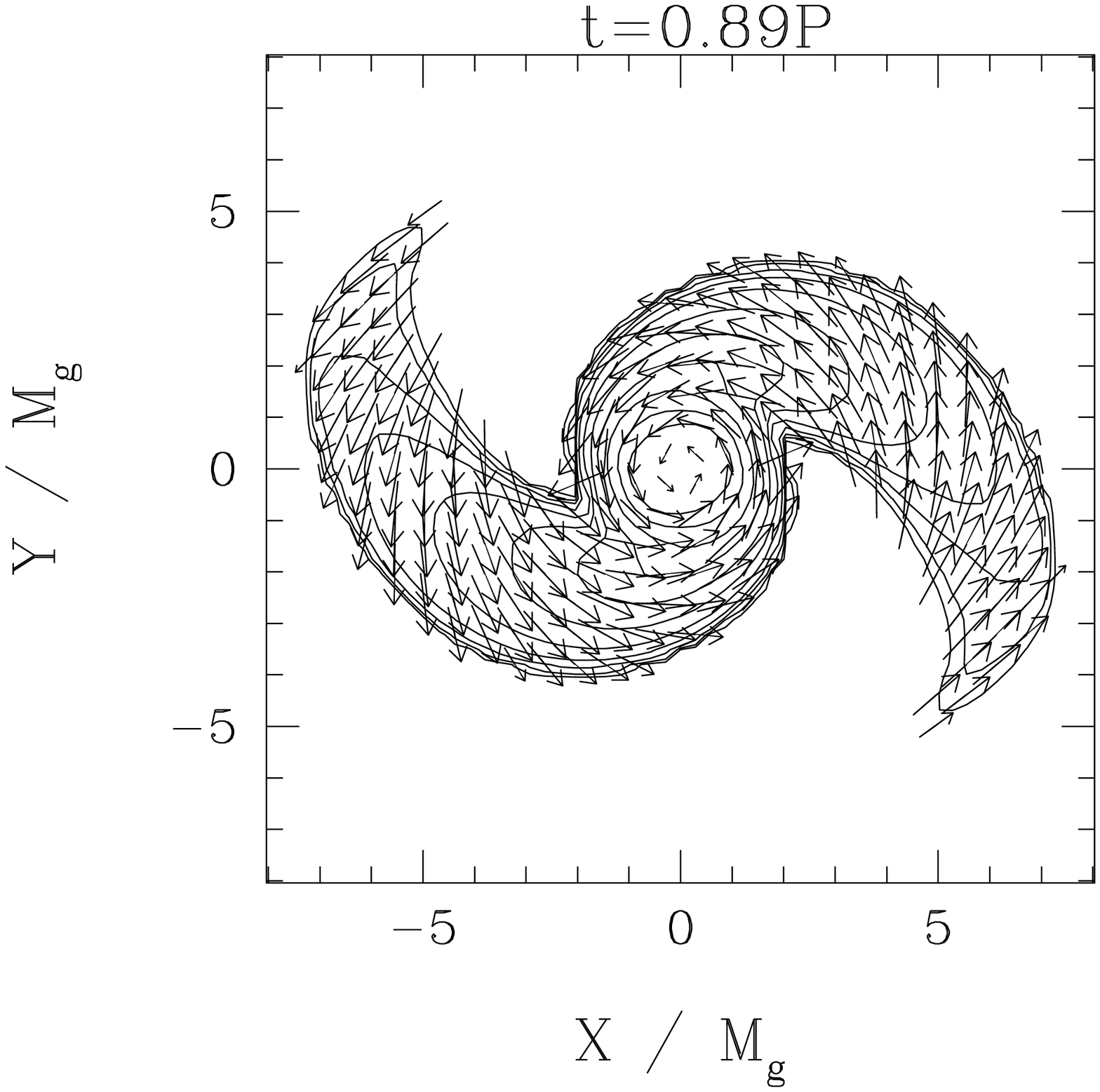} \\
\epsfxsize=2.6in
\leavevmode
\epsffile{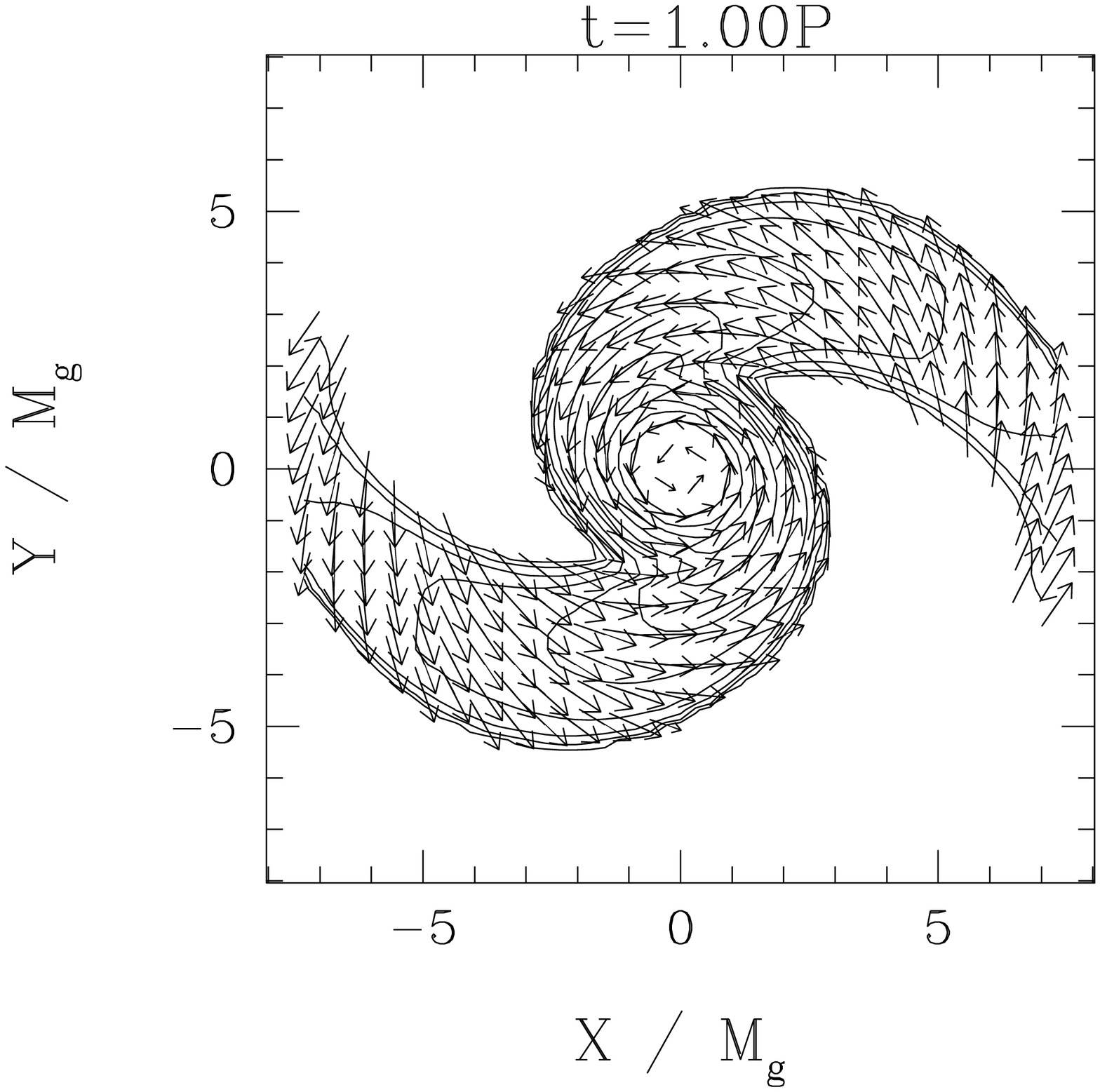}
\epsfxsize=2.6in
\leavevmode
\epsffile{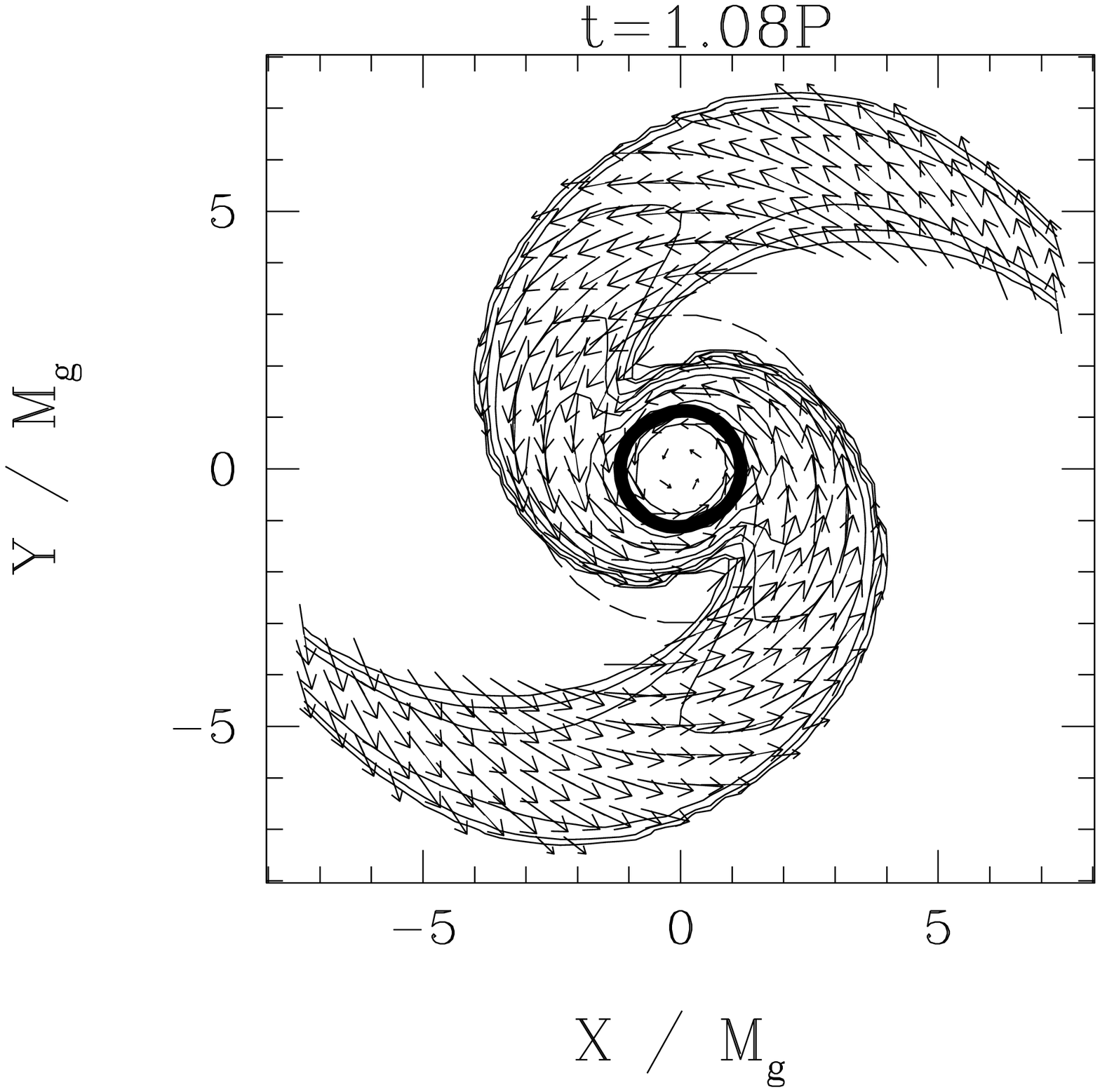} 
\caption{
The same as Fig. 2, but for model (C3). 
The contour lines are drawn for 
$\rho_*/\rho_{*~{\rm max}}=10^{-0.3j}$, 
where $\rho_{*~{\rm max}}=0.0171$, for $j=0,1,2,\cdots,10$. 
The dashed line in the last snapshot denotes the circle with 
$r=3M_{g0} $ 
within which $\sim 95\%$ of the total rest mass is included. 
The thick solid line for $r \sim M_{g0}$ in the last snapshot 
denotes the location of the apparent horizon. 
Note that there are $\sim 7$ grid points 
along the radius of the apparent horizon.  
}
\end{center}
\end{figure}

\clearpage
\begin{figure}[t]
\begin{center}
\epsfxsize=3.5in
\leavevmode
\epsffile{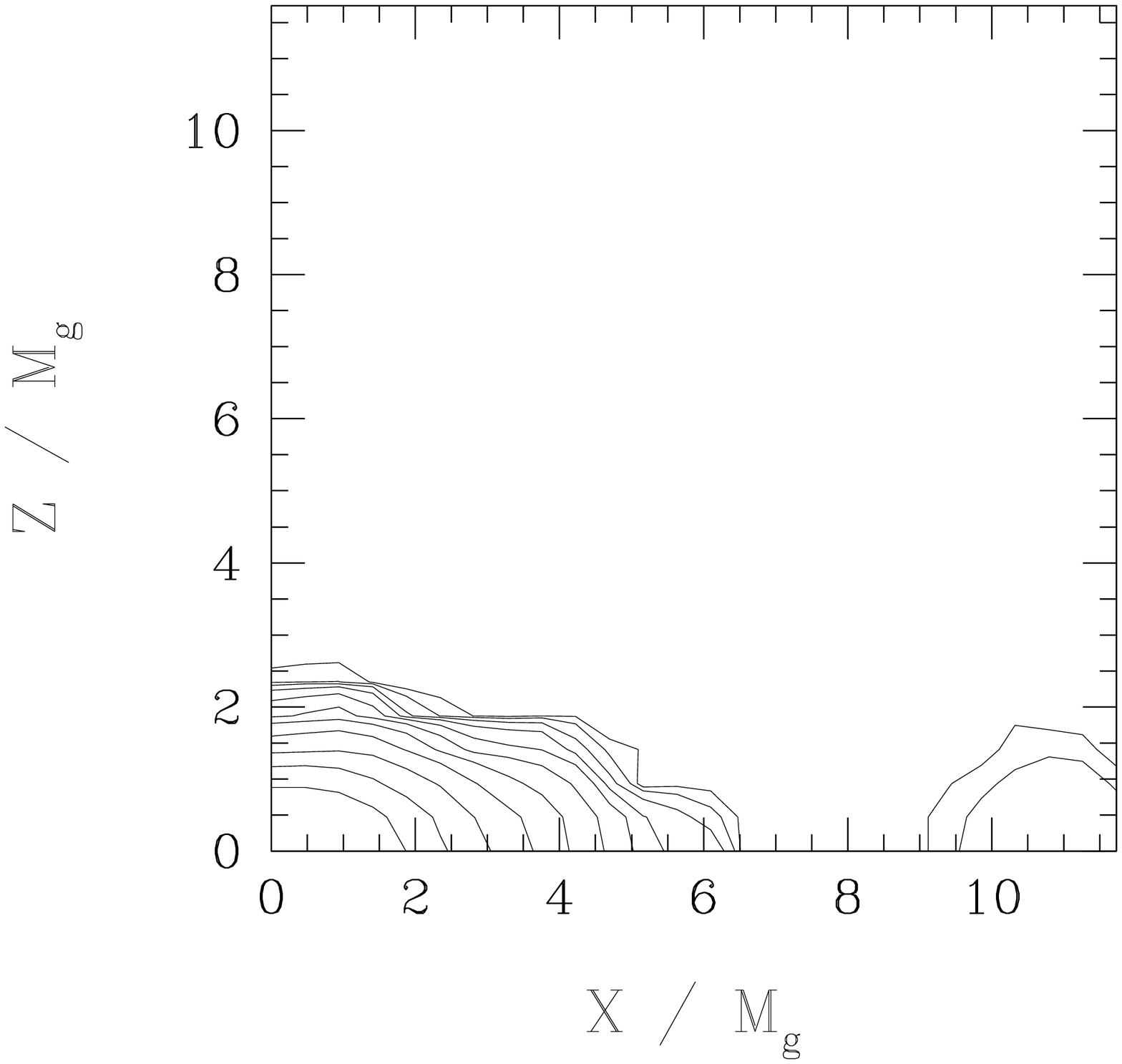}
\end{center}
\caption{
The density contour lines for $\rho_*$ 
in the $y=0$ plane at $t=2.07P_{\rm orb}$ for model (C1). 
The contour lines are drawn in the same way as in Fig. 2. 
The length scale is shown in units of $GM_{g0}/c^2$. }
\end{figure}

\begin{figure}[t]
\begin{center}
\epsfxsize=3.5in
\leavevmode
\epsffile{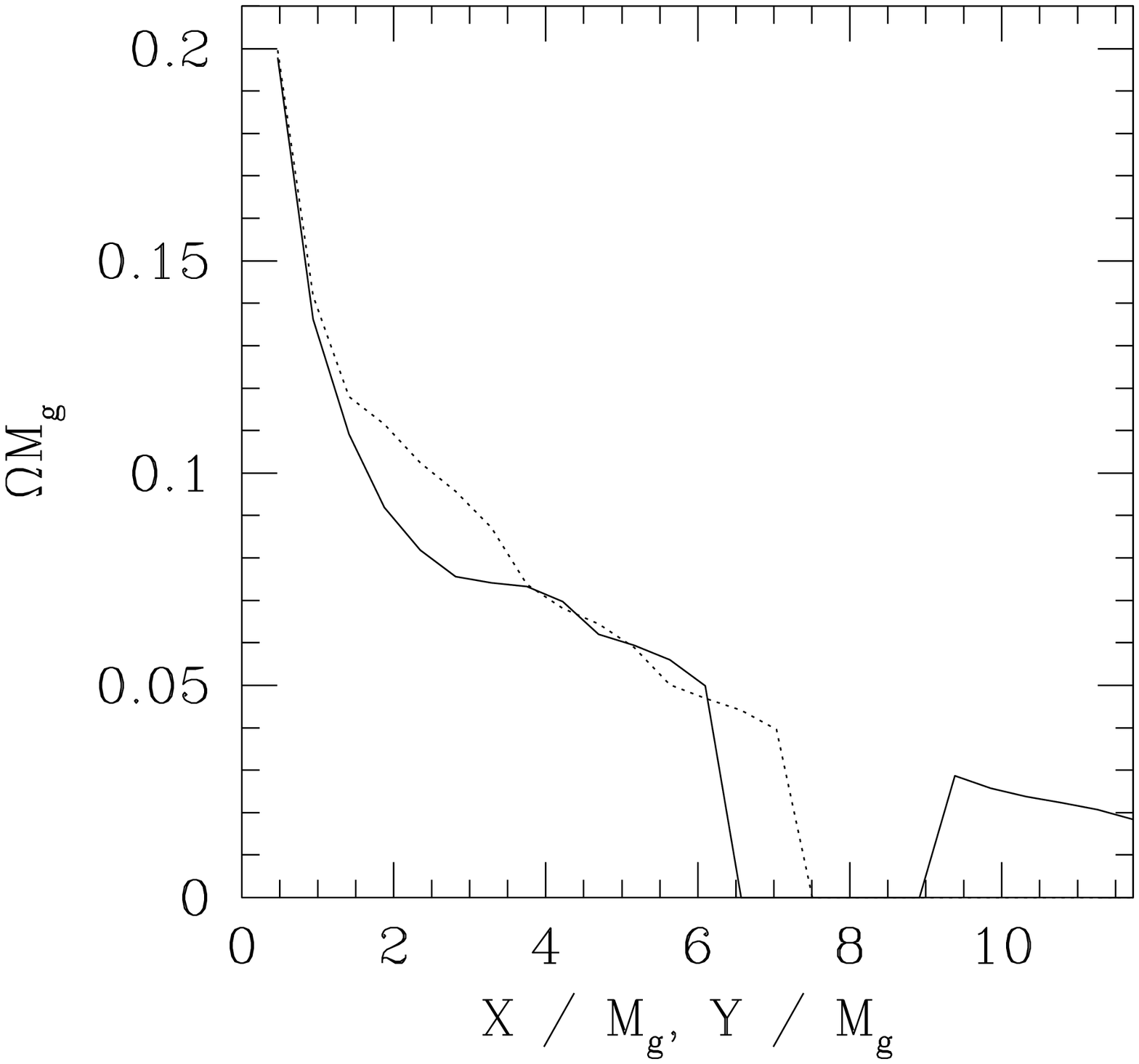}
\end{center}
\caption{The angular velocity $\Omega$ along the $x$-axis 
(solid line) and 
$y$-axis (dotted line) at $t=2.07P_{\rm orb}$ 
for model (C1). 
The length scale and $\Omega$ are shown in units of $GM_{g0}/c^2$  
and $c^3/GM_{g0}$, respectively. 
}
\end{figure}

\clearpage
\begin{figure}[t]
\begin{center}
\epsfxsize=3.5in
\leavevmode
\epsffile{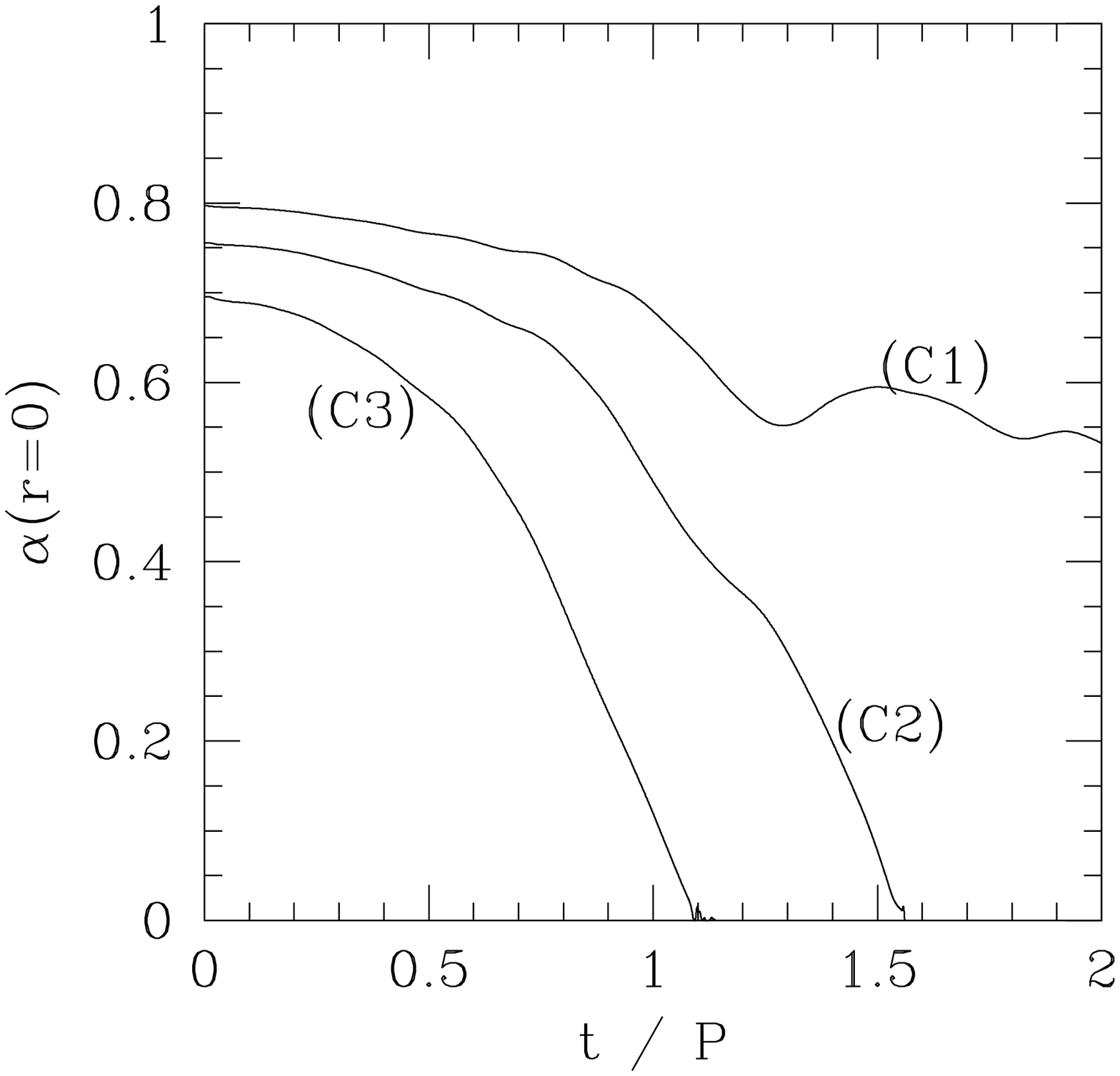}
\end{center}
\caption{$\alpha$ at $r=0$ as a function of $t/P_{\rm orb}$ 
for models (C1), (C2) and (C3). 
}
\end{figure}

\begin{figure}[t]
\begin{center}
\epsfxsize=3.5in
\leavevmode
\epsffile{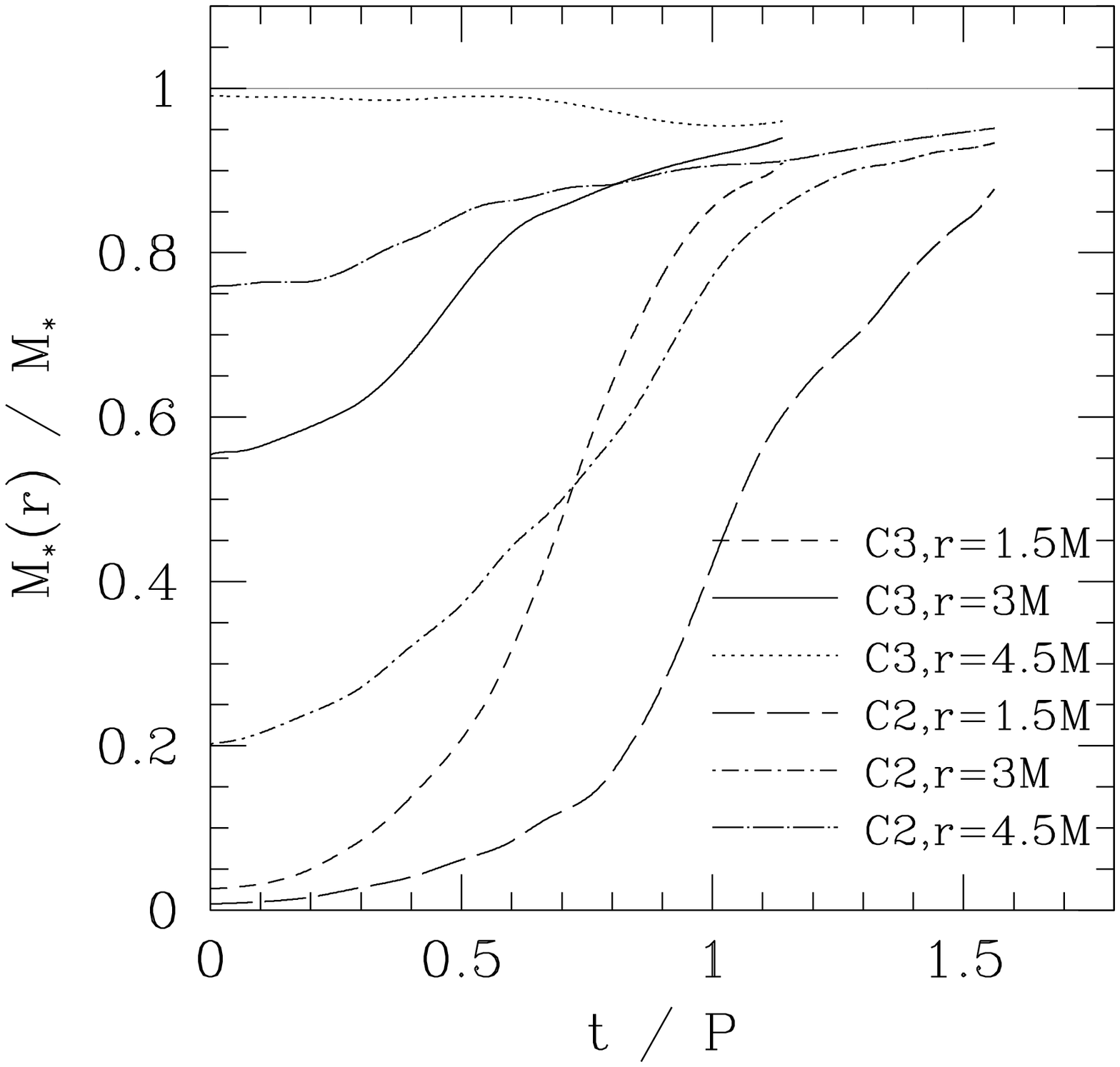}
\end{center}
\caption{Fraction of the rest mass inside a coordinate radius $r$ 
as a function of $t/P_{\rm orb}$ for models (C2) and (C3) in  which 
a black hole is formed after the merger. 
}
\end{figure}

\clearpage
\begin{figure}[t]
\begin{center}
\epsfxsize=2.6in
\leavevmode
\epsffile{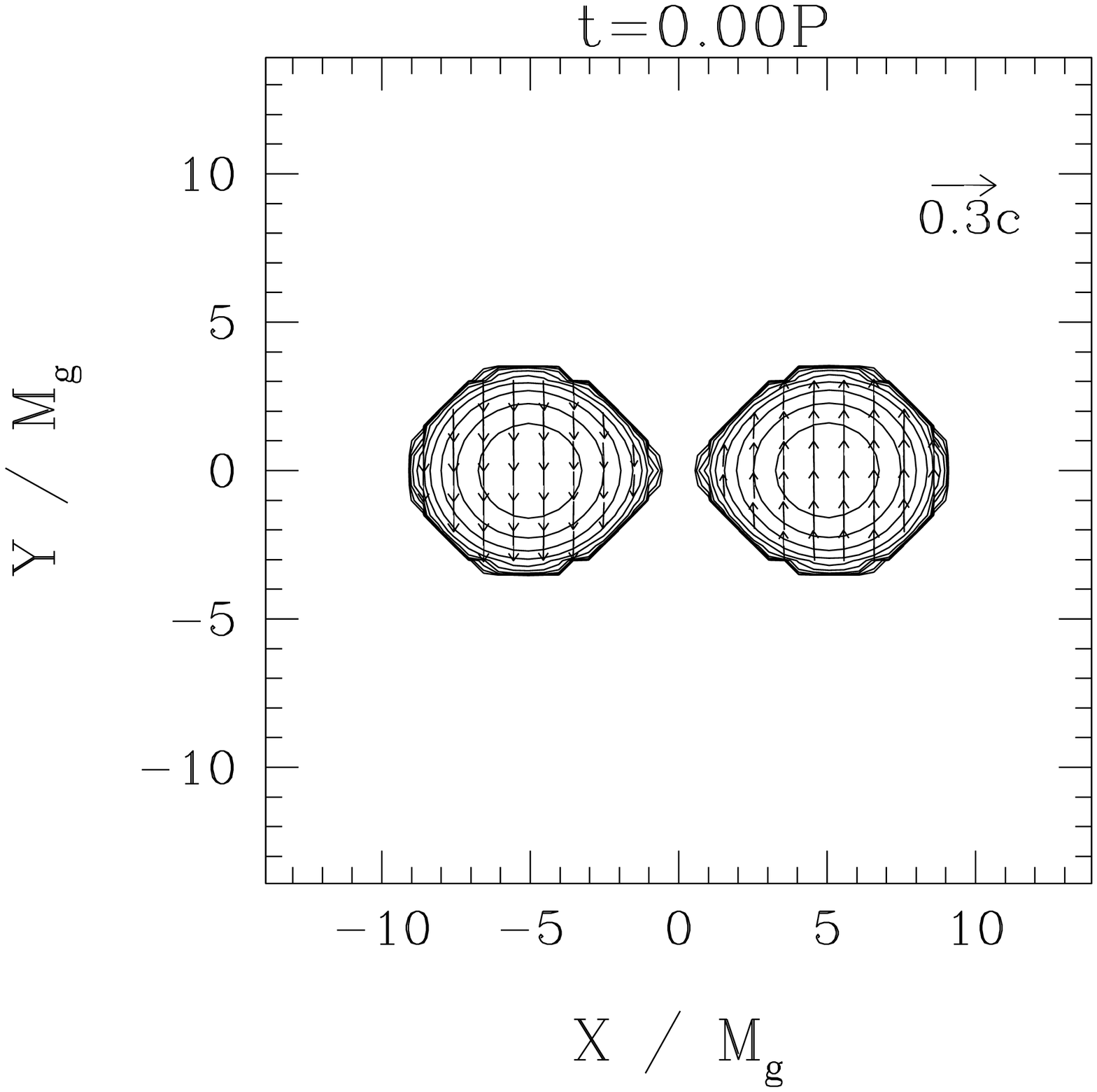}
\epsfxsize=2.6in
\leavevmode
\epsffile{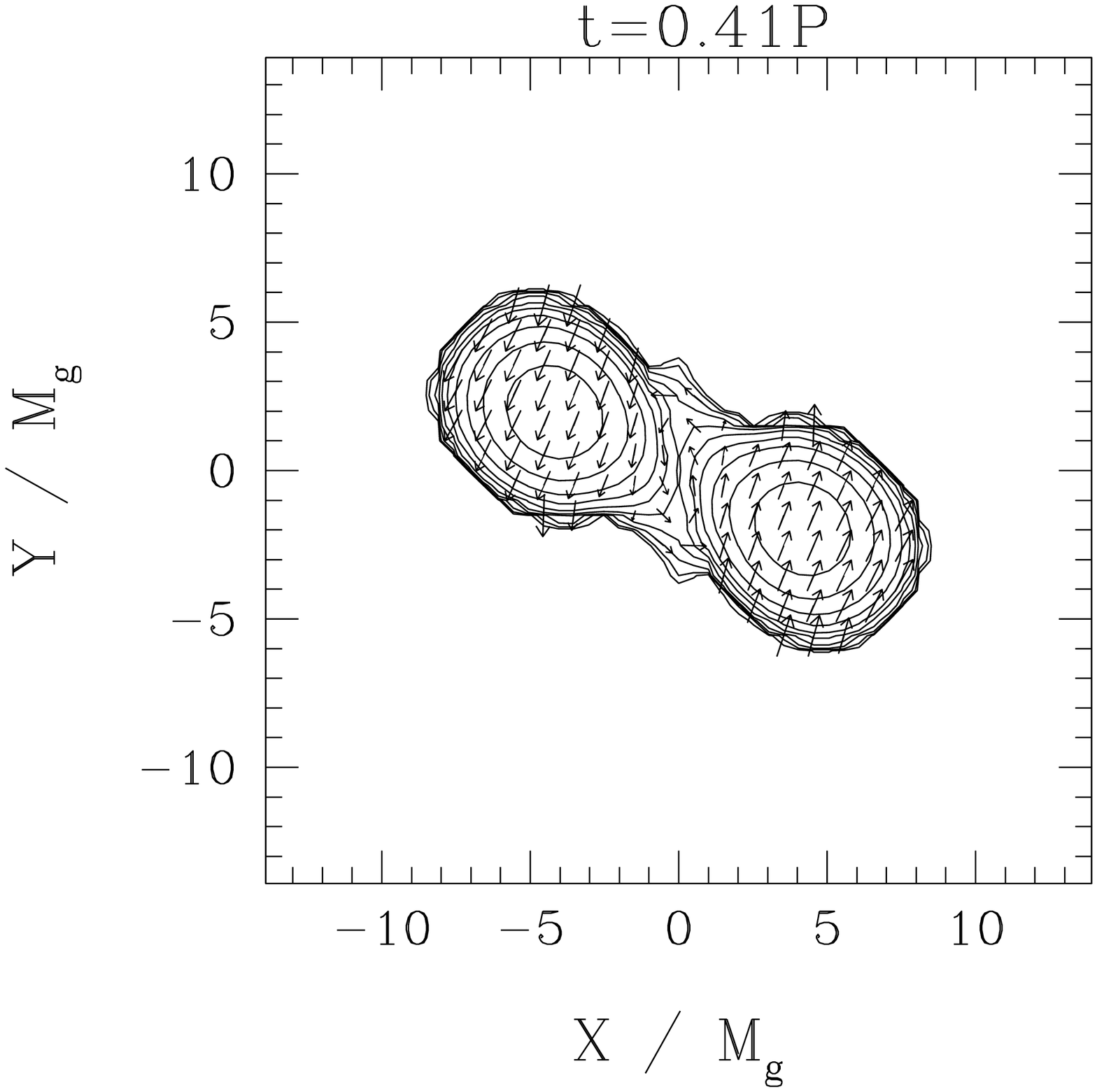}\\
\epsfxsize=2.6in
\leavevmode
\epsffile{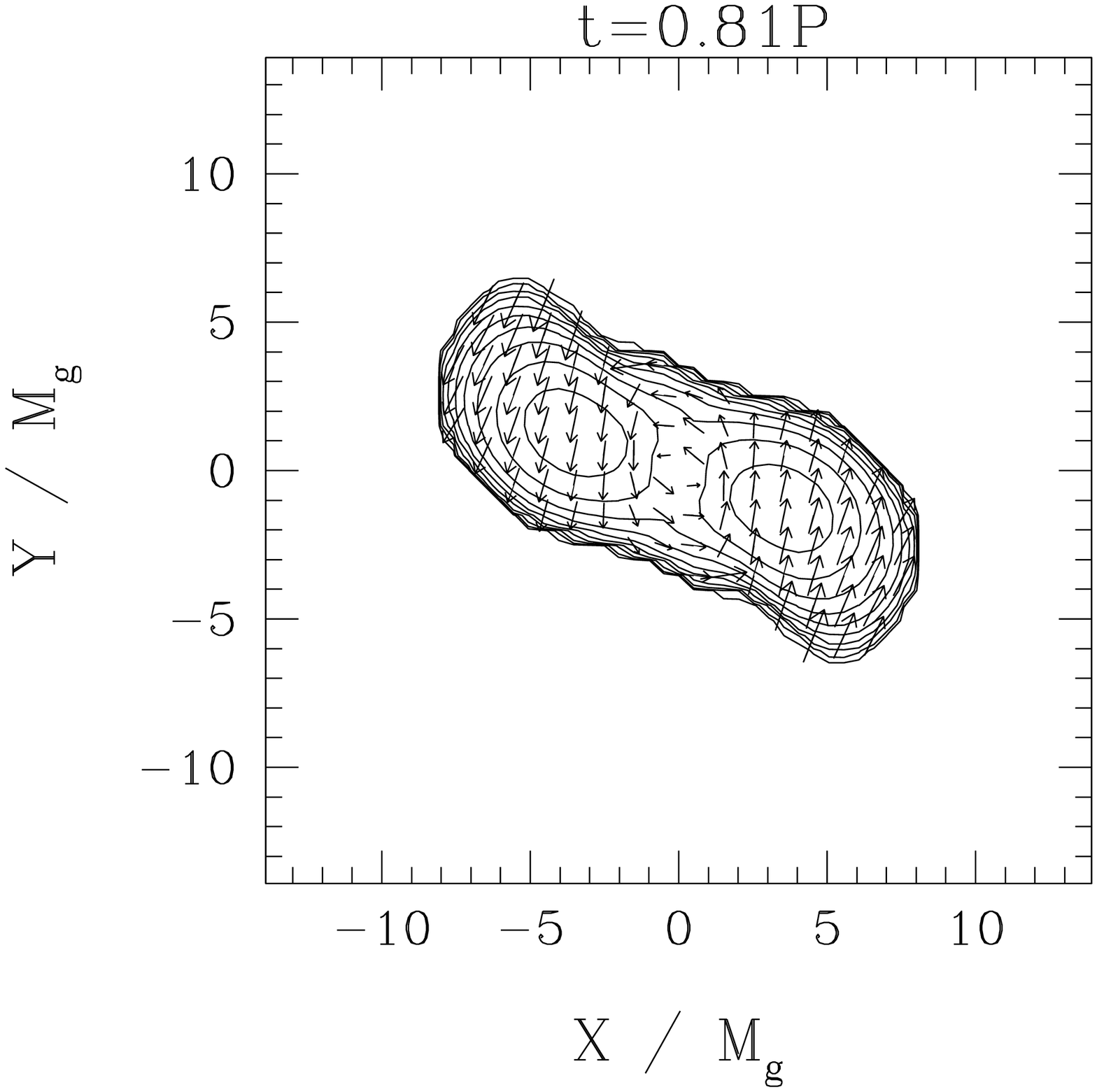}
\epsfxsize=2.6in
\leavevmode
\epsffile{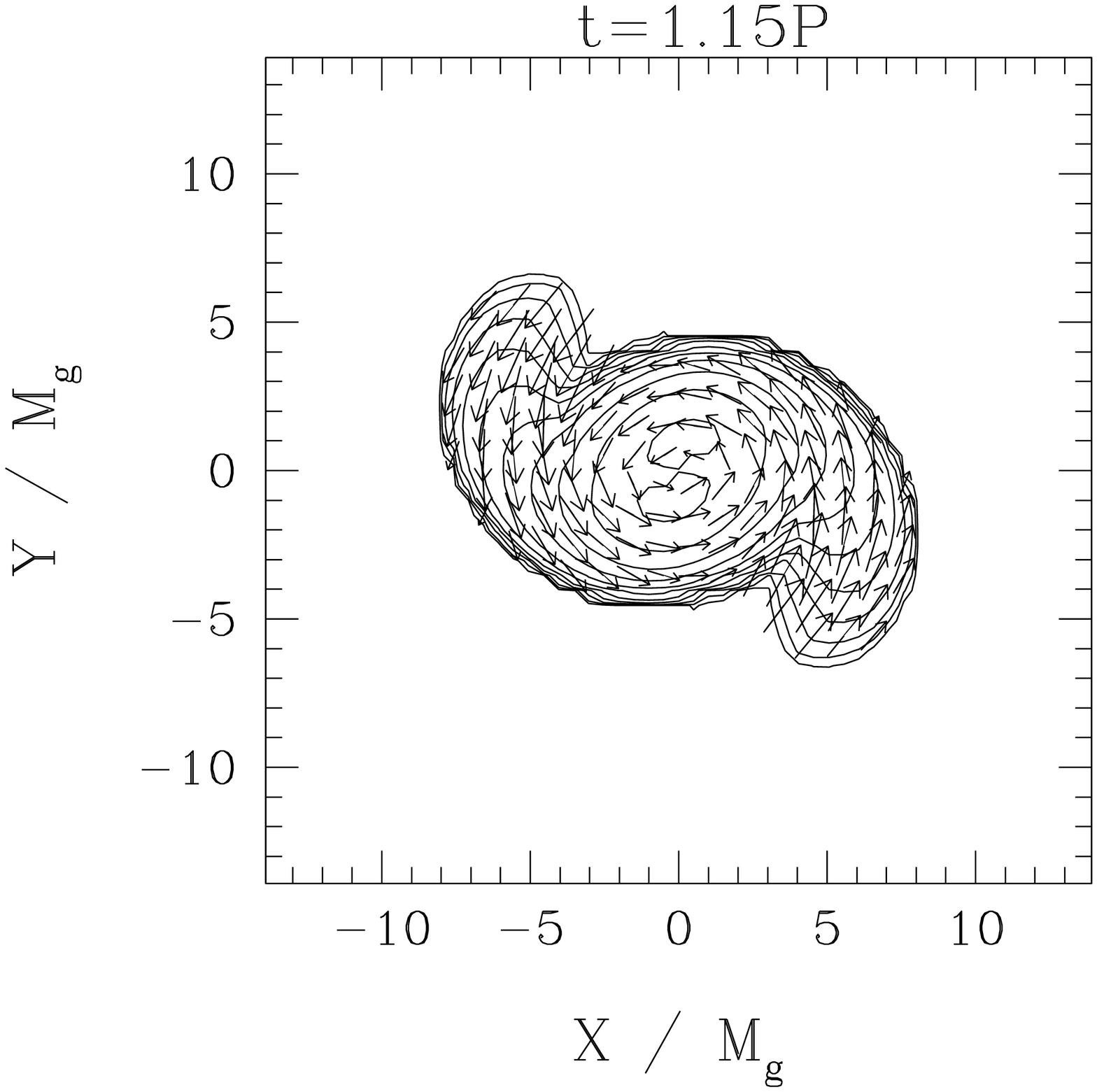} \\
\epsfxsize=2.6in
\leavevmode
\epsffile{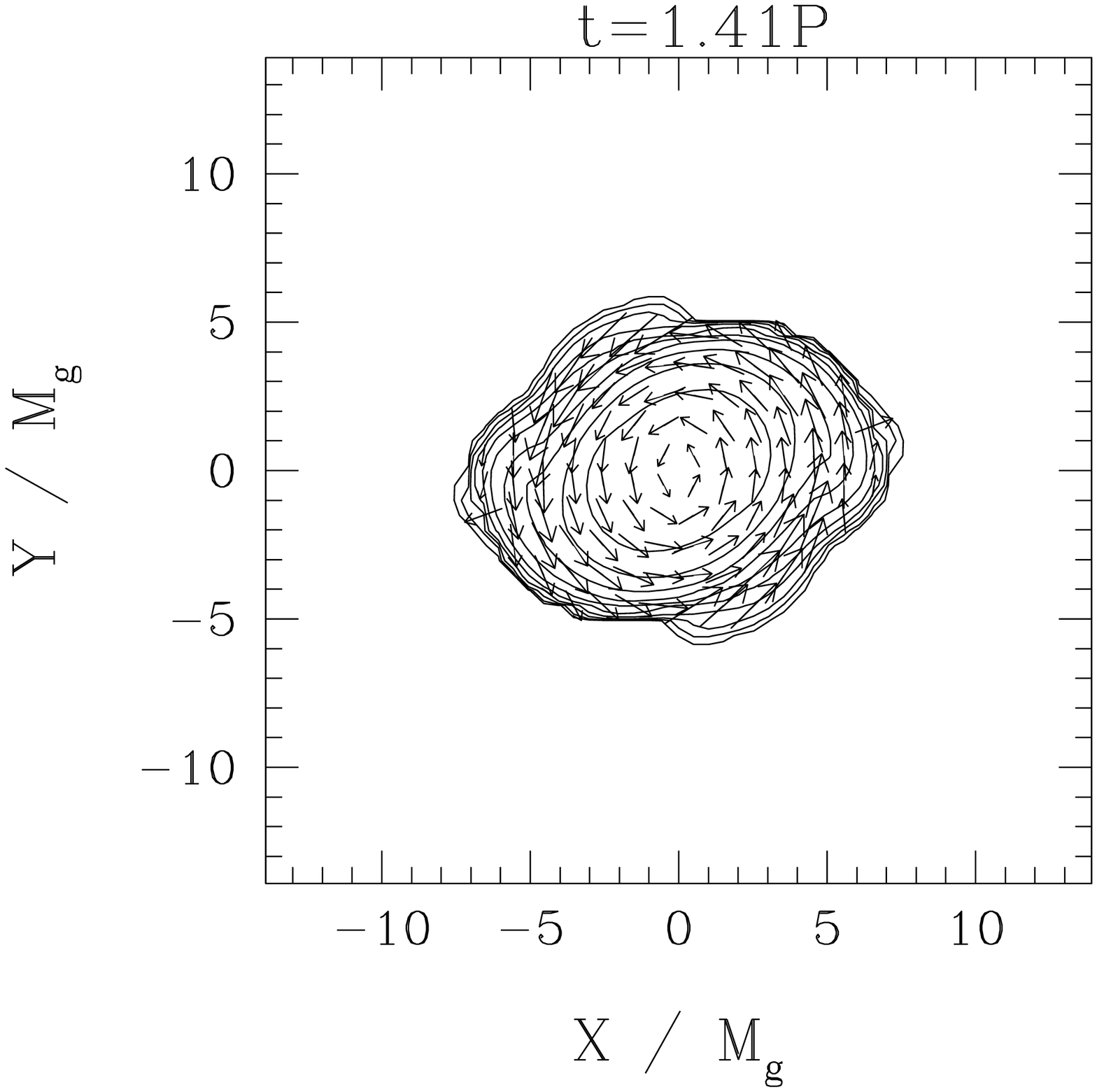}
\epsfxsize=2.6in
\leavevmode
\epsffile{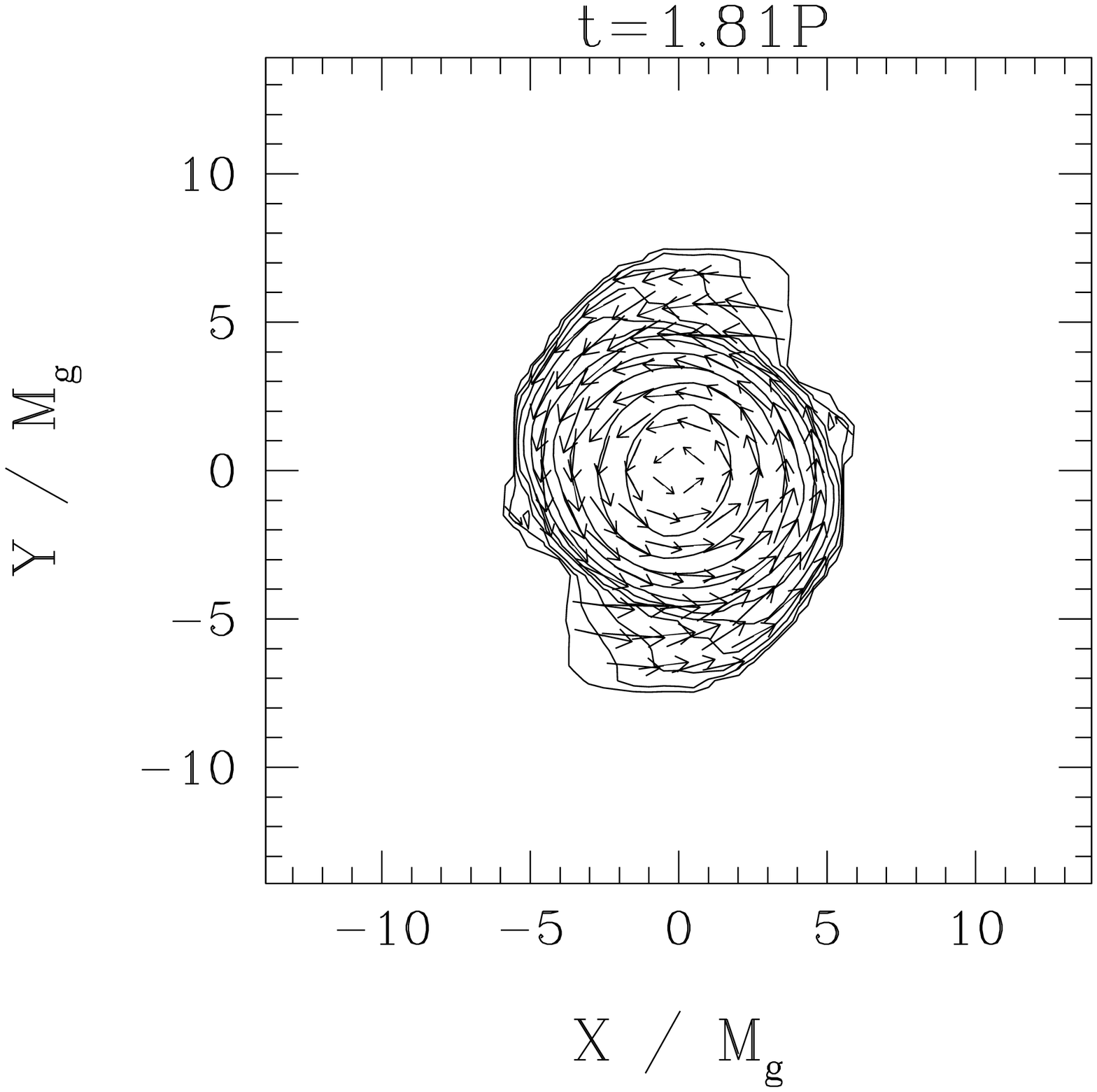} 
\caption{
The same as Fig. 2, but for model (I1). 
The contour lines are drawn for 
$\rho_*/\rho_{*~{\rm max}}=10^{-0.3j}$, 
where $\rho_{*~{\rm max}}=0.00401$, for $j=0,1,2,\cdots,10$. 
}
\end{center}
\end{figure}

\clearpage
\begin{figure}[t]
\begin{center}
\epsfxsize=2.6in
\leavevmode
\epsffile{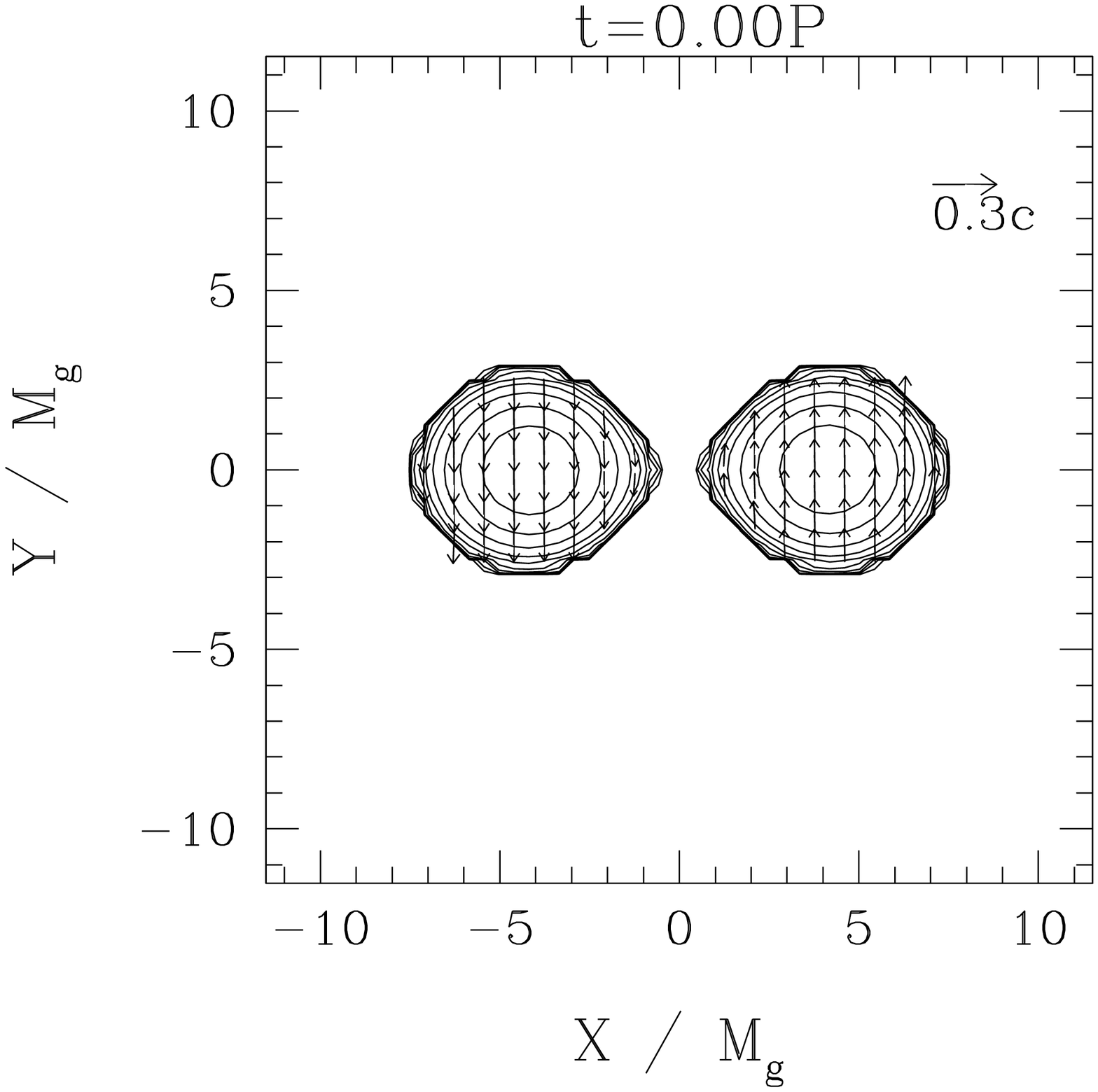}
\epsfxsize=2.6in
\leavevmode
\epsffile{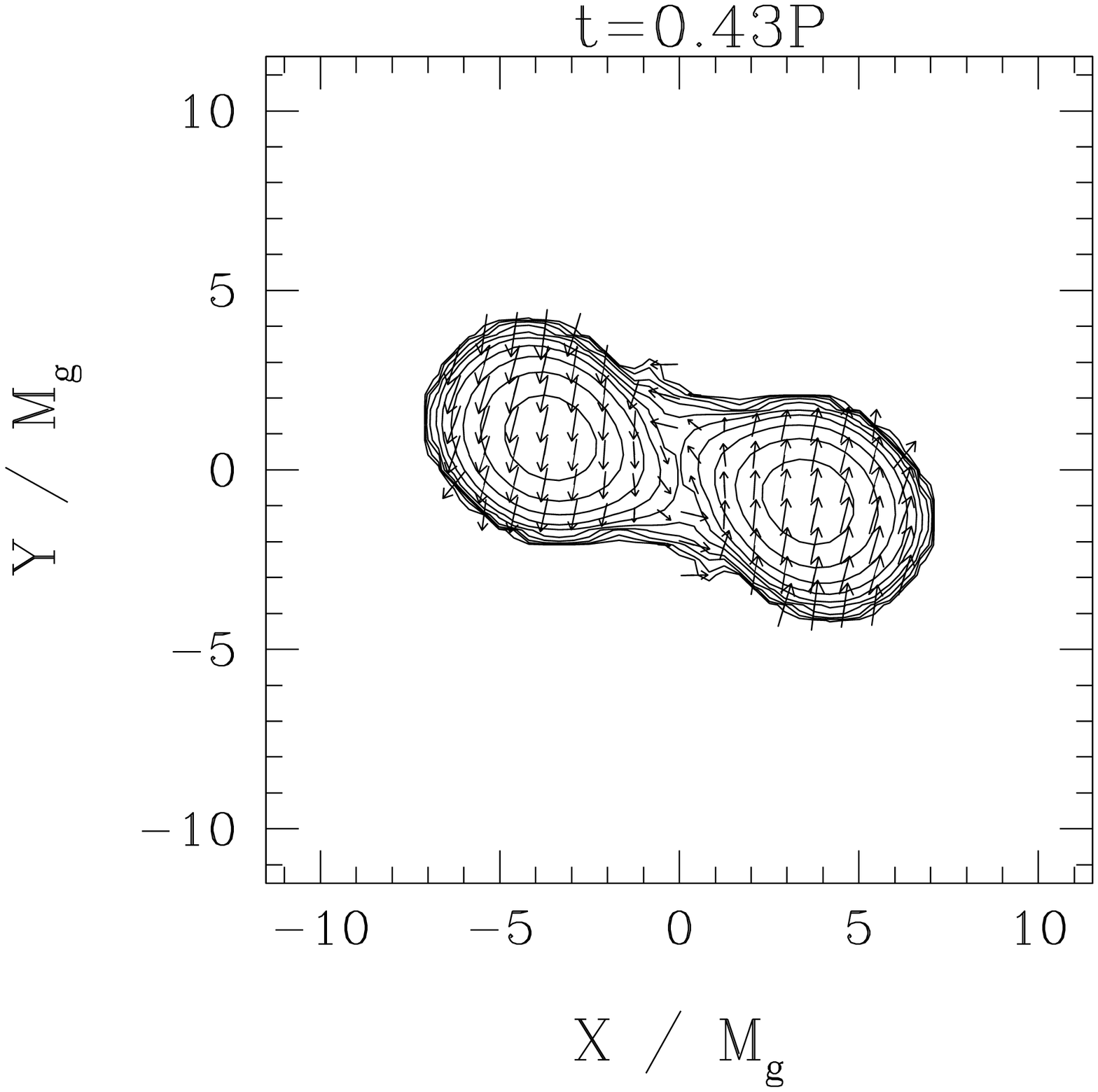}\\
\epsfxsize=2.6in
\leavevmode
\epsffile{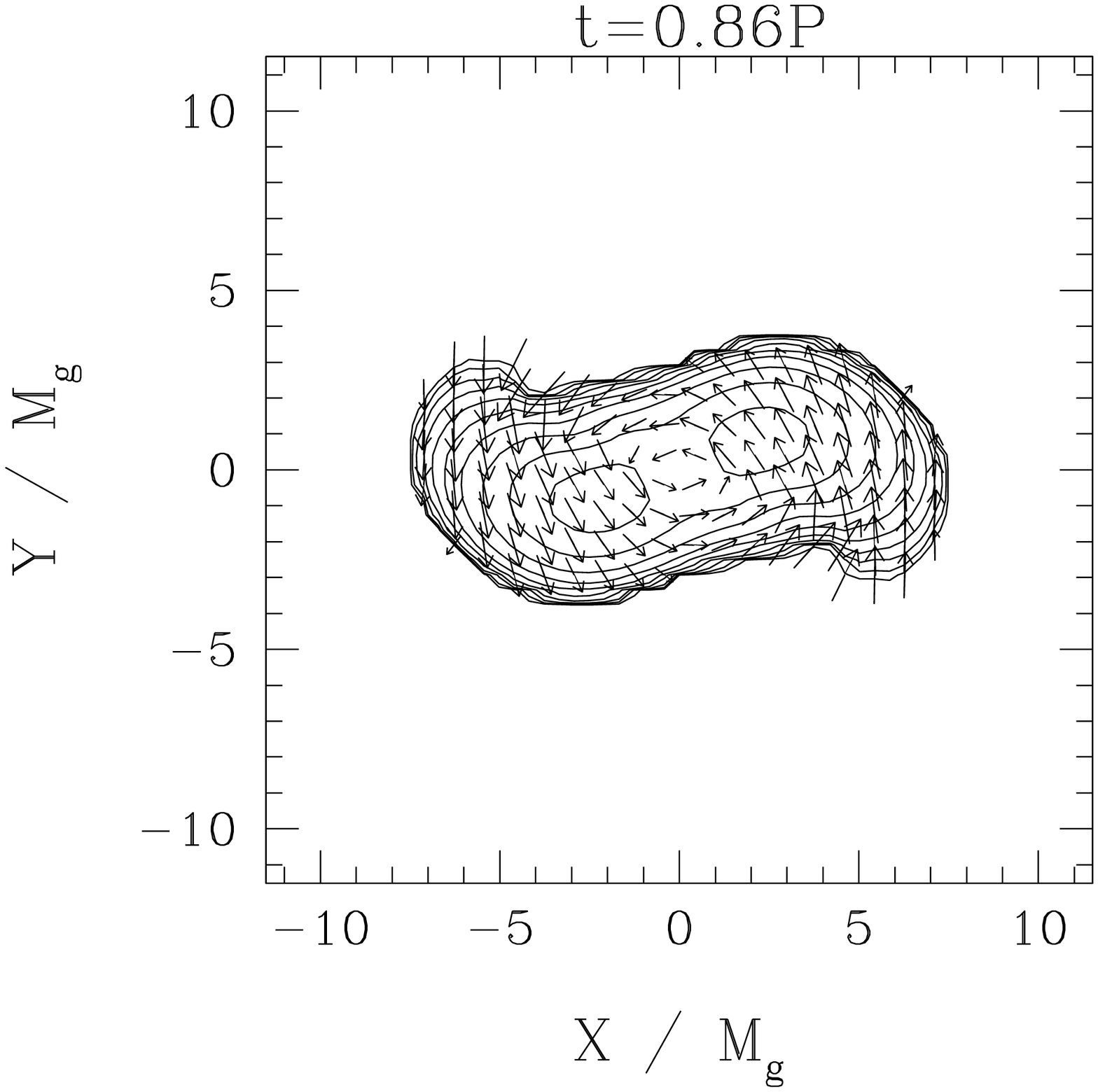}
\epsfxsize=2.6in
\leavevmode
\epsffile{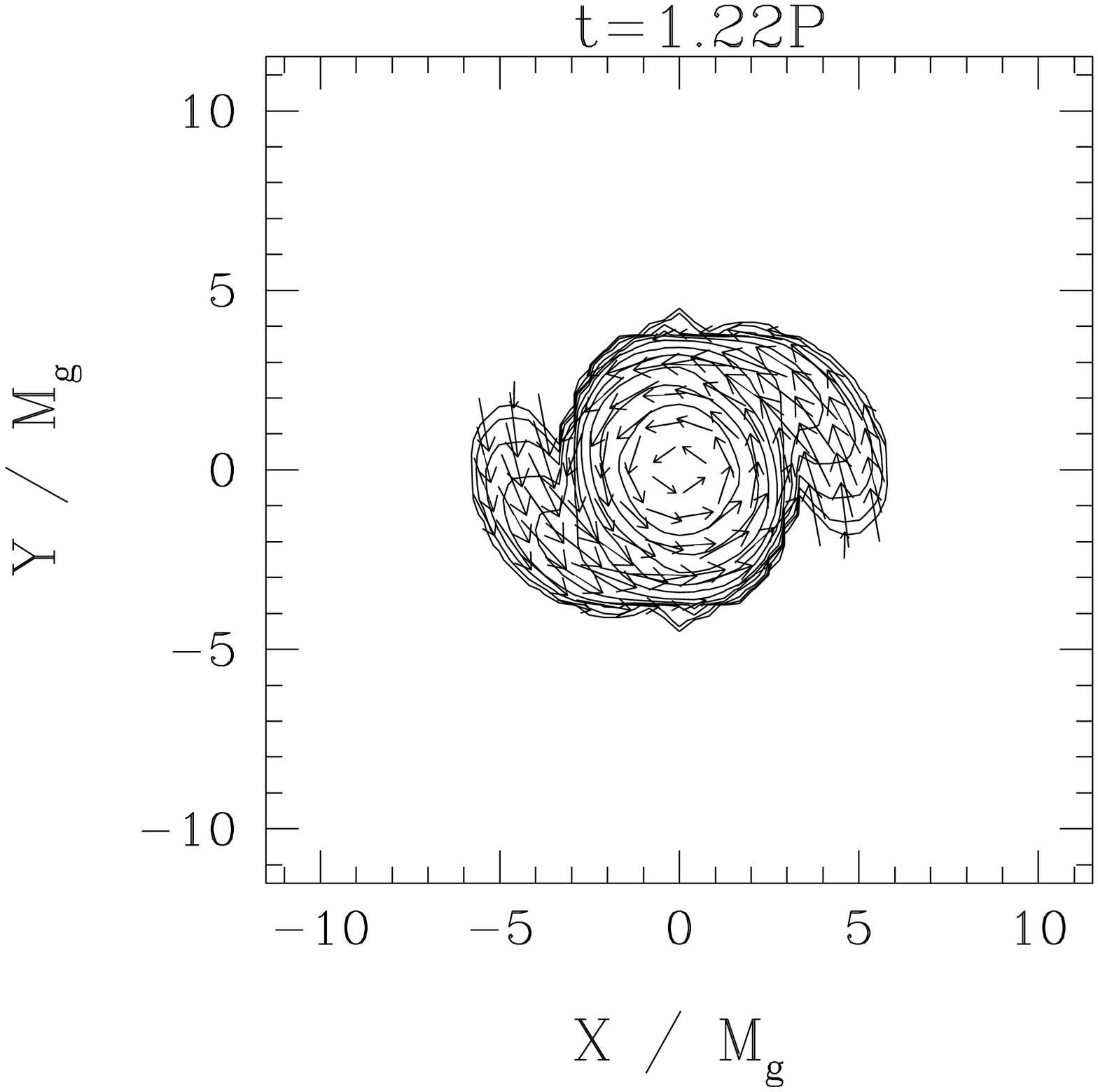} \\
\epsfxsize=2.6in
\leavevmode
\epsffile{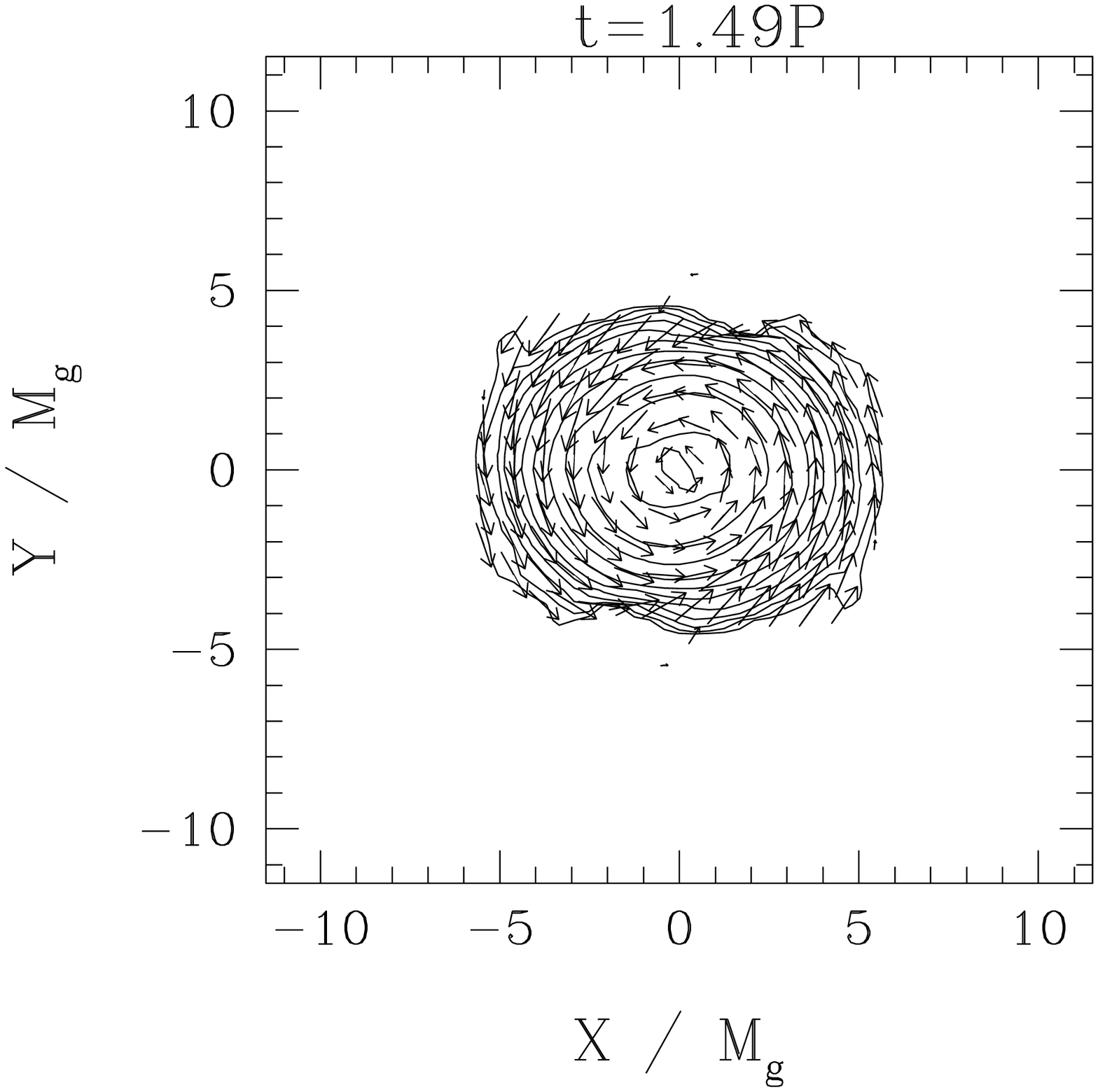}
\epsfxsize=2.6in
\leavevmode
\epsffile{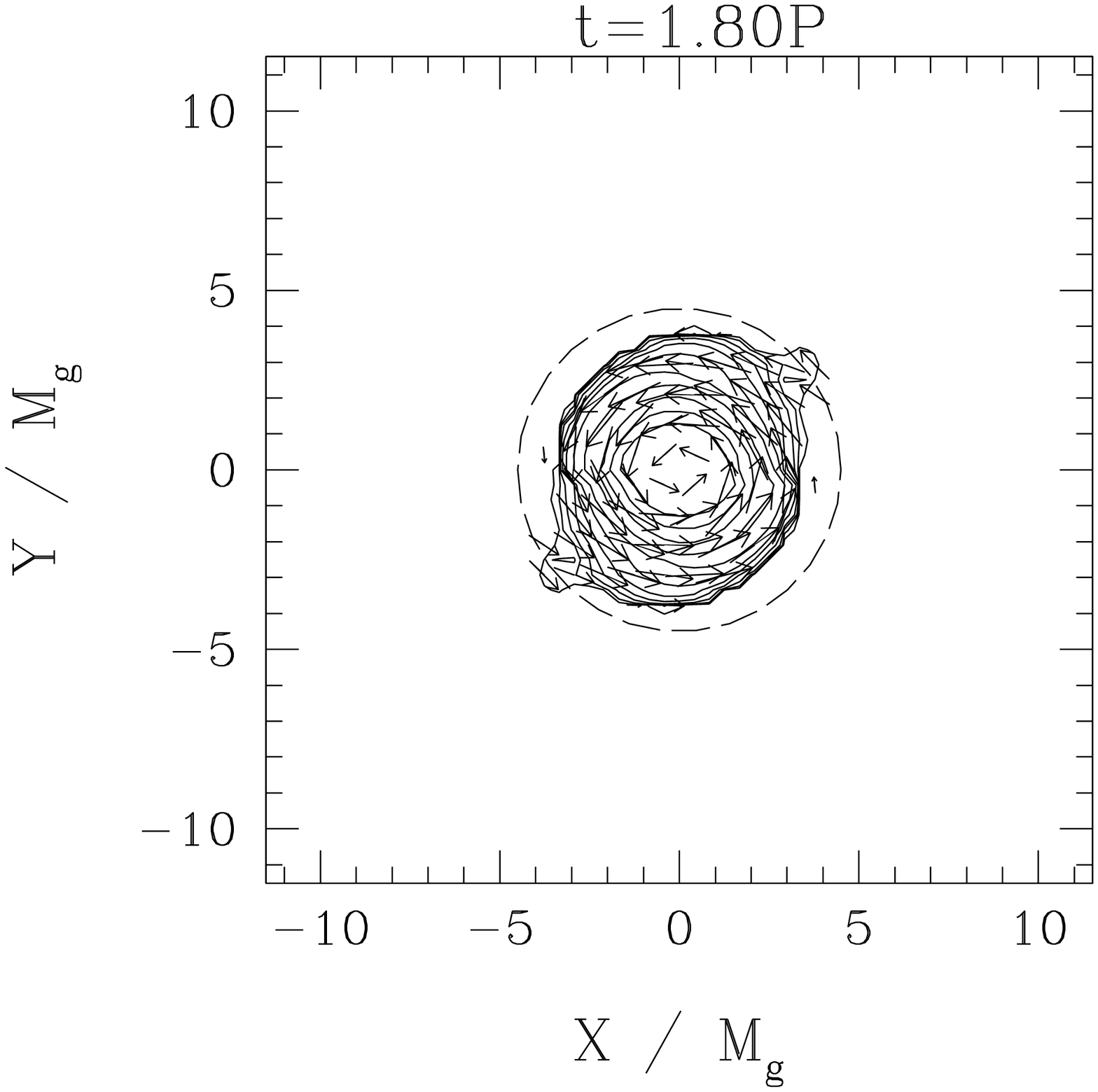} 
\caption{
The same as Fig. 2, but for model (I2). 
The contour lines are drawn for 
$\rho_*/\rho_{*~{\rm max}}=10^{-0.3j}$, 
where $\rho_{*~{\rm max}}=0.00642$, for $j=0,1,2,\cdots,10$. 
The dashed line in the last figure denotes the circle with 
$r=4.5M_{g0} $ 
within which more than $99\%$ of the total rest mass is included. 
}
\end{center}
\end{figure}

\clearpage
\begin{figure}[t]
\begin{center}
\epsfxsize=2.6in
\leavevmode
\epsffile{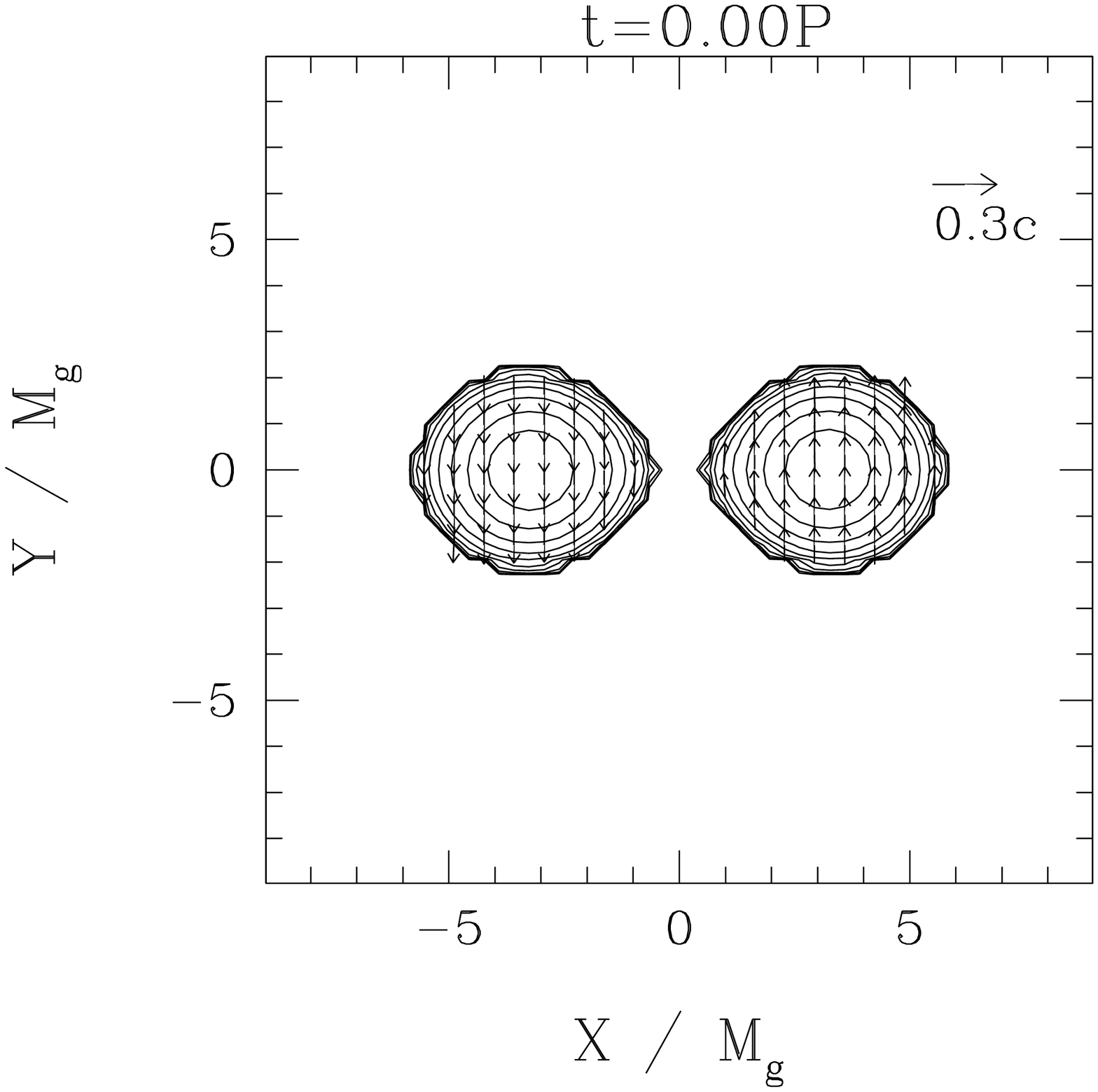}
\epsfxsize=2.6in
\leavevmode
\epsffile{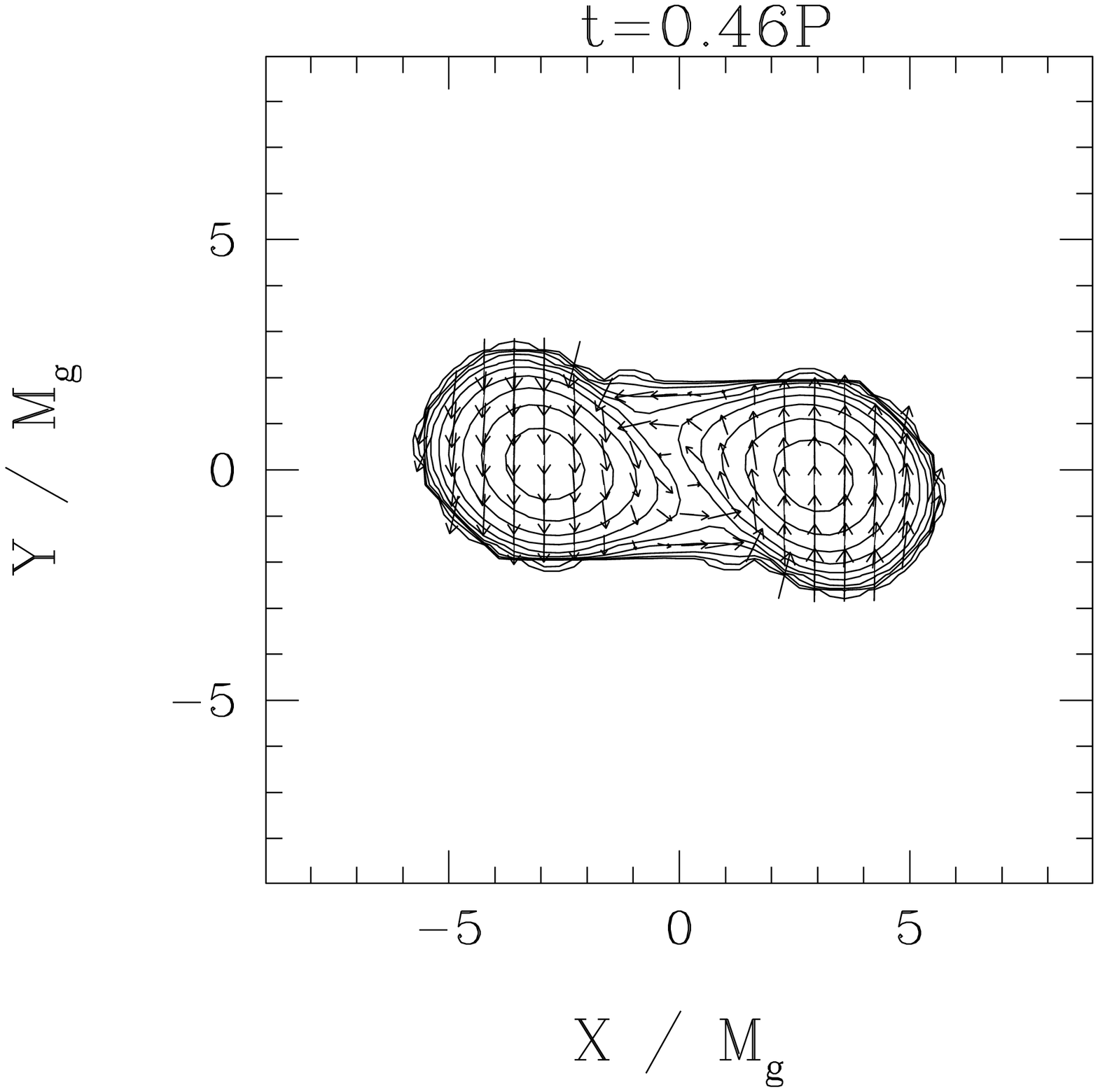}\\
\epsfxsize=2.6in
\leavevmode
\epsffile{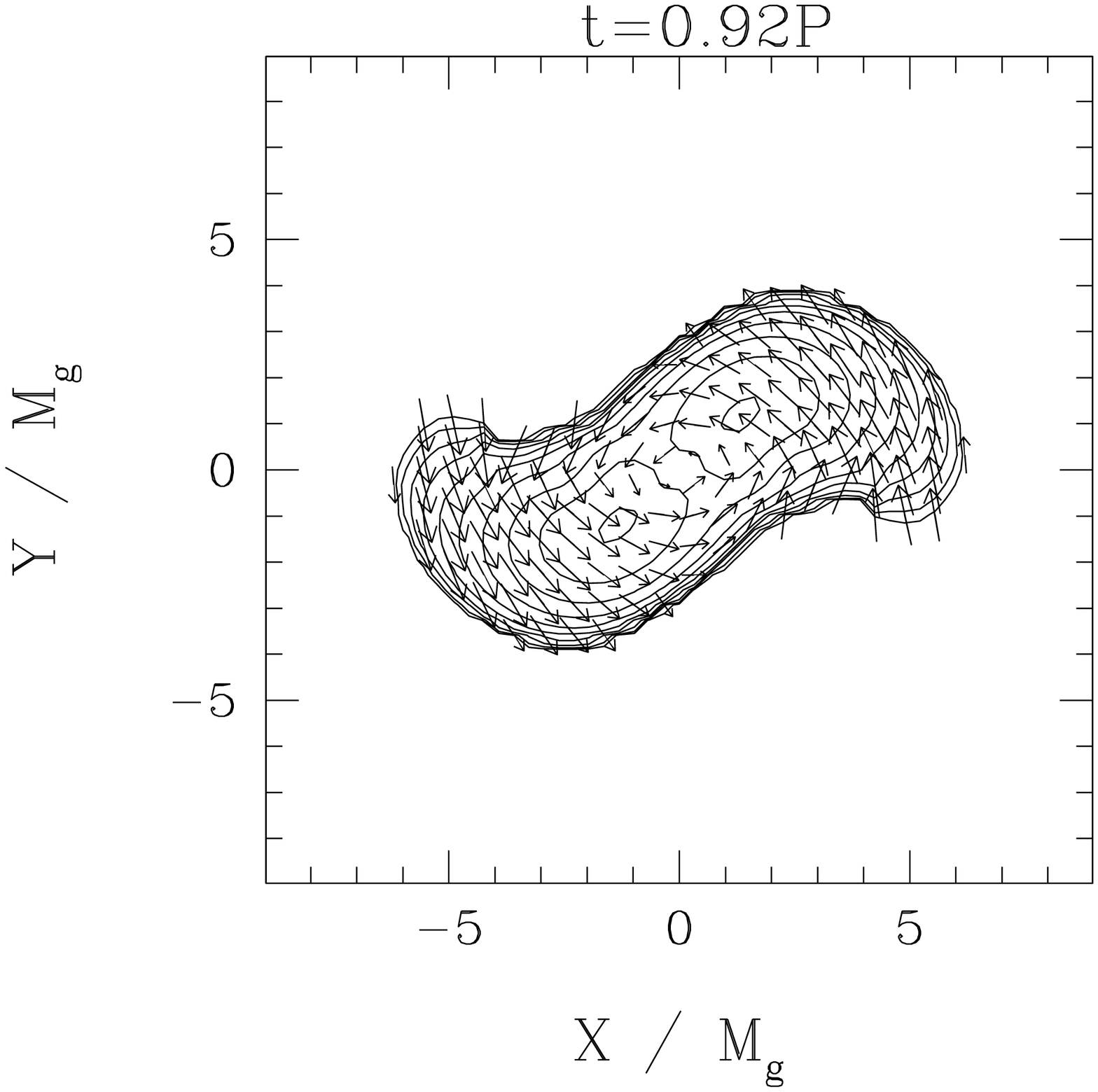}
\epsfxsize=2.6in
\leavevmode
\epsffile{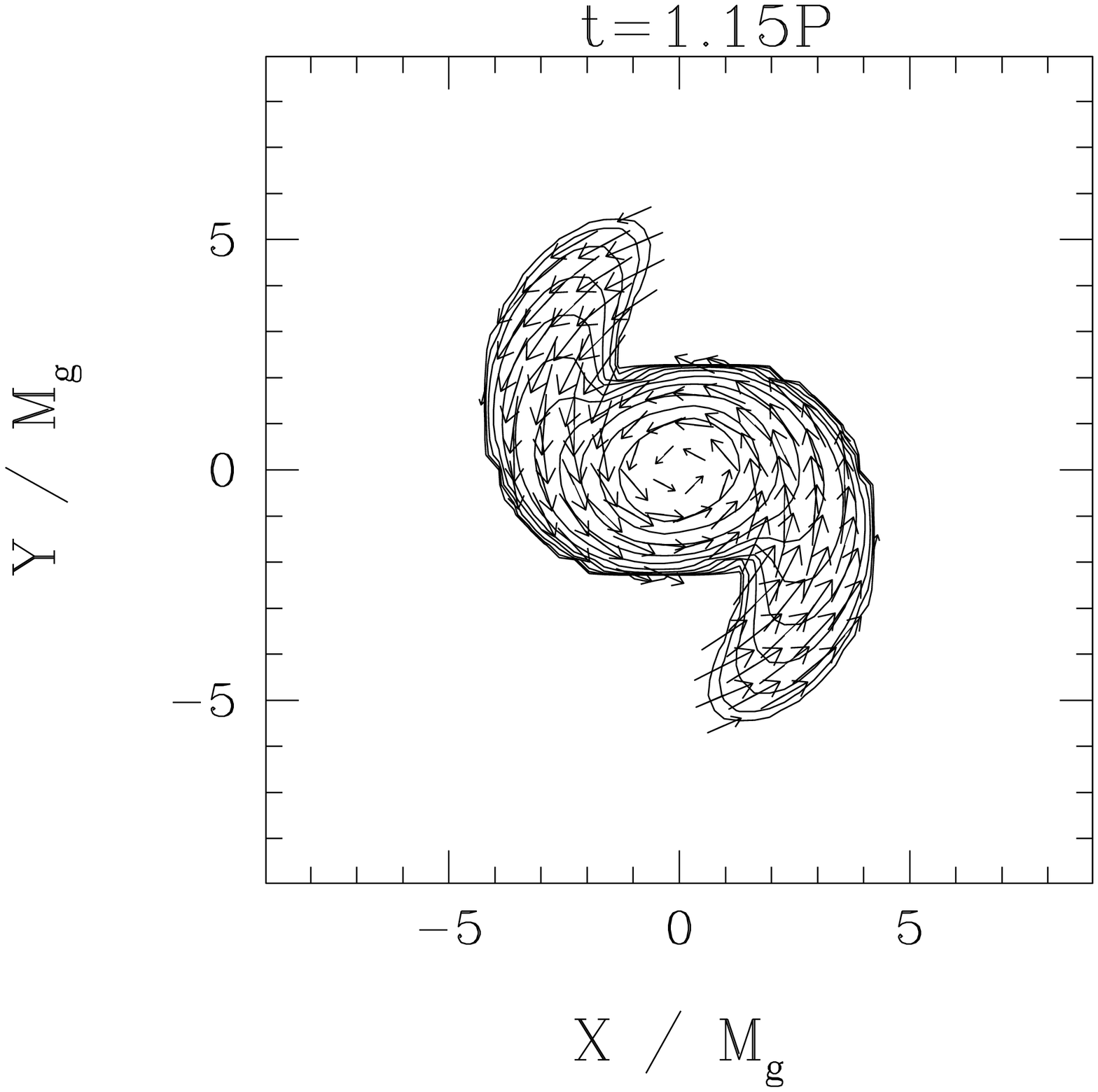} \\
\epsfxsize=2.6in
\leavevmode
\epsffile{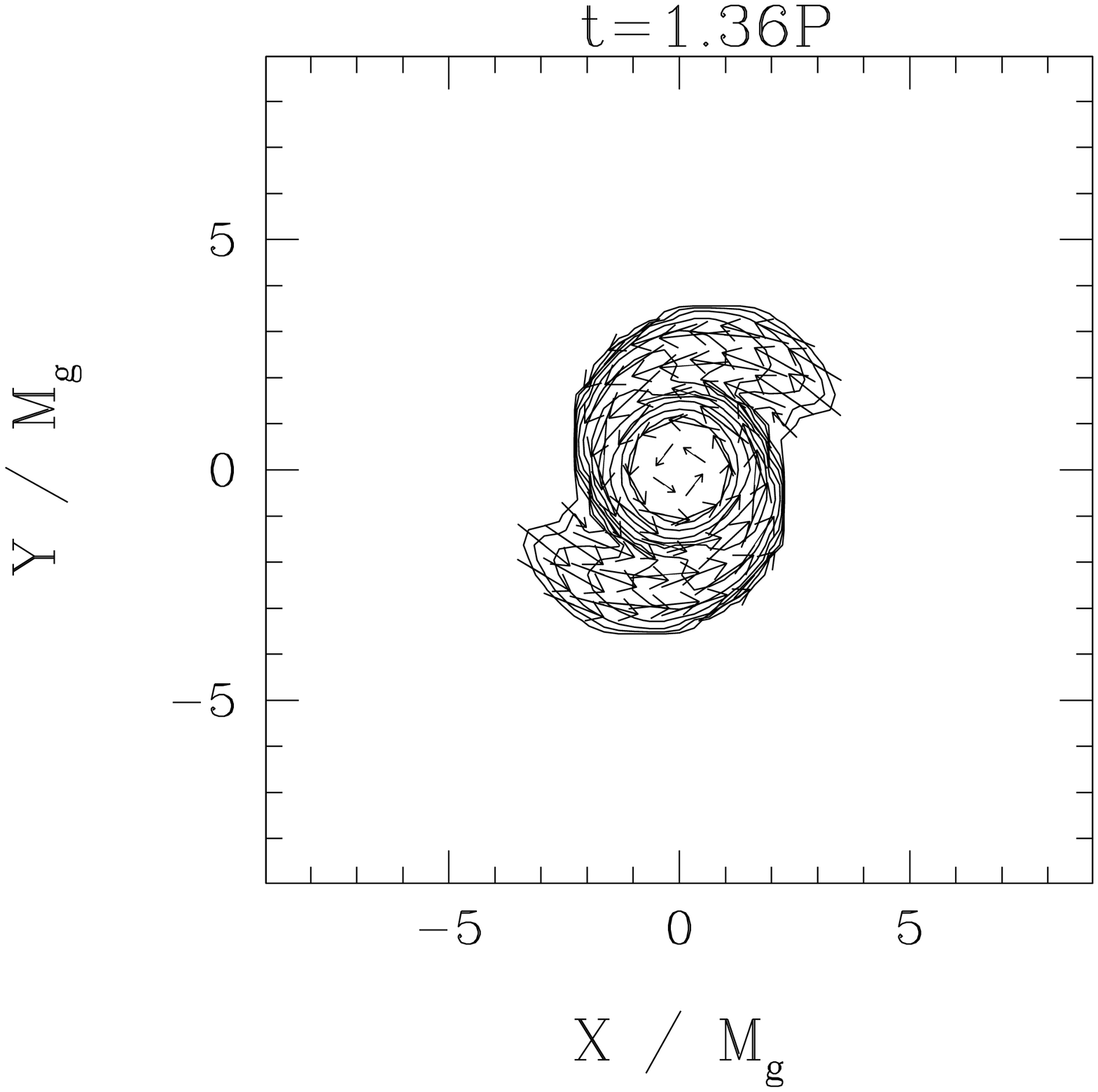}
\epsfxsize=2.6in
\leavevmode
\epsffile{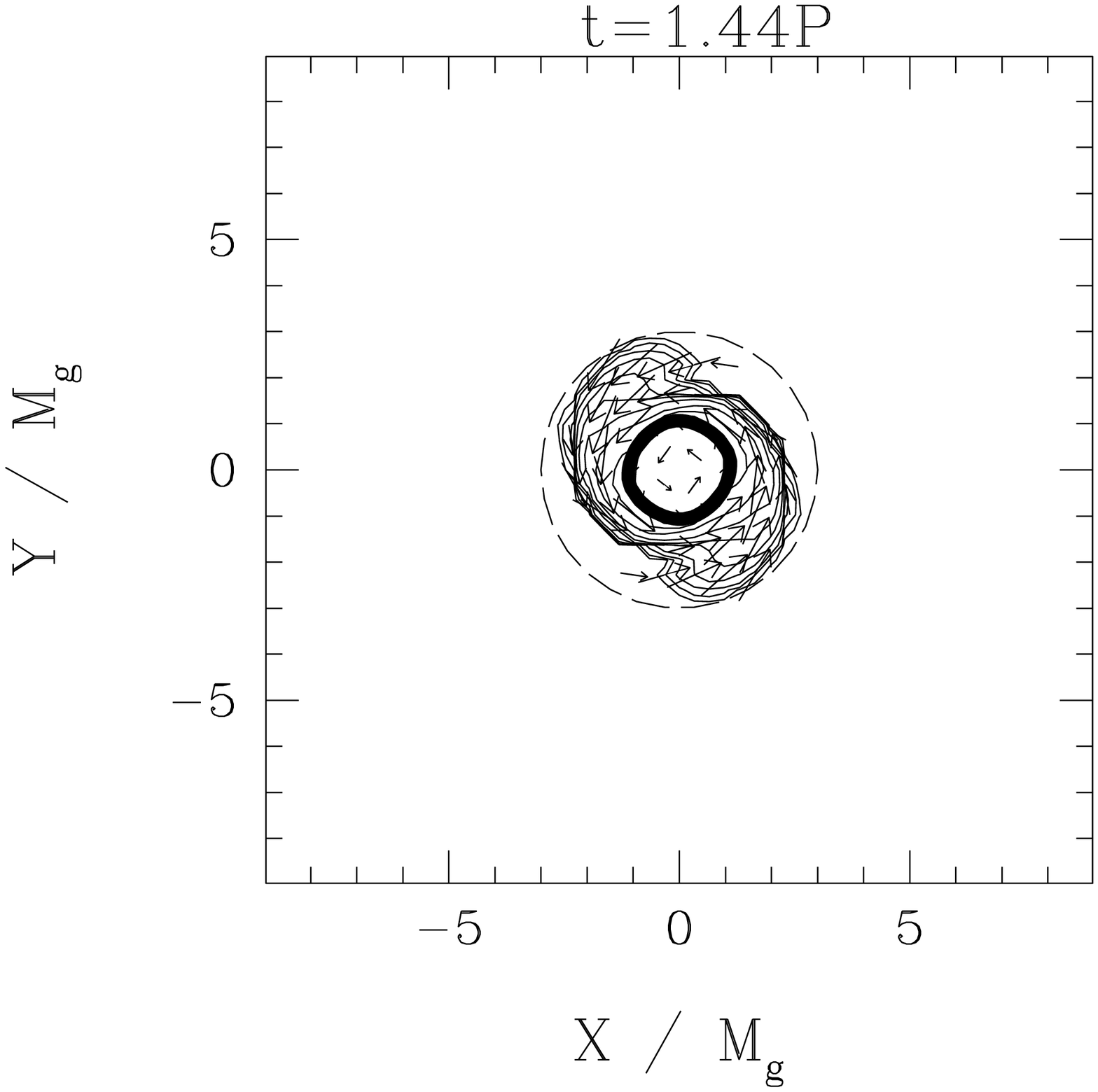} 
\caption{
The same as Fig. 2, but for model (I3). 
The contour lines are drawn for 
$\rho_*/\rho_{*~{\rm max}}=10^{-0.3j}$, 
where $\rho_{*~{\rm max}}=0.0136$, for $j=0,1,2,\cdots,10$. 
The dashed line in the last snapshot denotes the circle with 
$r=3M_{g0}$ 
within which more than $99\%$ of the total rest mass is included. 
The thick solid line for $r \sim M_{g0}$ in the last snapshot
denotes the location of the apparent horizon. 
Note that there are $\sim 7$ grid points 
along the radius of the apparent horizon.  
}
\end{center}
\end{figure}

\clearpage
\begin{figure}[t]
\begin{center}
\epsfxsize=3.5in
\leavevmode
\epsffile{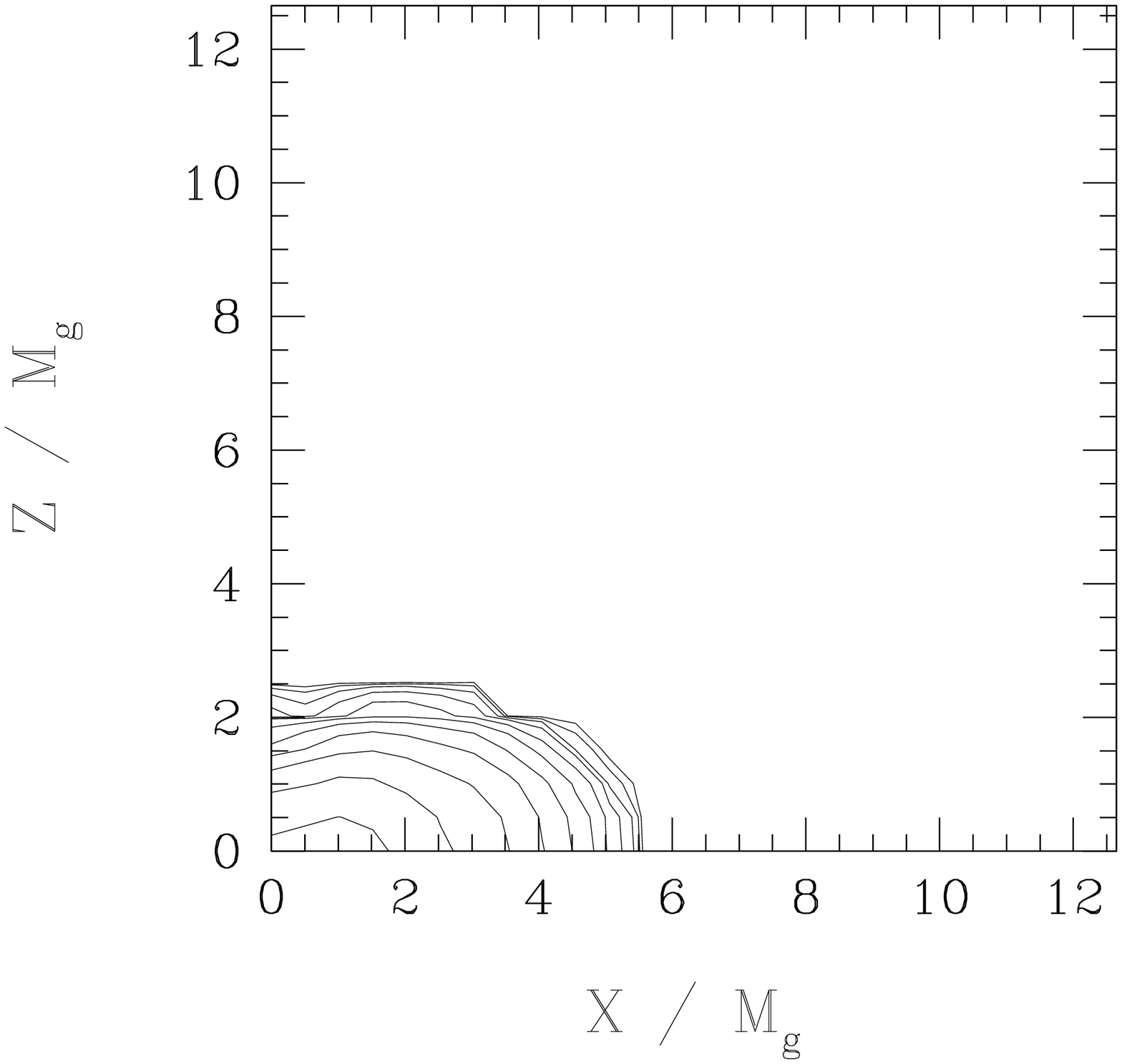}
\end{center}
\caption{
The density contour lines for $\rho_*$ 
in the $y=0$ plane at $t=1.81P_{\rm orb}$ for model (I1). 
The contour lines are drawn in the same way as in Fig. 9. 
The length scale is shown in units of $GM_{g0}/c^2$. 
}
\end{figure}

\begin{figure}[t]
\begin{center}
\epsfxsize=3.5in
\leavevmode
\epsffile{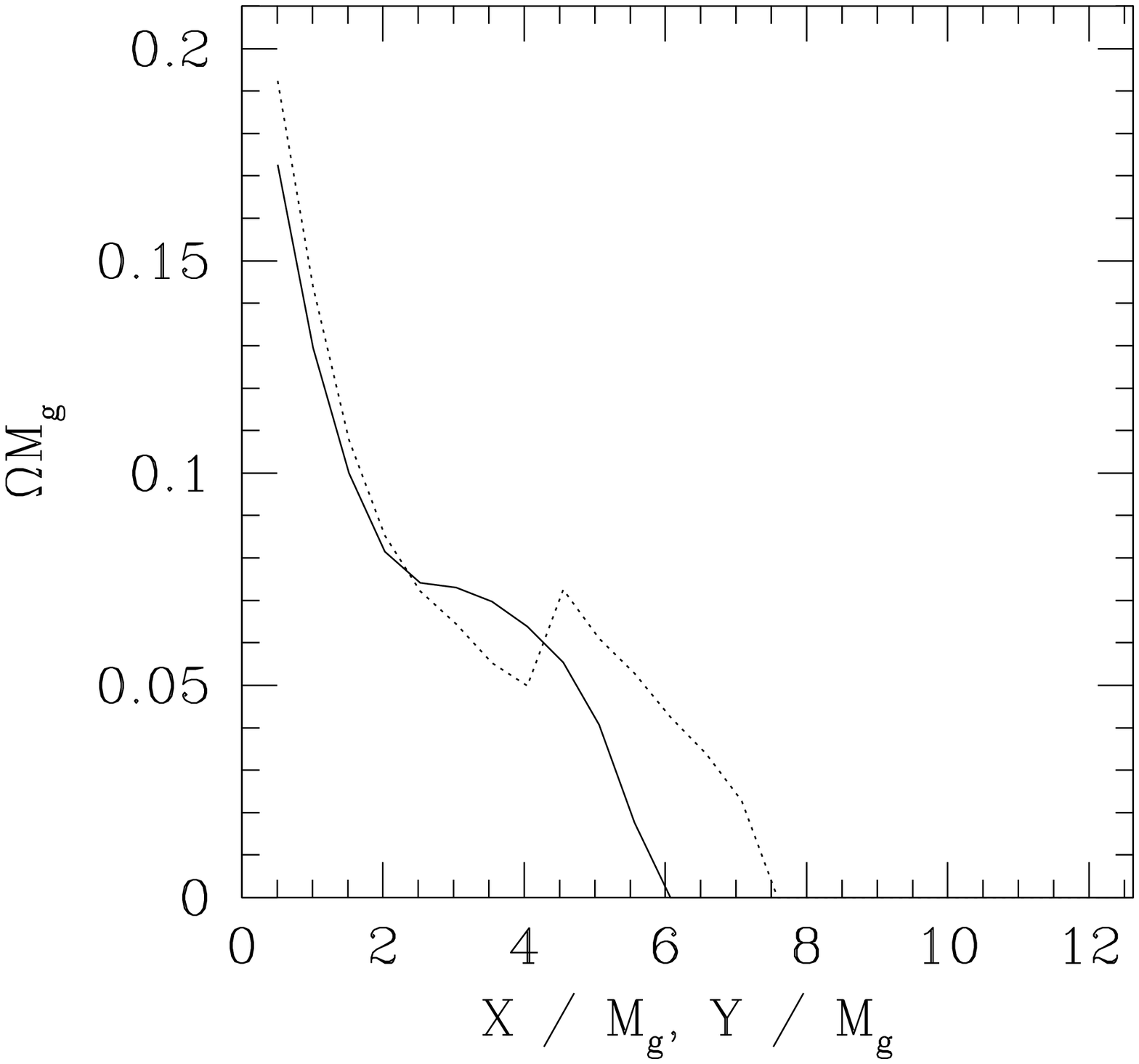}
\end{center}
\caption{The angular velocity $\Omega$ along the $x$-axis 
(solid line) and 
$y$-axis (dotted line) at $t=1.81P_{\rm orb}$ 
for model (I1). 
The length scale and $\Omega$ are shown in units of $GM_{g0}/c^2$  
and $c^3/GM_{g0}$, respectively. 
}
\end{figure}

\clearpage
\begin{figure}[t]
\begin{center}
\epsfxsize=3.5in
\leavevmode
\epsffile{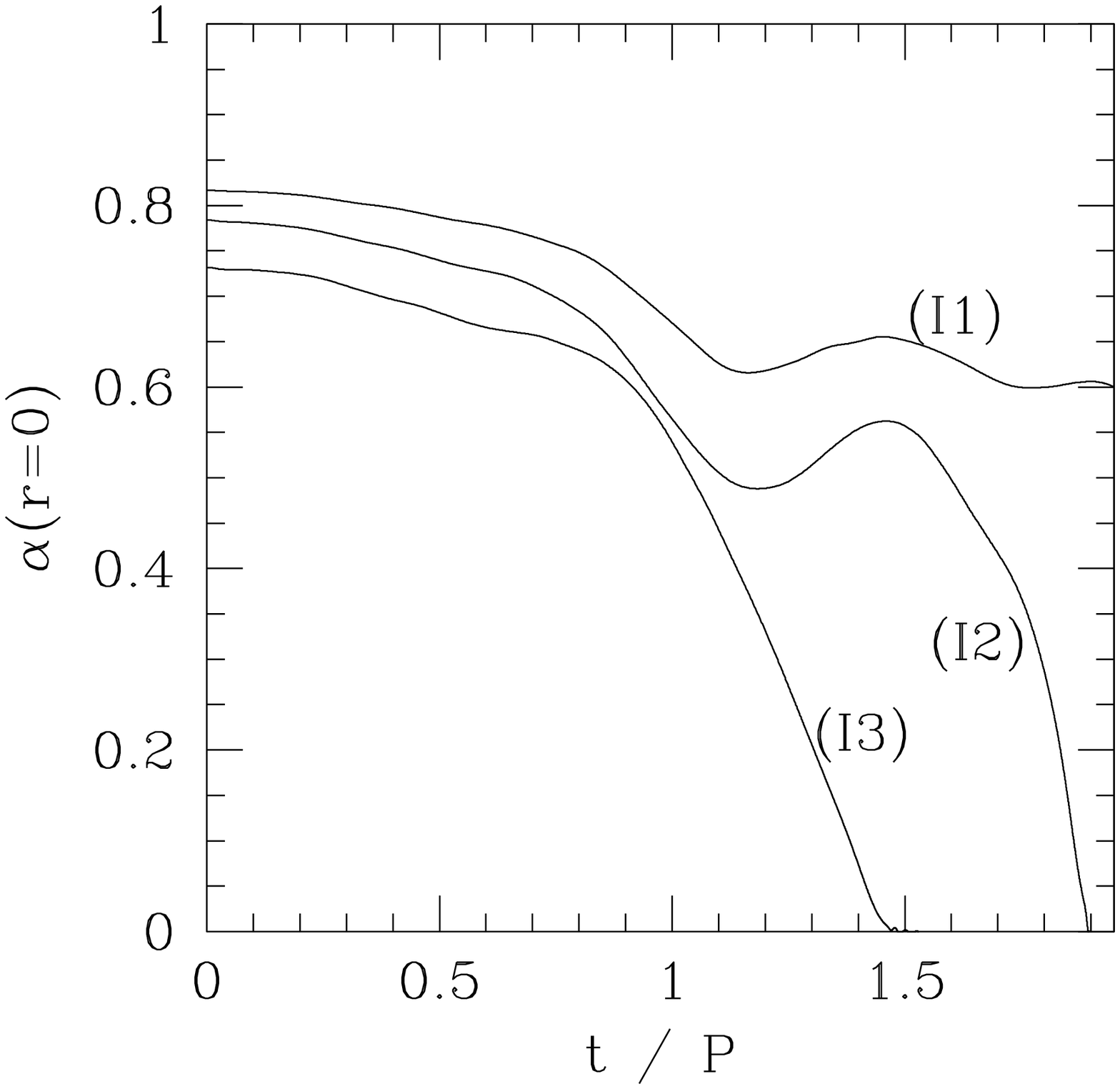}
\end{center}
\caption{$\alpha$ at $r=0$ as a function of $t/P_{\rm orb}$ 
for models (I1), (I2) and (I3). 
}
\end{figure}

\begin{figure}[t]
\begin{center}
\epsfxsize=3.5in
\leavevmode
\epsffile{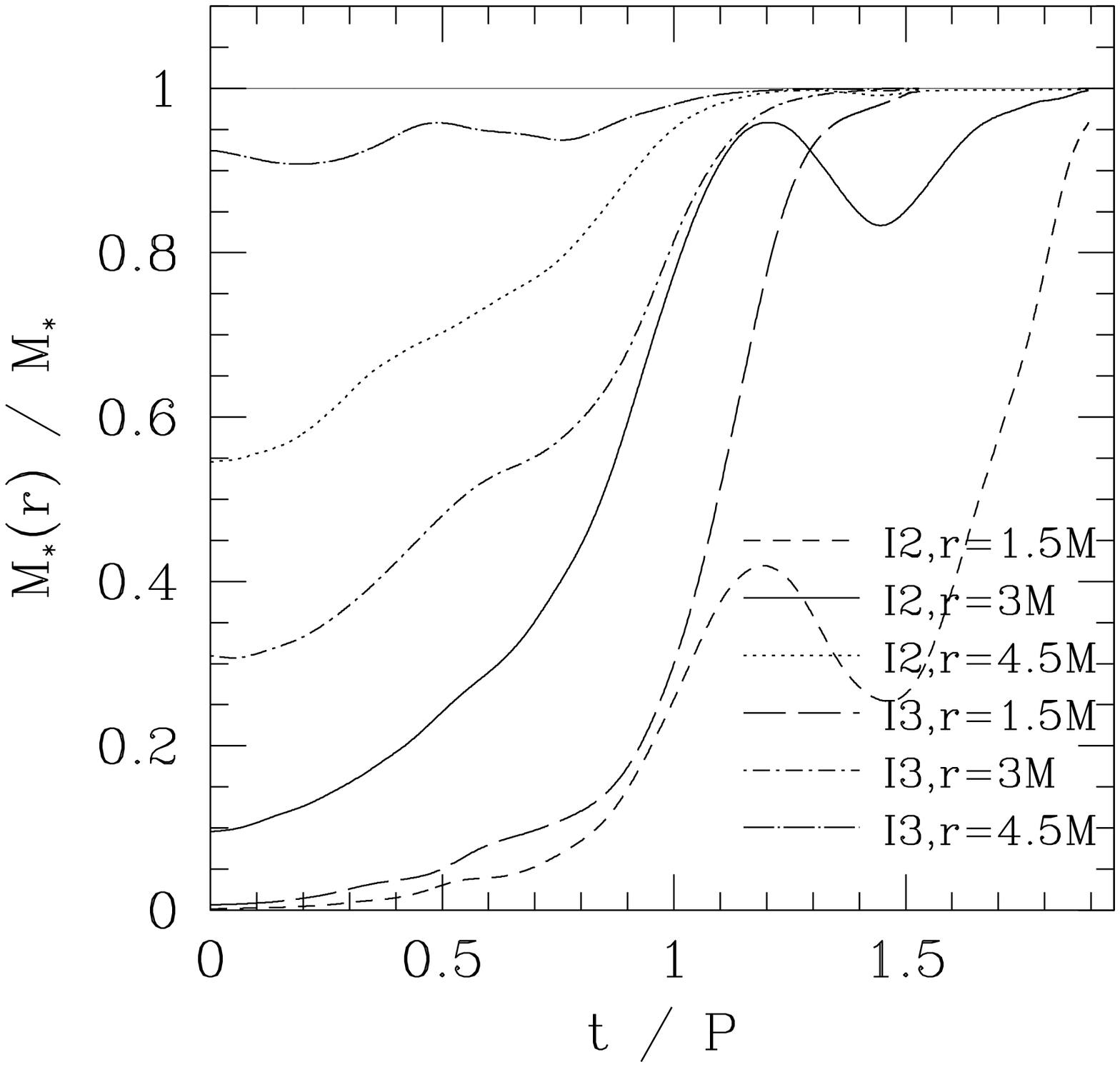}
\end{center}
\caption{Fraction of the rest mass inside a coordinate radius $r$ 
as a function of $t/P_{\rm orb}$ for models (I2) and (I3) in  which 
a black hole is formed after the merger 
}
\end{figure}

\clearpage

\begin{figure}[t]
\begin{center}
\epsfxsize=3.5in
\leavevmode
\epsffile{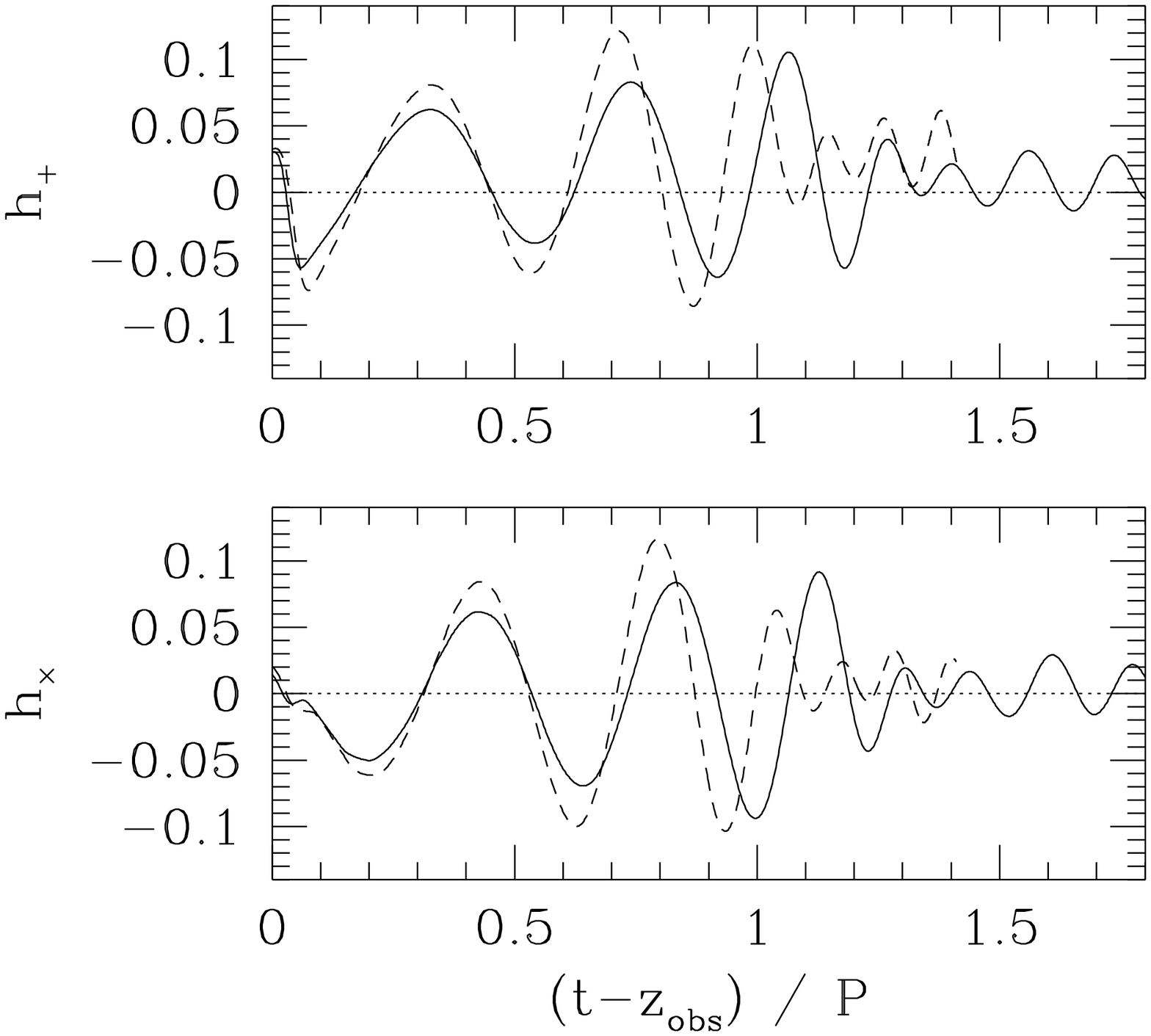}
\end{center}
\caption{$h_+$ and $h_{\times}$ as functions of retarded time 
for the corotational models (C1) (solid lines) 
and (C2) (dashed lines). 
}
\end{figure}

\begin{figure}[t]
\begin{center}
\epsfxsize=3.5in
\leavevmode
\epsffile{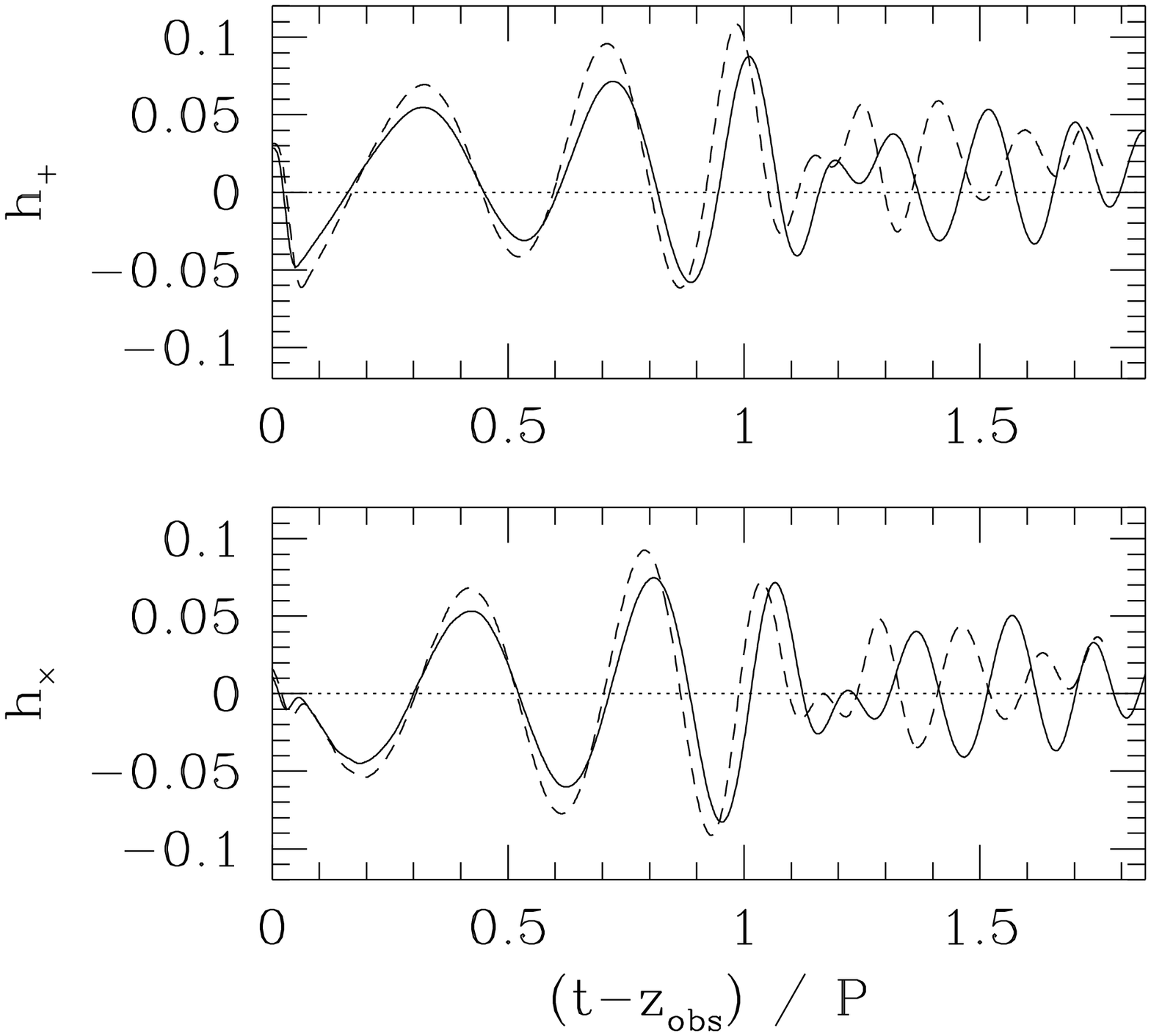}
\end{center}
\caption{$h_+$ and $h_{\times}$ as functions of retarded time 
for the irrotational models (I1) (solid lines) 
and (I2) (dashed lines). 
}
\end{figure}

\begin{figure}[t]
\begin{center}
\epsfxsize=3.5in
\leavevmode
\epsffile{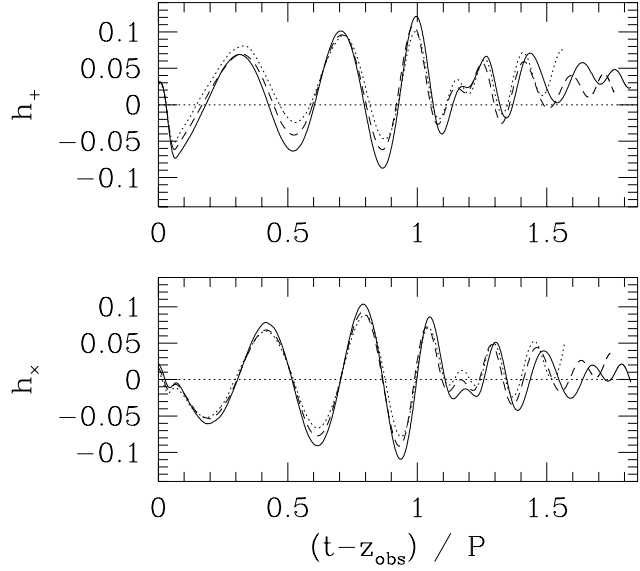}
\end{center}
\caption{$h_+$ and $h_{\times}$ as functions of retarded time 
for irrotational models (I2) with  
$293 \times 293 \times 147$ grid size (solid lines), 
$233 \times 233 \times 117$ grid size (dashed lines), and 
$193 \times 193 \times 97$ grid size (dotted lines). 
In each case, the outer boundaries (and the points where 
the waveforms are extracted) are located at 
$z\simeq 0.35$, 0.28 and $0.23 \lambda_{\rm gw}$, respectively. 
}
\end{figure}

\begin{figure}[t]
\begin{center}
\epsfxsize=5in
\leavevmode
\epsffile{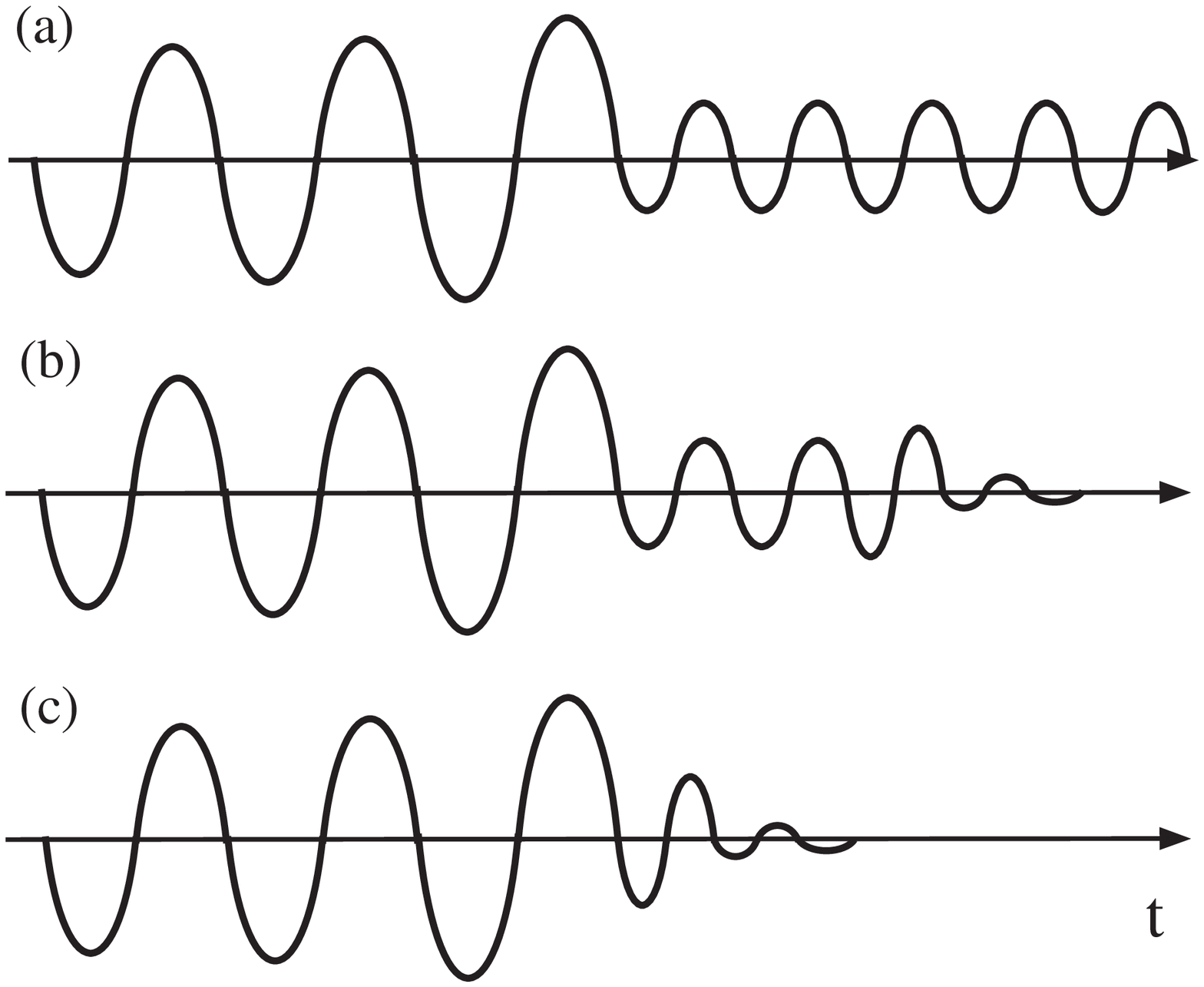}
\end{center}
\caption{Schematic pictures for expected gravitational waveforms during 
and after the merger for (a) the neutron star formation case; 
(b) the black hole formation case in which the compactness of 
the neutron stars 
before the merger is not very large and the formation timescale is 
fairly long; 
(c) the black hole formation case in which the compactness of 
the neutron stars 
before the merger is large enough that the formation timescale is 
short. 
}
\end{figure}


\begin{thebibliography}{99}

\bibitem{PULSAR} For example, J. H. Taylor, R. N. Manchester and 
G. Lyne, Astrophys. J. Suppl. {\bf 88}, 529 (1993). 

\bibitem{ST} For example, S. L. Shapiro and S. A. Teukolsky, {\em Black
Holes, White Dwarfs, and Neutron Stars}, Wiley Interscience
(New York, 1983). 

\bibitem{EOS} For example, H. Heiselberg and M. Hjorth-Jensen, nucl-th/
9902033. 

\bibitem{FIP} For example, J. L. Friedman, J. R. Ipser and 
	L. Parker, Astrophys. J. {\bf 304}, 115 (1986):\\
H. Komatsu, Y. Eriguchi and I. Hachisu, 
Mon. Not. R. Astron. Soc. {\bf 237}, 355 (1989):\\
G. Cook, S. L. Shapiro and S. A. Teukolsky, 
Astrophys. J. {\bf 422}, 227 (1994):\\
M. Salgado, S. Bonazzola, E. Gourgoulhon, 
and P. Haensel, Astron. Astrophys. {\bf 291}, 155 (1994). 


\bibitem{BSS0}
T. W. Baumgarte, S. L. Shapiro, and M. Shibata, 
Astrophys. J. Lett. (1999), in press. 

\bibitem{SF} 
J. L. Friedman and B. F. Schutz, Astrophys. J. {\bf 222}, 281 (1978) and 
references cited therein: \\
N. Stergioulas and J. L. Friedman, Astrophys. J. {\bf 444}, 306 (1995). 


\bibitem{phinney} E. S. Phinney, Astrophys. J. {\bf 380}, L17 (1991): \\
R. Narayan, T. Piran and A. Shemi, Astrophys. J. {\bf 379}, L17 (1991). 


\bibitem{LIGO}
A. Abramovici, et al. Science {\bf 256}, 325 (1992). 

\bibitem{KIP} For example, 
K. S. Thorne, in {\it Proceeding of Snowmass 
95 Summer Study on Particle and Nuclear Astrophysics and 
Cosmology}, eds. E. W. Kolb and R. Peccei (World Scientific, 
Singapore, 1995), p. 398, and references therein. 


\bibitem{piran} For example, 
R. Narayan, B. Paczynski, and T. Piran, Astrophys. J. {\bf 395}, L83
(1992): \\
M. J. Rees, in {\it proceedings of the eighteenth Texas symposium on 
relativistic astrophysics, and cosmology}, eds. A. V. Olinto, J. A. 
Frieman, and D. N. Schramm (World Scientific, 1998), p. 34:\\
H.-Th. Janka and M. Ruffert, Astron. Astrophys. {\bf 307}, L33 (1996):\\
P. Meszaros, astro-ph/9904038. 


\bibitem{grb} M. R. Metzger et al., Nature {\bf 387}, 878 (1997):\\
S. R. Kulkarni et al., Nature {\bf 393}, 35 (1998). 

\bibitem{NO} K. Oohara and T. Nakamura, Prog. Theor. Phys. 
{\bf 82}, 535, 1066 (1989);{\bf 83}, 906 (1990); {\bf 86}, 73 (1991);
{\bf 88}, 307 (1992). 

\bibitem{SNO} M. Shibata, T. Nakamura, and K. Oohara, 
Prog. Theor. Phys. {\bf 88}, 1079 (1992). 

\bibitem{RS} F. A. Rasio and S. L. Shapiro, 
Astrophys. J. {\bf 401}, 226 (1992); {\bf 432}, 242 (1994). 

\bibitem{Centrella} 
X. Zhunge, J. M. Centrella, and S. L. W. McMillan, Phys. Rev. D {\bf 50}, 
6247 (1994); {\bf 54}, 7261 (1996).

\bibitem{ruffert} 
M. Ruffert, H.-Th. Janka, and G. Sch\"afer, Astron. Astrophys. 
{\bf 311}, 532 (1996); M. Ruffert and H.-Th. Janka, {\it ibid}, 
{\bf 573} (1999). 

\bibitem{davis} M. B. Davis, W. Benz, T. Piran and F.-K. Thielemann, 
Astrophys. J. {\bf 431}, 742 (1994). 

\bibitem{supple} T. Nakamura, K. Oohara, and Y. Kojima, 
Prog. Theor. Phys. supplement {\bf 90}, 76 (1987): \\
T. Nakamura and K. Oohara, in {\it Frontiers in Numerical 
Relativity}, eds. C. R. Evans, L. S. Finn, and D. W. Hobill 
(Cambridge University Press, 1989), p. 254. 


\bibitem{ON} K. Oohara and T. Nakamura, in {\it Relativistic 
gravitation and gravitational radiation}, edited by J.-P. Lasota 
and J.-A. Marck (Cambridge University Press, Cambridge, 1997), 
p. 309. . 

\bibitem{waimo} J. A. Font, M. Miller, W.-M. Suen, and M. Tobias, 
gr-qc/9811015: M. Miller, W.-M. Suen and M. Tobias, gr-qc/9904041. 

\bibitem{gr3d} M. Shibata, Phys. Rev. D {\bf 60}, 104052 (1999). 

\bibitem{KBC} C. S. Kochanek, Astrophys. J. {\bf 398}, 234 (1992):\\
L. Bildsten and C. Cutler, Astrophys. J. {\bf 400}, 175 (1992).


\bibitem{BGM} S. Bonazzolla, E. Gourgoulhon and J.-A. Marck, 
Phys. Rev. Lett. {\bf 82}, 1892 (1999); 
in {\it  proceeding of 19th Texas symposium on relativistic 
astrophysics} to be published (gr-qc/9904040). 

\bibitem{UE} K. Ury\=u and Y. Eriguchi, submitted to Phys. Rev. D. 

\bibitem{MMW} P. Maronetti, G. J. Mathews and J. 
Wilson, Phys. Rev. D {\bf 60}, 087301 (1999). 

\bibitem{gw3p} M. Shibata, Prog. Theor. Phys. {\bf 101}, 251 (1999). 

\bibitem{gw3p2} M. Shibata, Prog. Theor. Phys. {\bf 101}, 1199 (1999). 

\bibitem{footH} The equation for the Hamiltonian constraint 
is written in the form $\tilde \Delta \psi = S_{\psi}$ where 
$\tilde \Delta $ and $S_{\psi}$ are the Laplacian with respect 
to $\tilde \gamma_{ij}$ and a function of $\tilde \gamma_{ij}$, 
$\psi$, $K_{ij}$, $\rho$, $P$ and $\alpha u^0$. We define a 
function $f_{\psi}=|\tilde \Delta \psi-S_{\psi}|/
(|\tilde \Delta \psi|+|S_{\psi}|)$ for measuring the 
violation of the Hamiltonian constraint. We have found that 
$f_{\psi}$ is typically less than $0.1$ for a 
 region in which $\rho_*$ is larger than 
$\sim 10^{-3} \rho_{*~{\rm max}}$. For the less dense region, however, 
$f_{\psi}$ often becomes $O(1)$ because such a low density region is 
not well resolved in our finite differencing scheme for the hydrodynamic 
equations. 


\bibitem{footJ} The rest mass is conserved because of no mass ejection. 
The angular momentum decreases by $5-10\%$ 
in the whole evolution, and the total amount of the 
decrease roughly agrees with the angular momentum emission 
in gravitational waves within $5\%$ error (see Eq. (4.2)). 

\bibitem{S} M. Shibata, Phys. Rev. D {\bf 55}, 2002 (1997).

\bibitem{foot2} The definition of differential rotation depends on the 
coordinate condition. Strictly speaking, we have found that the 
new massive neutron star is differentially rotating in our 
present gauge, and the rotation law found in this paper 
could change slightly 
if it is defined in the stationary axisymmetric gauge 
used in \cite{FIP}. 

\bibitem{footnote} 
Note that we recompute the constraint equations
whenever we modify the initial quasi-equilibrium configurations. 


\bibitem{BCSST} T. W. Baumgarte, G. B. Cook, M. A. Scheel, S. L. 
        Shapiro and S. A. Teukolsky, 
	Phys. Rev. D {\bf 57}, 6181 (1998).

\bibitem{irre} For example, 
M. Shibata, Phys. Rev. D {\bf 58}, 024012 (1998), 
and references therein. 

\bibitem{BSS} Actually, it is possible to construct differentially 
rotating neutron stars of such a large mass, which are 
dynamically (but not always secularly) stable 
against gravitational collapse and bar mode deformation \cite{BSS0}; 
M. Shibata, T. W. Baumgarte and S. L. Shapiro, in preparation.  

\bibitem{Stu} S. L. Shapiro, Phys. Rev. D {\bf 58}, 103002. 

\bibitem{NNS} 
The new, massive neutron stars may be secularly unstable 
to becoming a black hole 
on a long timescale even if the effect of 
gravitational radiation is small. The reason is that they are 
differentially rotating and supra-massive, 
which implies if we take into account the effects of viscosity or 
magnetic fields, angular momentum will be transported outward 
or dissipated, and eventually 
they may become unstable to gravitational collapse. 
See also \cite{BSS0}. 

\bibitem{Nakamura} 
T. Nakamura, Prog. Theor. Phys. {\bf 65}, 1876 (1981): \\
R. F. Stark and T. Piran, Phys. Rev. Lett. {\bf 55}, 891 (1985). 

\bibitem{foot5} If the time duration is fairly long, we may be 
able to observe a peak around the oscillation frequency in 
the Fourier spectrum of gravitational waves as pointed out 
in \cite{Centrella}. 
On the other hand, we will not find the peak if the time duration 
is short. Thus, we may say that the amplitude of the peak in 
the Fourier space also provides important information. 

\bibitem{CF} C. Cutler and E. E. Flanagan, Phys. Rev. D {\bf 49}, 2658 
(1994);
E. Poisson and C. M. Will, {\it ibid} {\bf 52}, 848 (1995). 

\bibitem{BIWW} L. Blanchet, B. R. Iyer, C. M. Will and 
A. G. Wiseman, Class. Quant. Grav. {\bf 13}, 575 (1996). 

\end{thebibliography}
\end{document}